\documentclass[iop, twocolappendix, numberedappendix]{emulateapj}
\usepackage[breaklinks,colorlinks,citecolor=blue,linkcolor=red]{hyperref}
\usepackage{etoolbox}

\makeatletter

\patchcmd{\NAT@citex}
  {\@citea\NAT@hyper@{%
     \NAT@nmfmt{\NAT@nm}%
     \hyper@natlinkbreak{\NAT@aysep\NAT@spacechar}{\@citeb\@extra@b@citeb}%
     \NAT@date}}
  {\@citea\NAT@nmfmt{\NAT@nm}%
   \NAT@aysep\NAT@spacechar\NAT@hyper@{\NAT@date}}{}{}

\patchcmd{\NAT@citex}
  {\@citea\NAT@hyper@{%
     \NAT@nmfmt{\NAT@nm}%
     \hyper@natlinkbreak{\NAT@spacechar\NAT@@open\if*#1*\else#1\NAT@spacechar\fi}%
       {\@citeb\@extra@b@citeb}%
     \NAT@date}}
  {\@citea\NAT@nmfmt{\NAT@nm}%
   \NAT@spacechar\NAT@@open\if*#1*\else#1\NAT@spacechar\fi\NAT@hyper@{\NAT@date}}
  {}{}

\makeatother
\usepackage{booktabs}
\usepackage{xspace}
\usepackage{apjfonts}
\usepackage{amssymb, amsmath}
\usepackage{natbibspacing, natbib}
\usepackage{aas_macros}
\usepackage{graphics,graphicx}
\newcommand{\pmOne}{\mbox{$^{-1}$}\xspace}
\newcommand{\rarr}{$\rightarrow$}
\newcommand{\sig}{$\sigma$}

\newcommand{\aco}{\mbox{CO\,($J$\,=\,1\,\rarr\,0) }}

\newcommand{\cco}{\mbox{CO\,($J$\,=\,3\,\rarr\,2) }}

\newcommand{\eco}{\mbox{CO\,($J$\,=\,5\,\rarr\,4) }}
\newcommand{\jco}{\mbox{CO\,($J$\,=\,10\,\rarr\,9) }}

\newcommand{\cii}{[C{\scriptsize II}]\xspace}

\newcommand{\ciiline}{[C{\scriptsize II}]~($^2$P$_{3/2}$\,\rarr\,$^2$P$_{1/2}$)\xspace}

\newcommand{\mgas}{\mbox{$M_{\rm gas}$}\xspace}
\newcommand{\mdyn}{\mbox{$M_{\rm dyn}$}\xspace}

\newcommand{\mstar}{\mbox{$M_*$}\xspace}
\newcommand{\Msun}{\mbox{$M_{\odot}$}\xspace}

\newcommand{\Lsun}{\mbox{$L_{\odot}$}\xspace}
\newcommand{\LIR}{\mbox{$L_{\rm IR}$}\xspace}
\newcommand{\LFIR}{\mbox{$L_{\rm FIR}$}\xspace}
\newcommand{\kms}{km\,s$^{-1}$\xspace}
\newcommand{\bmm}{beam$^{-1}$\xspace}
\newcommand{\LpU}{\mbox{K\,\,km\,\,s$^{-1}$\,\,pc$^2$}\xspace}
\newcommand{\alphaU}{\mbox{$M_{\odot}~($K\,\,km\,\,pc$^2)^{-1}$}\xspace}

\newcommand{\Lp}{\mbox{$L^{\prime}_\textrm{CO(1-0)}$}\xspace}
\newcommand{\Lcii}{\mbox{$L_{\rm [CII]}$}\xspace}
\newcommand{\alphaco}{$\alpha_{\rm CO}$}

\newcommand{\E}[1]{$\times10^{#1}$}
\newcommand{\petm}[2]{$^{+#1}_{-#2}$}
\newcommand{\eq}{\,=\,}
\newcommand{\ssim}{\,$\sim$\,}
\newcommand{\pmm}{\,$\pm$\,}
\newcommand{\Fig}[1]{Figure~\ref{fig:#1}}

\newcommand{\Tab}[1]{Table~\ref{tab:#1}}
\newcommand{\Sec}[1]{\S\ref{sec:#1}}
\newcommand\tna{\,\tablenotemark{a}}

\newcommand\tnb{\,\tablenotemark{b}}
\newcommand\tnc{\,\tablenotemark{c}}
\newcommand\tnd{\,\tablenotemark{d}}
\newcommand\tne{\,\tablenotemark{e}}
\newcommand\tnf{\,\tablenotemark{f}}
\newcommand\tng{\,\tablenotemark{g}}
\newcommand\tnh{\,\tablenotemark{h}}
\newcommand\tni{\,\tablenotemark{i}}
\newcommand\tnj{\,\tablenotemark{j}}
\newcommand\tnk{\,\tablenotemark{k}}

\def\pdbi     {Plateau de Bure Interferometer\xspace}
\def\carma    {Combined Array for Research in Millimeter-wave Astronomy\xspace}

\newcommand{\ncode}[1]{{\sc #1}}

\def\casa {\ncode{casa}\xspace}
\newcommand{\z}{$z$\xspace}
\newcommand{\obs}{observations\xspace}

\newcommand{\highz}{high-$z$\xspace}

\newcommand{\SF}{star formation\xspace}

\newcommand{\SB}{starburst\xspace}

\newcommand{\xa}{HXMM05\xspace}

\slugcomment{Accepted to the ApJ}

\makeatletter
\renewcommand\normalsize{\@setfontsize\normalsize\@xpt{12.5}}
\makeatother

\shorttitle{The ISM Properties and Gas Kinematics of a $z$\ssim3 DSFG}
\shortauthors{Leung et al.}

\begin{document}
\title{
The ISM Properties and Gas Kinematics of a Redshift 3 Massive Dusty Star-forming Galaxy
}

\author{T. K. Daisy Leung\altaffilmark{1, 2}}
\author{Dominik A.\ Riechers\altaffilmark{1}}
\author{Andrew J. Baker\altaffilmark{3}}
\author{Dave L. Clements\altaffilmark{4}}
\author{Asantha Cooray\altaffilmark{5}}
\author{Christopher C. Hayward\altaffilmark{2}}
\author{R. J. Ivison \altaffilmark{6, 7}}
\author{Roberto Neri\altaffilmark{8}}
\author{Alain Omont\altaffilmark{9}}
\author{Ismael P\'erez-Fournon\altaffilmark{10, 11}}
\author{Douglas Scott\altaffilmark{12}}
\author{Julie L. Wardlow\altaffilmark{13, 14}}

\email{tleung@astro.cornell.edu}
\altaffiltext{1}{Department of Astronomy, Space Sciences Building, Cornell University, Ithaca, NY 14853, USA}
\altaffiltext{2}{Center for Computational Astrophysics, Flatiron Institute, 162 Fifth Avenue, New York, NY 10010, USA}
\altaffiltext{3}{Department of Physics and Astronomy, Rutgers, the State University of New Jersey,
136 Frelinghuysen Road, Piscataway, NJ, 08854-8019}
\altaffiltext{4}{Astrophysics Group, Imperial College London, Blackett Laboratory, Prince Consort Road, London SW7 2AZ, UK}
\altaffiltext{5}{Department of Physics and Astronomy, University of California, Irvine, CA 92697, USA}
\altaffiltext{6}{European Southern Observatory, Karl-Schwarzschild-Stra{\ss}e 2, D-85748 Garching, Germany}
\altaffiltext{7}{Institute for Astronomy, University of Edinburgh, Royal Observatory, Blackford Hill, Edinburgh EH9 3HJ, UK}
\altaffiltext{8}{Institut de Radioastronomie Millim\'etrique (IRAM), 300 Rue de la Piscine, Domaine Universitaire de Grenoble,
38406 St. Martin d'H\`eres, France}
\altaffiltext{9}{Institut d'Astrophysique de Paris, Centre national de la recherche scientifique (CNRS) \& Universit\'e Pierre et Marie Curie (UPMC),
98 bis boulevard Arago, 75014 Paris, France}
\altaffiltext{10}{Instituto de Astrofisica de Canarias, E-38200 La Laguna, Tenerife, Spain}
\altaffiltext{11}{Departamento de Astrofisica, Universidad de La Laguna, E-38205 La Laguna, Tenerife, Spain}
\altaffiltext{12}{Department of Physics and Astronomy, University of British Columbia, 6224 Agricultural Road, Vancouver, BC V6T 1Z1, Canada}
\altaffiltext{13}{Centre for Extragalactic Astronomy, Department of Physics, Durham University, South Road, Durham, DH1 3LE, UK}
\altaffiltext{14}{Physics Department, Lancaster University, Lancaster, LA1 4YB, UK}

\begin{abstract}
We present CO\,($J$\eq1\rarr0; 3\rarr2; 5\rarr4; 10\rarr9) and 1.2-kpc resolution \cii line \obs
of the dusty star-forming galaxy (SFG) \xa\ ---
carried out with the Karl G. Jansky Very Large Array, the Combined Array for Research in Millimeter-wave Astronomy,
the Plateau de Bure Interferometer, and the Atacama Large Millimeter/submillimeter Array, measuring an unambiguous redshift of \z\eq2.9850\pmm0.0009.
We find that \xa is a hyper-luminous infrared galaxy (\LIR$=$\,(4\pmm1)\E{13}\,\Lsun)
with a total molecular gas mass of (2.1\pmm0.7)\E{11}(\alphaco/0.8)\,\Msun.
The \aco and \cii emission are extended over $\sim$9\,kpc in diameter, and the CO line FWHM exceeds 1100\,\kms.
The \cii emission shows a monotonic velocity gradient consistent with a disk,
with a maximum rotation velocity of $v_{\rm c}$\eq616\pmm100\,\kms and a dynamical
mass of (7.7\pmm3.1)\E{11}\,\Msun.
We find a star formation rate (SFR) of 2900\petm{750}{595}\,\Msun\,yr\pmOne.
\xa is thus among the most intensely star-forming galaxies known at high redshift.
Photo-dissociation region modeling suggests physical conditions similar to nearby SFGs, showing
extended \SF, which is consistent with our finding that the gas and dust emission are co-spatial.
Its molecular gas excitation resembles the local major merger Arp\,220.
The broad CO and \cii lines and a pair of compact dust nuclei
suggest the presence of a late-stage major merger at the center of the extended disk, again reminiscent of Arp\,220.
The observed gas kinematics and conditions together with the presence of a companion and the pair of nuclei
suggest that \xa is experiencing multiple mergers as a part of the evolution.
\end{abstract}

\section{Introduction}   \label{sec:intro}
Most of the stellar mass in the Universe is assembled in the first few billion years of cosmic time, in the redshift range
1$\lesssim$\,$z$\,$\lesssim$3
 \citep[see e.g., review by][]{Madau14a}.
 Galaxies at this epoch typically have higher star formation rates (SFRs) compared to the present day.
 Among the \highz galaxy populations discovered, dusty star-forming galaxies (DSFGs) represent
 the most IR-luminous
 systems at this peak epoch.
 They are typically gas-rich, with molecular gas masses exceeding \mgas$=$10$^{10}$\,\Msun
 and IR luminosities exceeding those of nearby (ultra-)luminous infrared galaxies
 (U/LIRG; \LIR$>$10$^{11-13}$\,\Lsun; see reviews by \citealt{CW13, Casey14a}).
 Given the differences found between nearby ULIRGs and \highz DSFGs \citep[e.g.,][]{Younger10a,
 Rujopakarn11a, Rujopakarn13a}, studying their interstellar medium (ISM) properties, gas dynamics, and star-forming environments directly are
 essential to understanding how galaxies are initially assembled at early epochs.

 In the classical model of disk galaxy formation \citep{Fall80a},
 disk galaxies form out of the gas that is cooling off from the hot halos associated with
 dark matter (DM) potential wells while maintaining the specific angular momentum as the gas
 settles into rotationally supported disks \citep{Mo98a}.
 The structure and dynamics of disk galaxies are therefore closely correlated
 with the properties of their parent DM halos. Probing the structure and dynamics of disk galaxies at high redshift
 can thus inform us about the processes driving the assembly history of galaxies at early cosmic times.
 For instance, by tracing the gas dynamics, the Tully-Fisher relation \citep{Tully77a},
 which links the angular momentum of the parent DM halo of a disk
 galaxy with the luminosity/mass of its stellar populations, can be studied out to earlier epochs.
 Past \obs have led to two physical pictures for the nature and origin of DSFGs: compact
 irregular starbursts resulting from major mergers (of two or more disks) and extended disk-like galaxies with high SFRs
 \citep[e.g.,][]{Tacconi06a, Tacconi08a, Shapiro08a, Engel10a, Riechers10a, Ivison10d, Ivison11a, Riechers11a, Riechers11b, Hodge12a, Riechers13a,
 Ivison13a, Bothwell13a, Riechers14a, Hodge15a, Oteo16b, Riechers17a}
 resulting from minor mergers and/or cold gas accreted
 from the intergalactic medium (IGM; also known as cold mode accretion; CMA; e.g., \citealt{Keres05a, Dekel09a, Dave10a}).
 However, as individual DSFGs can fall into either physical picture, a third interpretation is
 that DSFGs are a heterogeneous population composed of both compact
 starbursts and extended disks (e.g., \citealt{Hayward13a}), presumably observed at different stages of evolution.
 Determining their gas kinematics is therefore key to better understanding their formation mechanisms and shedding light on
 whether major mergers or continuous accretion dominate and sustain their intense \SF.
 However, such studies require high spatial resolution and sensitivity in order to
 image their gas reservoirs, and thus,
 are relatively expensive to carry out.
 To date, only a handful of \highz galaxies have been mapped in their molecular gas at high resolution,
 revealing a mixture of rotating disks and galaxy mergers \citep[e.g.,][]{Swinbank11a, Hodge12a, Ivison13a, Oteo16a, Oteo17a, Oteo18a}.

 With the goal to better understand the star-forming conditions and the gas dynamics of \highz DSFGs,
 we observed multi-$J$ CO and
 \cii line emission in HerMES~J022547-041750 (\xa; RA, Dec \eq02$^{\rm h}$25$^{\rm m}$47$^{\rm s}$, $-$04\degr17\arcmin50\arcsec; J2000),
 one of the brightest DSFGs known, at $\lesssim$\,0\farcs15 resolution.
 Line emission from different rotational transitions of CO is useful for determining molecular gas mass and physical properties
 of the ISM. The \ciiline fine-structure line at rest-frame157.7\,$\micron$ is one of the brightest emission lines in star-forming galaxies, and can contribute up to
 1\% of the FIR luminosity of galaxies \citep{Malhotra97a, Nikola98a, Colbert99a}.
 In addition, \cii and \aco line emission trace similar gas kinematics in nearby star-forming galaxies
 (e.g., \citealt[][]{Mittal11a, Braine12a, Kramer13a, Pineda13a}),
 making the former a powerful probe of \highz gas kinematics,
 especially when paired with the exceptional capabilities of the Atacama Large Millimeter/submillimeter Array (ALMA).

 The target \xa was discovered in the
 {\it Herschel} Multi-tiered Extragalactic Survey (HerMES; \citealt{Oliver12a})
 as one of 29 \highz strongly-lensed galaxy candidates identified \citep{Wardlow13a, Bussmann15a}.
 The parent sample was selected based on a flux density threshold of
 $S_{\rm 500}\geq$80\,mJy at 500\,$\micron$.
 The surface density of such bright DSFGs is (0.31\pmm0.06)\,deg$^{-2}$ (\citealt{Wardlow13a}).
 Previous high-resolution imaging obtained with the {\it Hubble Space Telescope (HST)} and ALMA
 and lens modeling of $0\farcs4$ resolution dust continuum data at 870\,$\micron$
 show that \xa is at most weakly lensed, with magnification factor
 $\mu_{870} \lesssim1.4$
 \citep{Bussmann15a}\footnote{The orientation of the {\it HST} image of \xa shown in Figure 3 of \citet{Calanog14a, Bussmann15a}
 is incorrect (i.e., North is down instead of up), but the correct locations of all galaxies were used in the lens modeling.}.
 \xa is therefore intrinsically extremely IR-luminous, unlike other typically strongly-lensed DFSGs in the parent sample with
 similar sub-millimeter flux densities.
 \citet{Bussmann15a} find a total of three unlensed, intrinsically-bright DSFGs out of the parent sample of 29.
 This yields a surface density of $\sim$0.03\,deg$^{-2}$ for such sources, which makes them even rarer than
 strongly-lensed DSFGs.
 \xa therefore belongs to a rare and understudied luminous/massive \highz galaxy population.
 Currently, the general consensus is these unlensed DSFG with $S_{\rm 500}\gtrsim$\,100\,mJy appear to be
 predominantly
 major galaxy mergers (e.g., HXMM01 and G09v124; \citealt{Fu13a, Ivison13a}).
 In this work, we investigate the nature of \xa\ --- to examine whether it is a dispersion-dominated merger, or an isolated HyLIRG.
 We securely determine its redshift to be $z$\eq2.9850 through multi-$J$ CO and \cii line \obs,
 indicating that \xa is near the peak epoch of cosmic star formation.

 This paper is structured as follows.
 In \Sec{obs}, we summarize the observations and procedures used to reduce the data.
 We also briefly describe the ancillary data used in our analysis.
 In \Sec{results}, we present the main results.
 In \Sec{anal}, we present the results from spectral energy distribution (SED) modeling and
 dynamical modeling of the \cii line data using the tilted-ring and ``envelope''-tracing methods.
 In \Sec{diss}, we discuss the properties of \xa and
 compare them to those of other galaxy populations.
 We discuss the key implications of our findings in \Sec{implication}, and
 summarize the main results and present our conclusions in \Sec{sum}.
 Throughout this paper, we use a concordance
 $\Lambda$CDM cosmology, with
 parameters from the WMAP9 results:
 $H_0$ = $69.32$ \kms Mpc\pmOne, $\Omega_{\rm M}$ = $0.29$, and
 $\Omega_{\Lambda}$ = $0.71$ \citep{Hinshaw13a}.

 \section{Observations and Ancillary Data}  \label{sec:obs}
 \begin{turnpage}
 \def\arraystretch{1.2}
 \begin{deluxetable*}{cccccccccc}
 \tabletypesize{\scriptsize}
 \tablecaption{\xa Observations Summary.
 \label{tab:obssum}
 }
 \tablehead{
 Observation                     &
 Telescope                       &
 Date                            &
 $t_{\rm on}$                   &
 \multicolumn{3}{c}{Calibrators} &
 $\Omega_{\rm beam} $\tna        &
 Array config.                   &
 $\nu$
 \\
 \cline{5-7}  
 \\
                                      &
                                      &
                                      &
 (hr)                                 &
 Bandpass                     &
 Flux                              &
 Gain                             &
 (major $\times$ minor, PA)      &
                                      &
 (GHz)
     \\
 (1) & (2) & (3) & (4) & (5) & (6) & (7) & (8) & (9) & (10)
 }
 \startdata
 870\,$\micron$  &  SMA      & 25-Aug-2010 & 1.08 & 3C84 & Uranus & J0238$+$166, J0217$+$017  & 0\farcs99\,$\times$\,0.78, $-$68.2$\degr$   & Subcompact & 342.017    \\
             & & 25-Sep-2010  & 2.06 & 3C454.3 & Uranus, Callisto & --- &  ---   & Extended & 342.003     \\
             & & 05-Aug-2011& 2.93 & 3C454.3 & Uranus, MWC349A & ---  & ---  & Extended & 340.224    \\
 \cco  & CARMA &  02-Sep-2010 & 1.57 & 3C84     & Uranus              & J0239$-$025  & 7\farcs68\,$\times$\,5\farcs00, $-$53$\degr$   & D    &  89.9117  \\
              & & 03-Sep-2010 & 2.43 &      ---      &  ---   &  ---   &        ---             &  --- &   93.5887    \\
              & & 04-Sep-2010 & 2.09 &      ---      &  ---   &  ---   &        ---             &  --- &   103.3739  \\
              & & 05-Sep-2010 & 2.33 &      ---      &  ---   &  ---   &         ---             & ---  &   89.9115   \\
              & & 12-Sep-2010 & 1.65 &      ---      &  ---   &  ---   &        ---            &  --- &   107.0499   \\
 2.3\,mm   &  PdBI & 21-Sep-2010 & 0.8 &   B0215+015 & MWC349               & B0215+015 & 14\farcs85\,$\times$\,2\farcs59, $-$35.8$\degr$   &  D  &  131.1 \\
 \eco      & &  23-Sep-2010 & 0.6 & 3C454.3  & ---               & --- &    ---       &                                                                    ---   & 145.4 \\
          & & 26-Sep-2010 & 0.6 &       ---  &  ---      & --- &  7\farcs43\,$\times$\,4\farcs08, 111.2$\degr$ &  ---   &  145.4  \\
 \aco   & VLA & 20-Sep-2014  & 1.40 & 3C147 &  3C147        &  J0215$-$0222  &  1\farcs21$\times$\,0\farcs80, 36$\degr$   &  DnC  & Ka-band  \\
         && 17-Nov -- 11-Dec-2014 & 8.96 &    J0542$+$4951 & J0542$+$4951 & --- & ---   & C  & --- \\
 \cii  & ALMA 12\,m\tnb & 15-Jun-2015 & 0.15 & J2232$+$1143 & Ceres        & J0241$-$0815 & 0\farcs18\,$\times$\,0\farcs14, 61.3$\degr$   &  21$-$784 [m]  &  472.661 \\
         &  & 27-Aug-2015 & 0.15 &  J0224$+$0659 & J0224$+$0659 & J0241$-$0815 &        ---             &  15$-$1574 [m] &   472.665    \\
 \jco  & ALMA ACA\tnc & 11-Sep-2017 & 0.45 & J0006$-$0623 & Uranus & J0217$+$0144 & 5\farcs35\,$\times$\,3\farcs65, $-$85$\degr$ & ACA & 289.995  \\
          & & 16-Sep-2017 & 0.45 & J0522$-$3627 & --- & --- & --- & --- & ---
 \enddata
 \tablenotetext{a}{Synthesized beam size obtained with ``natural'' weighting and after combining all tracks of the same spectral setup.}
 \tablenotetext{b}{Cycle-2 data. For observation details of the ALMA cycle-0 data, see \citet{Bussmann15a}.}
 \tablenotetext{c}{Cycle-4 data. }
 \tablecomments{Columns: (1) line or continuum wavelength observed; (2) telescope; (3) date of \obs; (4) on-source time; (5$-$7)
 calibrators; (8) clean beam size (untapered); (9) array configuration or baseline range; (10) local oscillator frequency for \obs obtained with the SMA, CARMA, and ALMA, or
 observed frequency for \obs made with the PdBI and VLA.}
 \end{deluxetable*}
 \end{turnpage}

 \subsection{\carma (CARMA) \cco}
 Based on the {\it Herschel}/SPIRE multi-band colors of $S_{\rm 500} < S_{\rm 250} < S_{\rm 350}$, we expected
 the redshift of \xa to be 2\,$\lesssim z \lesssim$\,3.5, and its \cco line --- at rest-frame frequency $\nu_{\rm rest}$\eq345.79599\,GHz ---
 to be redshifted into the 3\,mm receiver window of CARMA.
 We therefore performed a blind CO line search in \xa with CARMA
 in the D array configuration.
 Five tracks were executed under excellent
 weather conditions between 2010 September 02 and 21 (Program ID: cx310; PI: D. Riechers).
 A total of 10.1\,hours of on-source time was obtained after combining all data.
 We scanned the 3\,mm window using four distinct frequency setups, covering a frequency range of $\nu_{\rm obs}$\eq84.98$-$111.97\,GHz.
 For each setup, the correlator provided sixteen spectral windows, each with a bandwidth of 494.792\,MHz
 and 95 channels, resulting in an effective bandwidth of 3.75\,GHz per sideband after accounting for overlapping edge channels.
 This correlator setup provides a
 spectral resolution of $\Delta\nu$\eq5.208\,MHz (i.e., $\Delta v$\eq18\,\kms at $\nu_{\rm obs}$\eq86.8\,GHz).
 All tracks used the same calibrators, as summarized in \Tab{obssum}.
 We estimate a flux calibration accuracy of $\sim$15\%.

 The \ncode{miriad} package was used to calibrate the visibility data.
 The calibrated visibility data were
 imaged and deconvolved using the CLEAN algorithm with natural weighting, yielding a synthesized
 beam size of 7\farcs68\,$\times$\,5\farcs00, at a position angle (PA) of $-$53\degr.
 The final rms noise is typically $\sigma_{\rm ch}$\eq2.26\,mJy\,\bmm over a
 channel width of 90\,\kms.
 We form four continuum images at $\nu_{\rm cont}$\eq90, 93.4, 103, and 107\,GHz,
 by averaging across the line-free channels in each setup (i.e., one per spectral tuning).
 The final rms of the continuum images are $\sigma_{\rm cont}$\eq0.17, 0.37, 0.33, and 0.43\,mJy\,\bmm, respectively.

 \subsection{\pdbi (PdBI) \eco and 131 GHz Continuum} \label{sec:pdbi}
 We detected a single line in the CARMA data (see \Sec{co}).
 Based on the SPIRE colors, the line is most likely CO($J$\eq3\rarr2),
 suggesting a redshift of $z\approx$\,2.985 for \xa.
 This redshift was spectroscopically confirmed through the detection of a second CO line, which was observed with IRAM
 PdBI (Program ID: U--3; PI: N. Fiolet).
 Based on the redshift suggested by the CARMA data, we expected the \eco line
 ($\nu_{\rm rest}$\eq576.26793\,GHz) to be redshifted to an observed frequency of $\nu_{\rm obs}$\eq144.6093\,GHz.
 Observations were carried out in good weather conditions in the
 D array configuration with six antennas on
 2010 September 23 and 26.
 A total on-source time of 1.4\,hours was obtained in the combined tracks.
 The 2\,mm receivers were used to cover the expected frequency of the \eco line and the underlying continuum.
 The WideX correlator was used,
 providing a spectral resolution of 1.95\,MHz (about 4\,\kms at $\nu_{\rm obs}$) over an effective
 bandwidth of 3.6\,GHz, in dual polarization mode.
 Calibrators used for bandpass, flux, and complex gain calibrations are listed in \Tab{obssum}.
 We estimate a flux calibration accuracy of 15\%.

The \ncode{gildas} package was used to calibrate and analyze the visibility data.
The calibrated visibility data were imaged and deconvolved using the CLEAN algorithm with natural
weighting, yielding a synthesized beam of 7\farcs43\,$\times$\,4\farcs08 at PA\eq111$\degr$.
The final rms noise is 5.53\,mJy\,\bmm over 20\,MHz (41.3\,\kms).
A continuum image at an average frequency of $\nu_{\rm cont}$\eq145.4\,GHz 
was produced by averaging over the line-free channels ($\Delta \nu$\eq3.12\,GHz), yielding
an rms noise of 0.44\,mJy\,\bmm.

We also observed the $\nu_{\rm obs}$\eq131.1\,GHz continuum emission in \xa with the PdBI (Program ID: U--3; PI: N. Fiolet) to rule out an alternative redshift option.
Observations were carried out on 2010 September 21 under good weather conditions in the D array configuration
for 0.6 hours of on-source time (\Tab{obssum}).
The visibility data were calibrated using \ncode{gildas}. Imaging and deconvolution were performed using
the CLEAN algorithm with natural weighting.
We formed a continuum image by averaging across all channels within an effective bandwidth of 3.6\,GHz,
reaching an rms of $\sigma_{\rm cont}$\eq0.21\,mJy\,\bmm and a beam size of 14\farcs85\,$\times$\,2\farcs59 at PA\eq$-$36$\degr$.

\subsection{NSF's Karl G. Jansky Very Large Array (VLA) \aco}
Based on the redshift determined from the CO($J$\eq3\rarr2) and \eco lines,
we targeted the \aco line ($\nu_{\rm rest}$\eq115.27120\,GHz) in \xa using the
the VLA, for a total of ten observing sessions (Program ID: 14B-302;
PI: S. Bussmann).
One session was carried out on 2014 September 20 in the DnC array configuration and the remaining 
nine sessions were carried out between
2014 November 17 and December 11 in the C array configuration,
A total of 10.5\,hours of on-source time was obtained in the combined ten sessions.
The Ka-band receivers were used to cover the redshifted \aco line.
The WIDAR correlator was used in full polarization mode, providing a total bandwidth of
2\,GHz covered by sixteen sub-bands, each with a bandwidth of 128\,MHz
and a channel spacing of 2\,MHz (29\,\kms).
Calibrators are listed in \Tab{obssum}.
We estimate a flux calibration accuracy of $\lesssim$15\%.

Visibility data were calibrated and analyzed using version 4.7.1 of
the \casa package. We combined all calibrated data and imaged
the visibilities using the CLEAN algorithm with natural weighting to maximize sensitivity, yielding
a synthesized beam size of 1\farcs21$\times$\,0\farcs80 at PA\eq36$\degr$.
The final rms noise is 0.041\,mJy\,\bmm over 6\,MHz (62\,\kms), or 0.028\,mJy\,\bmm
per $\Delta v$\eq145\,\kms velocity bin.  
A continuum image at $\nu_{\rm cont}$\eq31.27\,GHz
was produced by averaging over all the line-free channels, yielding an rms
noise of $\sigma_{\rm cont}$\eq3.19\,$\mu$Jy\,\bmm. 
To examine the kinematics of the \aco line emission at higher resolution,
we made an additional line cube using Briggs weighting with robustness $R$\eq0.5.
An rms noise of $\sigma_{\rm ch}$\eq0.031\,mJy\,\bmm per velocity bin ($\Delta v$\eq145\,\kms) 
is reached in the resulting line cube, with a beam size of 0\farcs94\,$\times$\,0\farcs71 at PA\eq31$\degr$.

\subsection{ALMA \cii}
We observed the \cii fine-structure line ($\nu_{\rm rest}$\eq1900.536900\,GHz) in
\xa with ALMA on 2015 June 15 and August 27 during Cycle 2
(ID: 2013.1.00749.S, PI: D. Riechers).
The \cii line is redshifted to Band 8 at the redshift of \xa determined from our CO data ($z$\eq2.9850).
We employed the frequency division mode (FDM) correlator setup with dual polarization, providing
an effective bandwidth of 7.5\,GHz and a spectral resolution of 1.95\,MHz (1.2\,\kms).
The on-source time, baseline coverage, and calibrators used in each track are listed in \Tab{obssum}.
All data were calibrated manually due to the uncertain
flux scale of Ceres, which was used as the flux calibrator in one
of the two tracks. The calibrated amplitudes of both the phase and bandpass calibrators are consistent with those
found in the ALMA Calibrator Source Catalogue.
The flux scale was also verified by comparing the calibrated
amplitudes of the same phase calibrator across the two tracks.
We estimate a flux calibration accuracy of 15\%.

All data were calibrated using \casa version 4.5.0 and were then combined,
imaged, and deconvolved using the CLEAN algorithm with
natural weighting,
yielding a synthesized beam of 0\farcs18$\times$0\farcs14 at PA\eq61.3\degr.
To obtain an optimal balance between sensitivity and spectral resolution,
we binned the data cubes to spectral resolutions of $\Delta v$\eq25\,\kms and 300\,\kms,
reaching typical rms noise values of $\sigma_{\rm ch}$\eq2.36 and 0.75\,mJy\,beam\pmOne per channel, respectively.
A continuum image was obtained by averaging across the line-free channels and
excluding any channels that were affected by atmospheric features.
The bandwidth used to form the continuum images is 5.47 GHz, yielding
an rms noise level of $\sigma_{\rm cont}$\eq0.22\,mJy\,beam\pmOne.

We also imaged the visibilities
with $uv$-tapering applied at 500\,k$\lambda$ (311.5\,m) to recover potential diffuse low surface brightness
emission and structure on larger spatial scales.
After tapering, a line cube binned to a spectral resolution of
$\Delta v$\eq150\,\kms was imaged and deconvolved using the CLEAN algorithm and natural weighting.
We used the tapered data cube and image to define the apertures used for extracting the line
and underlying continuum fluxes, and the line spectrum (see \Sec{results}).
The beam size for the tapered data is 0\farcs31\,$\times$\,0\farcs26 at PA\eq69.5$\degr$, which is roughly twice
the untapered beam size.
The final rms noise is
$\sigma_{\rm cont}$\eq0.33\,mJy\,beam\pmOne for the tapered continuum map,
and $\sigma_{\rm ch}$\eq1.25\,mJy\,beam\,\pmOne per 150\,\kms bin for the data cube.

\subsection{ALMA CO($J$\eq10\rarr9)}
In ALMA Cycle 4, we observed the CO($J$\eq10\rarr9) line ($\nu_{\rm rest}$\eq1151.98545200\,GHz) in
\xa on 2017 September 11 and 16 (ID: 2016.2.00105.S, PI: D. Riechers) using the 7\,m Atacama Compact Array (ACA).
The \jco line is redshifted to Band 7 for \xa.
We employed the time division mode (TDM) correlator setup with dual polarization, providing
an effective bandwidth of 7.5\,GHz
and a spectral resolution of 15.6\,MHz (16.2\,\kms).
The on-source time, baseline coverage, and calibrators of each track are listed in \Tab{obssum}.
We conservatively estimate a flux calibration accuracy of 15\%.

All data were calibrated using version 5.1.1 of \casa, and were then combined,
imaged, and deconvolved using the CLEAN algorithm with natural weighting.
This yields a clean beam of 5\farcs35$\times$3\farcs65 at PA\eq$-$85\degr.
We binned the data cube to a spectral resolution of $\Delta v$\eq49\,\kms, reaching a typical rms noise
of $\sigma_{\rm ch}$\eq1.20\,mJy\,beam\pmOne per channel.
A continuum image was obtained by averaging across the line-free channels
over a bandwidth of 5.61 GHz, yielding an rms noise of $\sigma_{\rm cont}$\eq0.37\,mJy\,beam\pmOne.

\subsection{Ancillary Data}
\begin{deluxetable}{lcccc}[!htbp]
\centering
\tabletypesize{\scriptsize}
\tablecolumns{5}
\tablecaption{Photometry obtained for \xa.}
\tablehead{
\colhead{Wavelength/Band} &
\colhead{Frequency} &
\multicolumn{2}{c}{Flux Density} &
\colhead{Instrument/Band}
\\
\colhead{($\micron$)} & \colhead{(GHz)} & \colhead{} & \colhead{} & \colhead{ } \vspace{0.05in}
}
\startdata
0.15     & 2000000 & $<$\,2.29         & $\mu$Jy & {\it GALEX}/FUV \\   
0.23     & 1300000 & $<$\,2.29         & $\mu$Jy & {\it GALEX}/NUV \\  
0.38     & 780000  & $<$\,0.19         & $\mu$Jy & CFHT/$u^*$ \\ 
0.49     & 610000  & $<$\,0.14         & $\mu$Jy & CFHT/$g'$ \\ 
0.63     & 480000  & $<$\,0.20         & $\mu$Jy & CFHT/$r'$ \\ 
0.76     & 390000  & $<$\,0.24         & $\mu$Jy & CFHT/$i'$ \\ 
0.88     & 340000  & $<$\,0.11          & $\mu$Jy & VISTA/$Z$-Band \\ 
0.89     & 340000  & $<$\,0.35         & $\mu$Jy & CFHT/$z'$ \\ 
1.02     & 290000  & $<$\,0.31          & $\mu$Jy & VISTA/$Y$-Band \\ 
1.16     & 260000  & $<$\,0.10                 & $\mu$Jy  & {\it HST}/F110W \\     
1.25     & 240000  & $<$\,0.35          & $\mu$Jy & VISTA/$J$-Band \\ 
1.65     & 180000  & $<$\,0.55          & $\mu$Jy & VISTA/$H$-Band \\ 
2.15     & 140000  & $<$\,0.78          & $\mu$Jy & VISTA/$Ks$-Band \\
3.4      & 88174   & $<$0.20          & mJy    & {\it WISE}/W1 \\ 
3.6      & 83275   & $<$1.25           & $\mu$Jy & {\em Spitzer}/IRAC \\  
4.5      & 66620   & $<$1.25           & $\mu$Jy & {\em Spitzer}/IRAC \\  
4.6      & 65172   & $<$0.19          & mJy    & {\it WISE}/W2 \\ 
5.8      & 51688   & 8.61\pmm1.54         & $\mu$Jy & {\it Spitzer}/IRAC \\ 
8.0      & 37474   & 8.14\pmm4.84         & $\mu$Jy & {\it Spitzer}/IRAC \\ 
12       & 24983   & $<$0.52          & mJy & {\it WISE}/W3 \\
22       & 13627   & $<$3.24           & mJy & {\it WISE}/W4 \\   
24       & 12491   & 1.08\pmm0.02      & mJy & {\it Spitzer}/MIPS \\  
70       & 4283    & <10.8      & mJy & {\it Spitzer}/MIPS \\ 
100      & 2998    & $<$17.3           & mJy & {\it Herschel}/PACS \\   
160      & 1874    & $<$90.0           & mJy & {\it Spitzer}/MIPS\\ 
160      & 1874    & 86.3\pmm17.9   & mJy & {\it Herschel}/PACS \\  
250      & 1200    & 106\pmm7         & mJy & {\it Herschel}/SPIRE \\ 
350      & 857     & 120\pmm10     & mJy & {\it Herschel}/SPIRE \\  
500      & 600     & 92.1\pmm7.6      & mJy & {\it Herschel}/SPIRE \\ 
635      & 472     & 52.5\pmm5.9       & mJy & ALMA\\
870      & 345     & 18.0\pmm0.4       & mJy & ALMA\\  
870      & 345     & 21.5\pmm3.1       & mJy & SMA \\  
1037    & 289     & 11.8\pmm0.8     & mJy & ALMA ACA \\
1200     & 250     & 8.9\pmm0.9        & mJy & MAMBO \\
2061.3   & 145.4   & $<$1.31          & mJy & PdBI \\   
2284.7   & 131.1   & $<$0.63          & mJy & PdBI  \\   
2801.8   & 107     & $<$1.30          & mJy & CARMA \\ 
2910.6   & 103     & $<$0.98          & mJy & CARMA \\ 
3000\tna & 100     & 0.50\pmm0.11        & mJy & CARMA \\ 
3209.8   & 93.4    & $<$1.11          & mJy & CARMA \\ 
3331.0   & 90      & $<$0.50          & mJy & CARMA \\ 
9586.8  & 31.3    & 0.0184\pmm0.00314 & mJy & VLA \\    
\enddata
\label{tab:photometry}
\tablecomments{
All upper limits are 3\sig\ limits.
Uncertainties on the SPIRE flux densities include those due to confusion noise.
Uncertainties quoted here for the radio and mm interferometric measurements
(i.e., with ALMA, CARMA, PdBI, SMA, and VLA) do not include those from absolute flux calibration ($\sim$15\%), which are
accounted for in the SED modeling.
}
\tablenotetext{a}{Continuum emission measured in an image obtained by combining all four spectral setups covering the 3\,mm window.}
\tablerefs{
{\it GALEX} limits are from XMM-LSS DIS \citep{Pierre04a, Martin05a}. CFHT limits are from CFHTLS-D1 \citep{Chiappetti05a}.
VISTA limits are from the VIDEO survey \citep{Jarvis13a}.
{\it HST} limit is taken from \citet{Calanog14a}.
Upper limits from {\it Spitzer}/IRAC and MIPS \obs are the survey depths of SWIRE and SERVS
\citep{Lonsdale03a, Nyland17a}.
{\it Herschel}/PACS limit at 100\,$\micron$ is obtained from Level 5 \obs of the XMM-VIDEO3 field \citep{Oliver12a}.
PACS 160\,$\micron$ flux density is obtained from the DR4 PACS catalog of the XMM-VIDEO3 field.
{\it Herschel}/SPIRE and MAMBO photometry are from \citet{Wardlow13a}.
ALMA 870\,$\micron$ flux density is from \citet{Bussmann15a}.
}
\end{deluxetable}

\subsubsection{{Herschel}/SPIRE and PACS, and MAMBO 1.2\,mm}
\xa was observed with {\it Herschel}/PACS and SPIRE at 100, 160, 250, 350, and 500\,$\micron$
as part of the HerMES project (\citealt{Oliver12a}).
\xa remains undetected at 100\,$\micron$ down to a 5$\sigma$ limit of
$S_{\rm 100}$\,$<$\,28.8\,mJy,
but is detected at 160\,$\micron$.
The 160\,$\micron$ photometry was
extracted from the Level 5 XMM-VIDEO3 data using a positional prior from the {\it Spitzer}/MIPS 24\,$\micron$ catalog
with aperture photometry, and with appropriate aperture corrections applied (PACS DR4).
For the SPIRE photometry, we adopted the fluxes reported by
\citet[][]{Wardlow13a}, which were extracted using
\ncode{StarFinder} \citep{Diolaiti00a}.  We also include the 1.2\,mm photometry obtained
with the IRAM 30-m telescope/MAMBO in modeling the SED of \xa  (\citealt{Wardlow13a}; \Tab{photometry}; see \Sec{sed}).

\subsubsection{SMA 870\,$\micron$} 
We also make use of 870\,$\micron$ continuum data obtained with
the Submillimeter Array (SMA; IDs: 2010A-S091 and 2011A-S068, PIs: A. Cooray and S. Bussmann; \citealt{Wardlow13a}).
Observations were carried out in the extended and subcompact array configurations on
2010 August 16 and September 25, and 2011 August 05,
with local oscillator frequencies of
342.224\,GHz and 342.003\,GHz (extended),
and 340.017\,GHz (subcompact), respectively.
The on-source time of each track is listed in \Tab{obssum}.
Uranus was used as the primary flux calibrator,
and the quasars J0238$+$166 and J0217$+$017 were used as
complex gain calibrators for all three tracks.
Quasars 3C454.3 and 3C84 were used for bandpass calibration.
MWC349A and Callisto were observed as secondary flux calibrators in the extended configuration tracks.

All visibility data were calibrated using the IDL-based \ncode{mir} package and imaged using \ncode{miriad}.
We combined all tracks to form a continuum image using the CLEAN algorithm with natural weighting, yielding
a synthesized beam of 0\farcs99\,$\times$\,0\farcs78 at PA\eq$-$68.2$\degr$ and
an rms noise of 0.92\,mJy\,\bmm over the full bandwidth of 7.5\,GHz.

\subsubsection{ALMA Cycle-0 870\,$\micron$}
We previously observed the 870\,$\micron$ continuum emission in \xa
with ALMA in Band 7 (ID: 2011.0.00539.S; PI: D. Riechers; also see \citealt{Bussmann15a}).
Visibilities were imaged using the CLEAN algorithm with Briggs weighting (robustness $R$\eq0.5), yielding
a synthesized beam of 0\farcs50\,$\times$\,0\farcs40 (PA\eq76.4$\degr$) and
an rms noise of $\sigma_{\rm cont}$\eq0.28\,mJy\,beam\pmOne.

\subsubsection{{\it Spitzer}/IRAC and MIPS Near- and Mid-IR}

\xa was observed with {\it Spitzer}/IRAC and MIPS as part of the {\it Spitzer}
Wide-area InfraRed Extragalactic Survey (SWIRE; \citealt{Lonsdale03a})
in the XMM-LSS field.
The survey depths (5$\sigma$) for point sources are
$S_\nu$\,$<$\,3.7, 5.4, 48, and 37.8 $\mu$Jy for the IRAC channels at 3.6, 4.5, 5.8 and 8.0\,$\micron$, respectively,
and 230\,$\mu$Jy, 18\,mJy, and 150\,mJy for the MIPS bands at 24, 70, and 160\,$\micron$,
respectively\footnote{\url{http://swire.ipac.caltech.edu/swire/astronomers/program.html}}.
In the MIPS bands, \xa is detected at 24\,$\micron$
(SWIRE catalog DR2)\footnote{\url{https://irsa.ipac.caltech.edu/data/SPITZER/SWIRE/docs/delivery_doc_r2_v2.pdf}}.
The 24\,$\micron$ photometry was extracted using aperture photometry and \ncode{SExtractor} \citep{Savage07a}.
Appropriate aperture corrections have been applied.
\xa remains undetected at 70 and 160\,$\micron$;
we adopt 3$\sigma$ levels as the upper limits for the non-detections (see \Tab{photometry}).

In the post-cryogenic period of {\it Spitzer}, more sensitive continuum images at 3.6 and 4.5\,$\micron$ were obtained
in the deeper {\it Spitzer} Extragalactic Representative Volume Survey (SERVS), which reaches 5\sig\ limits
of 1.25\,$\mu$Jy \citep{Mauduit12a, Nyland17a}.
For the two SWIRE images observed at longer wavelengths (IRAC 5.8 and 8.0\,$\micron$), 
we perform aperture photometry to extract the fluxes of \xa at the centroid position determined from the SMA 870\,$\micron$ map.
Final flux densities are reported in \Tab{photometry}.

\subsubsection{{\it Wide-field Infrared Survey Explorer (WISE)} Near- and Mid-IR} \label{sec:WISE}
\xa was observed with {\it WISE} as part of the ALLWISE program.
Its flux density limits are reported in the ALLWISE source catalog available on the NASA/IRAC Infrared Science Archive (IRSA)
and were extracted through profile-fitting.
In Vega magnitude units, we find
15.460\pmm0.040, 14.905\pmm0.065, $<$12.457, and $<$8.817
for the four WISE bands (at 3.4, 4.6, 12, and 22\,$\micron$, respectively). The latter two are 3\sig\ upper limits. Since a few sources with IR emission
near \xa are detected in the {\it Spitzer} images, we expect emission toward \xa to be unresolved and blended within the {\it WISE} beam.
As such, we adopt all the {\it WISE} fluxes as upper limits only, yielding 3\sig\ limits of 0.20, 0.19,
0.52, and 3.24\,mJy, respectively (\Tab{photometry}).

\subsubsection{Visible and Infrared Survey Telescope for Astronomy (VISTA) Near-IR } \label{sec:vista}

The XMM-LSS field was imaged with VISTA in the $Z$-, $Y$-, $J$-, $H$-, and $Ks$-bands
as part of the VISTA Deep Extragalactic Observations (VIDEO) Survey
\citep{Jarvis13a}, reaching 5$\sigma$ limits of 25.7, 24.6, 24.5, 24.0, and 23.5 AB mag
for a point source in a 2$^{\prime\prime}$ diameter aperture.
\xa is undetected in all bands.
In \Tab{photometry}, we report the corresponding 3$\sigma$ levels as upper limits.

\subsubsection{CFHT UV-optical-IR } \label{sec:cfht}
\xa was imaged with the CFHT/MegaCam
in $u^*$, $g'$, $r'$, $i'$, $z'$ bands as part of the CFHT Legacy Survey Deep-1 field (CFHTLS-D1).
In the final CFHTLS release (version T0007),
the sensitivity limits corresponding to 80\% completeness for a point source
are 26.3, 26.0, 25.6, 25.4, and 25.0 AB mag for the five bands, respectively,
or 3\sig\ point-source sensitivities of 0.19, 0.14, 0.20, 0.24, 0.35\,$\mu$Jy.
We show the $\sim$0.8$^{\prime\prime}$ resolution CFHT deep field images retrieved from
the CFHT Science Archive from the Canadian Astronomy Data Centre (CADC) in the Appendix.
\xa remains undetected in all bands according to the T0007 CFHTLS-Deep catalog (\citealt{Hudelot12a}; \Tab{photometry}).

\subsubsection{{\it Galaxy Evolution Explorer (GALEX)} Near and Far-UV}
UV emission in the \xa field was observed with {\it GALEX} in the FUV-1500 and NUV-2300 bands
as part of the XMM-LSS Deep Imaging Survey (DIS). 
\xa was covered in the XMMLSS\_00 tile, which was observed for
75262 and 60087 seconds in the NUV and the FUV bands, respectively\footnote{Based on the
images and catalog released in GR6.},
reaching 3$\sigma$ limits of 25.5 in AB mag \citep{Pierre04a, Martin05a}.

\subsubsection{XMM-Newton X-ray } 	\label{sec:xmm}
\xa is located in the CFHTLS-D1 field, which was observed with the European Photon Imaging Camera (EPIC)
onboard {\it XMM-Newton} for an integration time of around 20\,ks in the {\it XMM} Medium Deep Survey (XMDS;
\citealt{Chiappetti05a}), reaching 3$\sigma$ point source limits
of 3.7\E{-15}~erg\,s\pmOne\,cm$^{-2}$ and 1.2\E{-14}~erg\,s\pmOne\,cm$^{-2}$ in the soft (0.5$-$2 keV)
and hard (2$-$10\,keV) X-ray bands, respectively.
These limits correspond to
$L_{X{\rm, 0.5-2\,keV}}$\,$<$\,7.4\E{43}\,erg\,s\pmOne (soft) and
$L_{X{\rm, 2-10\,keV}}$\,$<$\,9.5\E{44} erg s\pmOne (hard) at $z$\eq2.9850,
which reach the levels
of 
powerful Seyfert galaxies \citep{Elvis78a}.
\xa remains undetected in these \obs.

\section{Results} \label{sec:results}
\def\arraystretch{1.2}
\begin{deluxetable*}{lcccccc}
\tabletypesize{\scriptsize}
\tablewidth{1.\textwidth}
\tablecolumns{7}
\tablecaption{Parameters from fitting single Gaussians to the CO and \cii line profiles and intensity maps shown in
Figures~\ref{fig:spectra},  \ref{fig:comom0}, \ref{fig:co10spec}, and \ref{fig:mom0}.
\label{tab:line}}
\tablehead{
Line           &
$S_{\rm peak}$ &
FWHM           &
$I$            &
\multicolumn{2}{c}{Deconvolved source size at FWHM} &
$S_{\rm cont}$
\\
                    &
(mJy)           &
(\kms)          &
(Jy \kms)      &
(arcsec$\times$arcsec, deg)               &
(kpc)          &
(mJy)
}
\startdata
CO($J$\eq1\rarr0)\tna &  0.55\pmm0.11 & 1118\pmm307 & 0.65\pmm0.22 & ---                                      & ---            & (0.07\pmm0.03)\tnc \\
\hspace{3mm} X-Main   & 0.44\pmm0.07 & 1100\pmm210 & 0.51\pmm0.13 & 1.12\pmm0.37\,$\times$\,0.81\pmm0.45, 173\pmm49\tnb & 8.8$\times$6.4 & (0.05\pmm0.02)\tnc \\     
\hspace{3mm} X-NE     & 0.26\pmm0.04 & 718\pmm130  & 0.20\pmm0.05 & 1.12\pmm0.41\,$\times$\,0.26\pmm0.42, 72\pmm37\tnb & 8.8$\times$2.0 & (0.02\pmm0.01)\tnc \\ [2.5pt]    
\cco                  & 6.21\pmm0.98 & 791\pmm157  & 5.21\pmm1.32 & ----                                     & ---            & (0.23\pmm0.26)\tnc \\ [2.5pt]
\eco                  & 9.38\pmm2.49 & 500\pmm159  & 4.97\pmm2.06 & ---                                      & ---            & (0.16\pmm0.43)\tnc \\ [2.5pt]
\jco                  & 3.72\pmm0.23 & 760\pmm55   & 3.01\pmm0.29 & ---                                      & ---            & ---\tnd \\ [2.5pt]
\cii                  & 183\pmm10    & 687\pmm53   & 133\pmm12    & 0.91\pmm0.08\,$\times$\,0.75\pmm0.07, 29\pmm17     & 7.2$\times$5.9 & ---\tnd  \\ [-4.5pt]
\enddata
\tablenotetext{a}{ Emission from both X-Main and X-NE.}
\tablenotetext{b}{Only marginally resolved.}
\tablenotetext{c}{Not detected.}
\tablenotetext{d}{Continuum emission was subtracted from the line cubes before extracting the spectrum.}
\tablecomments{The higher-$J$ CO lines are unresolved. }
\end{deluxetable*}

\def\arraystretch{1.2}
\begin{deluxetable*}{lccccc}
\tabletypesize{\scriptsize}
\tablewidth{1.\textwidth}
\tablecolumns{6}
\tablecaption{Continuum flux densities and deconvolved source sizes.
\label{tab:cont}}
\tablehead{
Instrument/Component           &
Wavelength &
$S_{\rm peak}$ &
$S_{\rm total}$           &
\multicolumn{2}{c}{Deconvolved source size at FWHM} \\
               &
($\micron$)  &
(mJy\,\bmm)          &
(mJy)         &
(arcsec$\times$arcsec, deg)               &
(kpc)
}
\startdata
ALMA Total       & 635 & 5.61\pmm0.22 & 52.5\pmm5.9 & --- & --- \\
\hspace{3mm} XD1 peak                  &  &  4.95\pmm0.38 & 28.1\pmm2.4 & 0.39\pmm0.05\,$\times$\,0.36\pmm0.05, 17\pmm87  & 3.1\,$\times$2.8 \\
\hspace{3mm} XD2 peak                  & & 2.75\pmm0.28 &  15.7\pmm1.8 &  0.39\pmm0.06\,$\times$\,0.35\pmm0.06, 174\pmm89 &  3.1\,$\times$2.8 \\  [2.5pt]
ALMA    & 870  &10.61\pmm0.35   & 17.96\pmm0.43  & 0.62\pmm0.02\,$\times$0.54\pmm0.02, 85\pmm10  & 4.9$\times$4.2 \\ 
SMA      & 870  &  12.8\pmm1.2    &     21.5\pmm3.1    & 0.75\pmm0.23\,$\times$\,0.66\pmm0.27, 112\pmm89 & 5.9\,$\times$\,5.2 \\ [-4.5pt]
\enddata
\end{deluxetable*}

\subsection{CO Line Emission and Redshift Identification}            \label{sec:co}

\begin{figure}[tbph]
\centering
\includegraphics[trim=0 -5 0 -5, clip, width=0.48\textwidth]{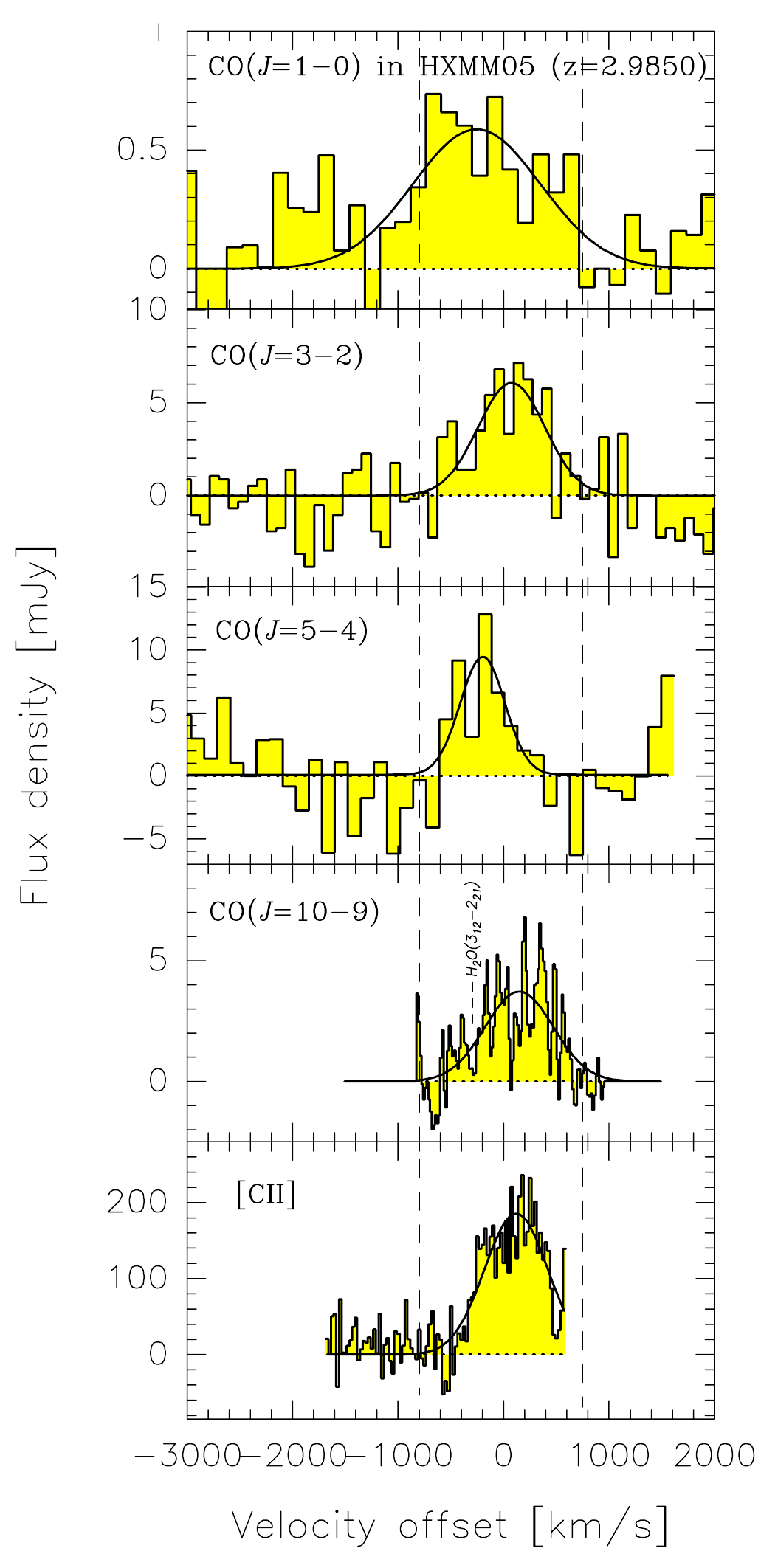}
\caption{Top to bottom: VLA \aco, CARMA \cco, PdBI \eco, and ALMA \jco and \cii line spectra (histograms) observed toward \xa.
\cii emission at $v\gtrsim500$\,\kms is dominated by noise near the edge of the spectral window, 
where a strong atmospheric feature is present.
The spectral resolutions are $\Delta v$\eq145, 90, 124, 16, and 25\,\kms from top to bottom.
\aco and \cii line spectra are same as those shown in \Fig{co10spec}.
Solid black lines show the best-fit single Gaussians.
Vertical dashed lines are shown to facilitate linewidth comparison across panels.
The redshifted frequency of the H$_2$O(3$_{12}$\rarr$2_{21}$) line is annotated on the \jco spectrum.
\label{fig:spectra}}
\end{figure}

\begin{figure*}[tbph]
 \centering
\includegraphics[width=1.\textwidth]{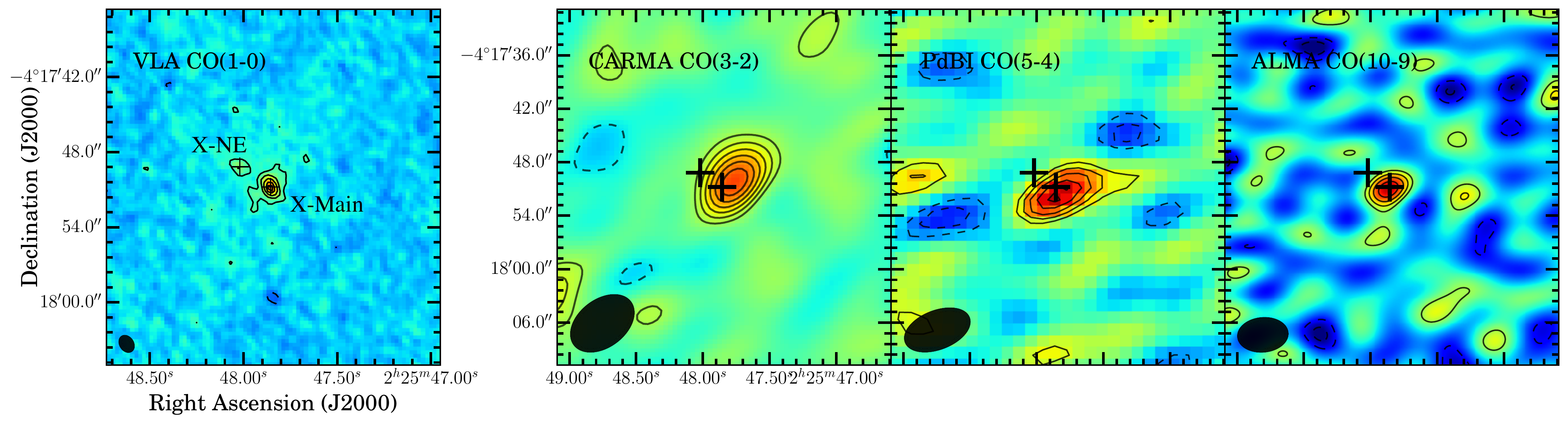}
\caption{
Left to right: Intensity maps of VLA \aco, CARMA \cco, PdBI \eco, and ALMA ACA \jco line emission.
Contours in the first panel are shown in steps of $[-$3, 3, 6, 9, 12, 14, 15$]$$\times$$\sigma_{\rm 1-0}$, where
$\sigma_{\rm 1-0}$\eq14\,mJy\,\kms\bmm.
For the remaining panels, contours are shown in steps of [$-$3, -2, 2, 3, 4, 5, 6, 7, 8]$\times$$\sigma$, where $\sigma_{\rm 3-2}$\eq0.53\,Jy\,\kms\bmm
for \cco, $\sigma_{\rm 5-4}$\eq0.73\,Jy\,\kms\bmm for \eco, and $\sigma_{\rm 10-9}$\eq0.63\,Jy\,\kms\bmm for \jco.
Black markers indicate the positions of X-Main and X-NE as observed in \aco emission.
Beam sizes are shown in the lower left corners and are summarized in \Tab{obssum}.
 \label{fig:comom0}}
\end{figure*}

From the first two CO lines we detected --- \cco and \eco with CARMA and the PdBI --- we spectroscopically determine
the redshift of \xa to be \z\eq2.9850\pmm0.0009.
The CO($J$\eq3\rarr2; 5\rarr4; 10\rarr9) lines remain spatially unresolved, and are
detected at $>$\,8$\sigma$, $>$\,6$\sigma$, and $>$5\sig\
significance, respectively (Figures \ref{fig:spectra} and \ref{fig:comom0}).
Due to the near-equatorial declination of \xa and the sparse $uv$-sampling of the data,
the PdBI synthesized beam is highly elongated,
and the image fidelity is heavily affected by strong sidelobes.
We fit single Gaussian profiles to the line spectra, as shown in \Fig{spectra}.
The resulting best-fit parameters are summarized in \Tab{line}.
We note that given the broad linewidths observed up to the $J$\eq10\rarr9 transition, the lack of emission at $v > 0 $\,\kms in the \eco line
may be attributed to the limited S/N of the data. The true \eco flux may be a factor of two higher.

Upon determining the redshift of \xa, we observed the \aco line with the VLA.
We detect marginally spatially-resolved \aco line emission at $>$14\,$\sigma$ peak significance (\Fig{comom0}).
The emission centroid is centered at the position of \xa, but shifts from NW to SE with increasing velocity.
A second peak is detected at 2\farcs6 NE of \xa,
at 6$\sigma$ significance in the blueshifted channels (see \Fig{comom0}), corresponding to
a projected separation of 20\,kpc.
In the subsequent sections of this paper, this NE component is referred to as X-NE, and the main component is referred to as ``X-Main''.

\begin{figure*}[tbph]
\centering
\includegraphics[trim=0 0 0 0 , clip, angle=270, width=0.95\textwidth]{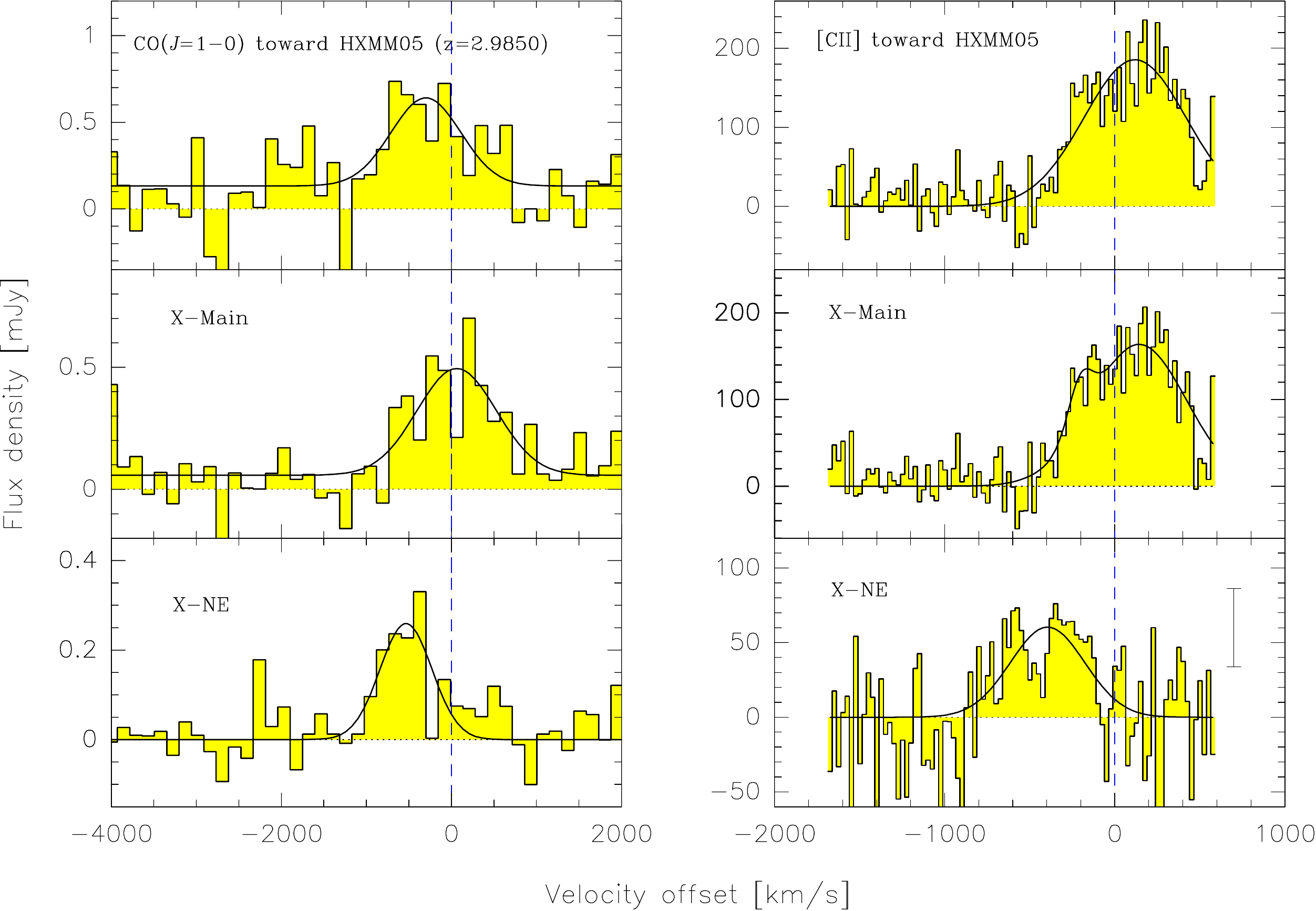} 
\caption{
VLA CO($J$\eq1\rarr0) (left column) and ALMA \cii (right column) spectra (histograms) of \xa.
Top panels: Spectra of \xa, including emission from X-Main (middle panels) and X-NE (bottom panels).
A typical error bar for the \cii spectrum extracted for X-NE is shown in the bottom right panel.
Vertical dashed lines mark a common $v$\eq0\,\kms to facilitate comparison of line shapes and widths across panels.
Spectral resolutions of the \aco and \cii lines are $\Delta v$\eq145 and 25\,\kms, respectively.
Continuum underlying the \cii line has been subtracted in the $uv$-plane.
To account for the weakly detected Ka-band continuum in the VLA data,
we fit models of a Gaussian and a zeroth order polynomial to the CO spectra (black lines).
For the \cii spectra, we fit single- (top and lower right) and double- (middle right) Gaussians.
The velocity scale is with respect to $z$\eq2.9850 (dashed lines).
X-NE is detected in both \aco and \cii lines at $\gtrsim$\,6\sig\ significance (see Figures~\ref{fig:comom0} and \ref{fig:mom0}).
\label{fig:co10spec}}
\end{figure*}

We extract a spectrum using an aperture defined by the
2$\sigma$ contours centered at the coordinates of \xa (middle left panel of \Fig{co10spec}),
and a spectrum for just X-NE (bottom left panel of \Fig{co10spec}).
The centroid of X-NE is blueshifted by $-$535\pmm55\,\kms with respect to X-Main.
Assuming that the line detected is \aco, the redshift of X-NE would be $z$\eq2.9779\pmm0.0007.
We also extract a spectrum for the \xa system as a whole, including emission from both X-Main and X-NE
(top panel of \Fig{spectra} and top left panel of \Fig{co10spec}).
The best-fit linewidths and intensities are listed in \Tab{line}.
The \aco line is remarkably broad ($>$1100\,\kms FWHM) and shows a hint of a double-horned profile, which
likely results from contributions from both X-Main and X-NE (see \Fig{co10spec}).

We fit 2D Gaussians to the two components detected in the velocity-integrated line intensity map, finding
a deconvolved source size of (1\farcs12\pmm0\farcs37)\,$\times$\,(0\farcs81\pmm0\farcs45) at PA\eq173\pmm49$\degr$
for \xa.
This corresponds to a physical diameter of 8.8\,kpc\,$\times$\,6.4\,kpc at $z$\eq2.9850. 
For the NE component, we find a deconvolved source
size of (1\farcs12\pmm0\farcs41)\,$\times$\,(0\farcs26\pmm0\farcs42) at PA\eq72\pmm37$\degr$, which corresponds to a physical size of 8.8\,kpc\,$\times$\,2.0\,kpc at $z$\eq2.9779. 
The extent of the cold molecular gas in both \xa and the NE component are consistent
with those observed in other DSFGs \citep[e.g.,][]{Ivison11a, Riechers11a}.

\subsection{\cii Line Emission}     \label{sec:cii}  
\begin{figure*}[htbp]
 \centering
\includegraphics[trim=15 0 0 5, clip, width=1\textwidth]{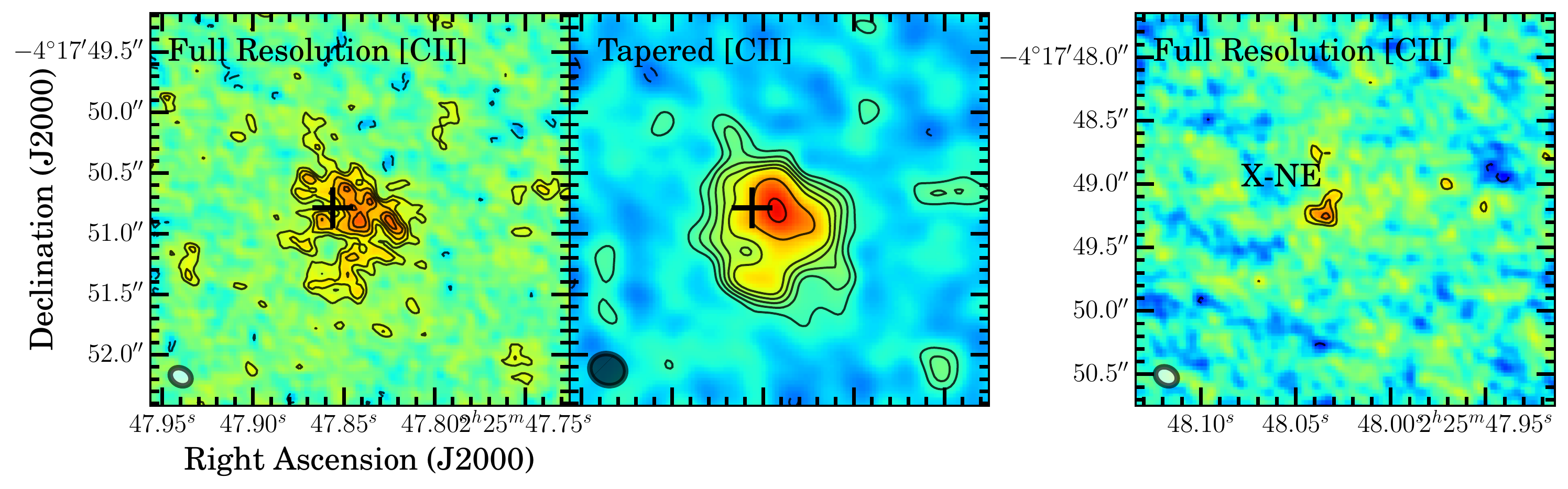}
\caption{\cii intensity maps at resolutions of 0\farcs15 (1.2\,kpc; left and right panels) and
at 0\farcs3 (tapered; middle panel).
The left and middle panels show maps formed by integrating over velocity channels between $\Delta v$\,$\in$\,[$-$226.9, 413.6] \kms. 
Contours are shown in steps of [$-$3, $-$2, 2, 3, 4, 5, 6, 9, 12, 15]$\times$\,$\sigma$,
where $\sigma$\eq0.63\,Jy\,\kms\bmm (full resolution) 
and $\sigma$\eq0.68\,Jy\,\kms\bmm (tapered).       
Black crosses indicate the centroid position of \aco emission detected in X-Main (see Figures \ref{fig:comom0} and \ref{fig:co10cii}).
The right panel shows a map of X-NE, formed by integrating over $\Delta v$\,$\in$\,[$-$654.0,$-$136.4] \kms.
Contours are shown in steps of [$-$3, 3, 4, 5, 6]$\times\sigma$, where $\sigma$\eq0.38\,Jy\,\kms\bmm.
Synthesized beam sizes of 0\farcs18\,$\times$\,0\farcs14, PA\eq61.3$\degr$ (untapered)
and 0\farcs31\,$\times$\,0\farcs26, PA\eq69.5$\degr$ (tapered) are shown in the lower left corners of the panels.
 \label{fig:mom0}}
\end{figure*}

We detect spatially resolved \cii line emission toward \xa at a peak significance of $>$13$\sigma$
(in a tapered intensity map).
At the full spatial resolution of the data (0\farcs15), \xa is resolved over $>$25 beams.
To better determine the line profile shape, we
create two \cii line cubes --- with and without $uv$-tapering (see \Sec{obs}).
The 635\,$\micron$ continuum emission has been subtracted from both line cubes in the $uv$-plane.
We collapse them to form velocity-integrated line intensity (i.e., zeroth moment)
maps 
as shown in \Fig{mom0}.
We show the \cii line spectrum of \xa in the last panel of \Fig{spectra} and the top right panel of \Fig{co10spec}.
The best-fit parameters obtained from fitting a single-Gaussian are
listed in \Tab{line}, together with those derived for the CO lines.

We extract separate spectra for X-Main and X-NE from the high resolution data cube using an aperture
defined by the 1\sig\ contours of the {\em tapered} intensity map.
The resulting spectrum of X-Main is shown in the middle right panel of \Fig{co10spec}.
Fitting a single Gaussian yields a peak flux density of $S_{\rm peak}$\eq172\pmm8\,mJy, a line FWHM of $\Delta v$\eq667\pmm46\,\kms,
and a line intensity of $I$\eq122\pmm10\,Jy\,\kms.
We also fit a double-Gaussian profile, yielding
best-fit peak fluxes of $S_{\rm peak}$\eq53\pmm30 and 164\pmm10\,mJy, and line FWHMs of
$\Delta v$\eq167\pmm85 and 659\pmm101\,\kms, respectively.
The peaks are separated by $\Delta v_{\rm sep}$\eq346\pmm124\,\kms. 
X-NE is detected at $\sim$6\sig\ significance (see bottom right panel of \Fig{co10spec} and also \Fig{mom0}).
We fit a 2D Gaussian to the tapered intensity map of X-Main, which yields a deconvolved FWHM source size of
(0\farcs91\pmm0\farcs08)$\times$\,(0\farcs75\pmm0\farcs07),  
or a physical size of (7.2\pmm0.6)\,$\times$\,(5.9\pmm0.6)\,kpc, consistent with the extent seen in the higher resolution image.

\begin{figure*}[tbph]
\includegraphics[trim=0 0 0 0, clip, width=\textwidth]{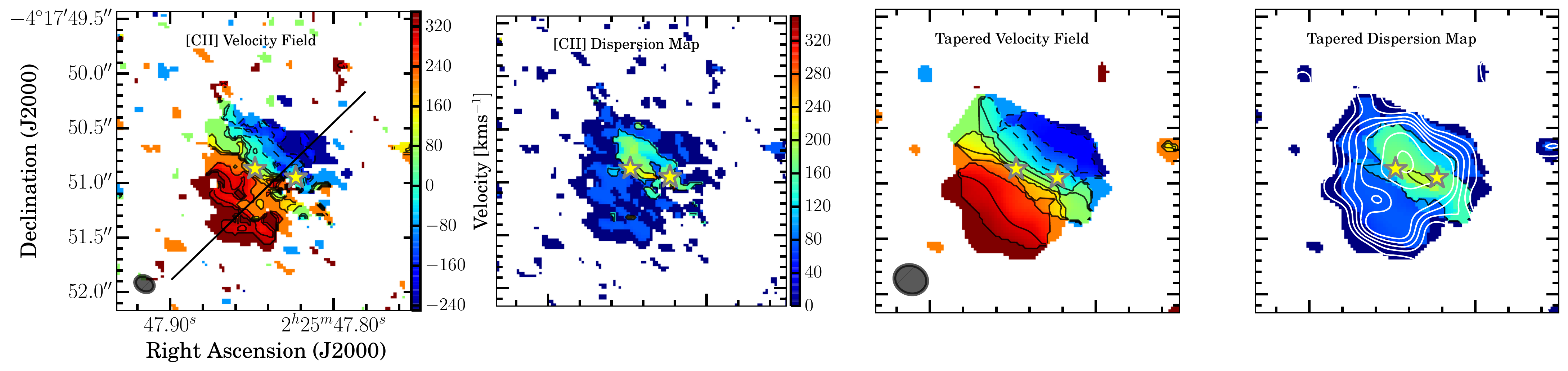}
\caption{
\cii velocity field and dispersion maps, centered at X-Main.
Contours in the velocity field maps start at $v$\eq$-$236\,\kms and increase to 364\,\kms in steps of 50\,\kms.
Black contours in the dispersion maps are shown in steps of $\Delta v$\eq100\,\kms.
Tapered velocity-integrated \cii line emission (white contours; same as right panel of \Fig{mom0}) is overlaid on the dispersion
map in the last panel.
 Synthesized beams are shown in the lower left corners of the first and third panels:
0\farcs18\,$\times$\,0\farcs14 for untapered (left) and
0\farcs31\,$\times$0\farcs26 for tapered \cii emission (right).
Black line in the first panel shows the kinematic major axis, along which the PV slice shown in \Fig{pvfit} is extracted.
Color-scales shown for the full resolution and tapered first and second moment maps are the same, respectively.
Yellow star symbols indicate dust peaks detected at 635\,$\micron$ (i.e., XD1 and XD2).
A velocity gradient along the NW-SE direction is seen.
The velocity dispersion is remarkably uniform across the galaxy, except at the central region,
where it peaks at $\sigma_{v}\simeq$\,200\,\kms.
X-NE is outside the field of view shown here.
\label{fig:mom1}}
\end{figure*}

The first and second moment maps of the \cii emission representing the
velocity
and the velocity dispersion of X-Main along the line-of-sight (LOS)
are shown in \Fig{mom1}.
Moment maps are created from the line cube after clipping at 3$\sigma_{\rm ch}$ per channel.
Structures on the scale of the angular resolution ($\lesssim1.2$\,kpc) are seen
in the channel maps (see Appendix~\Sec{channel}).
A velocity gradient along the NW-to-SE direction, varying over a velocity range of $\Delta v\simeq$\,600\,\kms,
is seen in the velocity field (\Fig{mom1}).
The dispersion map is remarkably uniform across the whole galaxy, with $\sigma_v\simeq$\,75\,\kms,
except in the central $\lesssim$0\farcs2 region,
where the dispersion reaches its peak at $\sigma_v\simeq$\,200\,\kms.

\begin{figure}[tbph]
\centering
\includegraphics[trim=25 20 15 15, clip, width=0.5\textwidth]{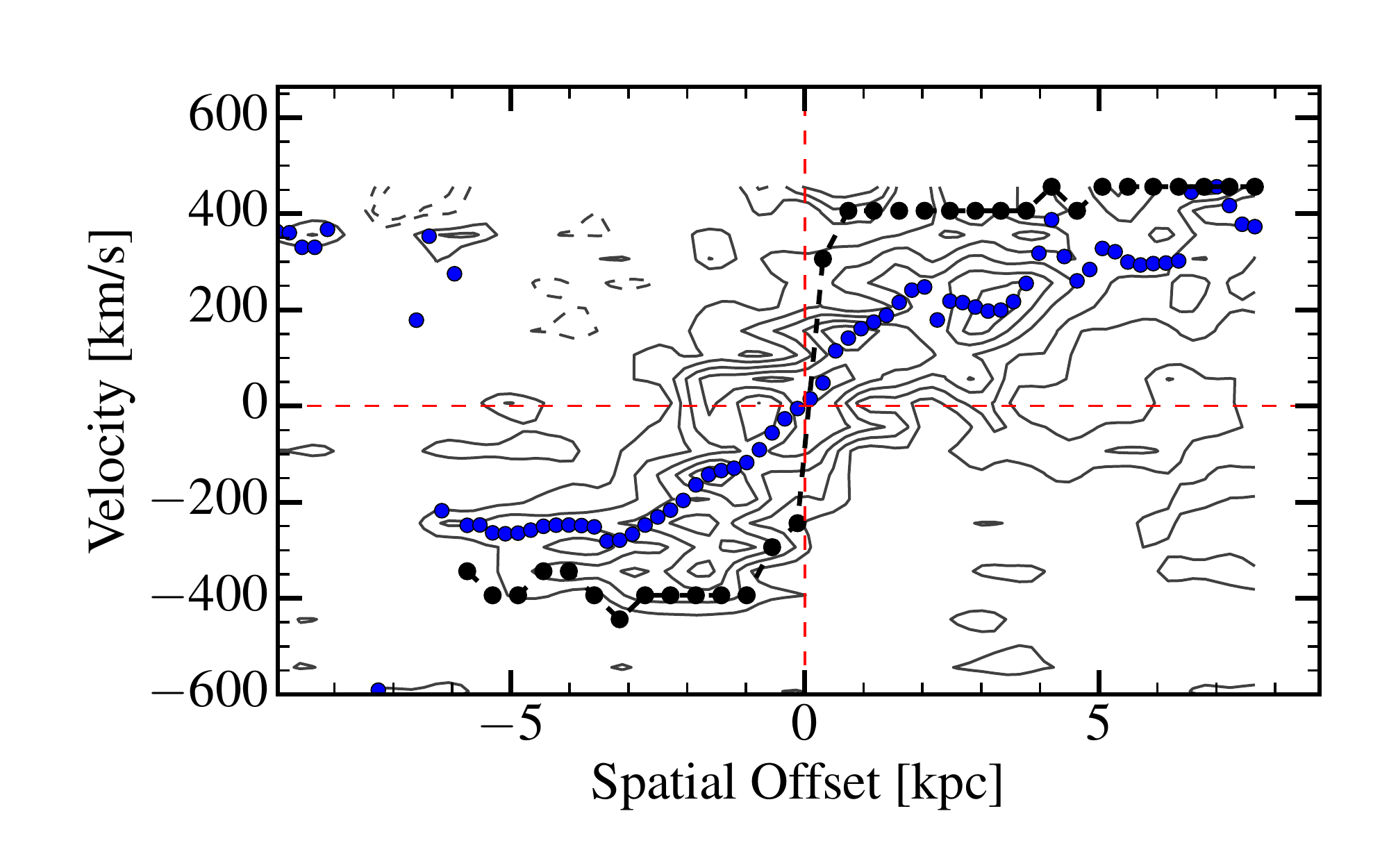}
\caption{
Rotation curve obtained from envelope-tracing (black dots and dashed line)
overplotted on a \cii PV diagram (contours) extracted along the major axis (see black line shown in the first panel of \Fig{mom1}).
Velocities shown on the $y$-axis are observed (i.e., uncorrected for inclination).
Blue markers show the centroid velocities of the spectra extracted at different
spatial positions (see text in \Sec{ET}).
Red dashed lines show the central position (vertical) and velocity (horizontal) determined from
\ncode{rotcur} and from fitting a double-Gaussian to the \cii spectrum.
\label{fig:pvfit}}
\end{figure}

A position-velocity (PV) diagram extracted along the major kinematic axis of X-Main
(see \Sec{rotcur}) is shown in \Fig{pvfit}.
The rising part of a rotation curve and the outer envelope are both detected. The latter is
usually more pronounced in more inclined disks (as seen in nearby galaxies; see review by \citealt{Sofue01a}).
The PV diagram is consistent with broad \cii line emission, which varies by $>$\,700\,\kms within about 9\,kpc.

\begin{figure*}[tbph]
\centering
\includegraphics[width=1\textwidth]{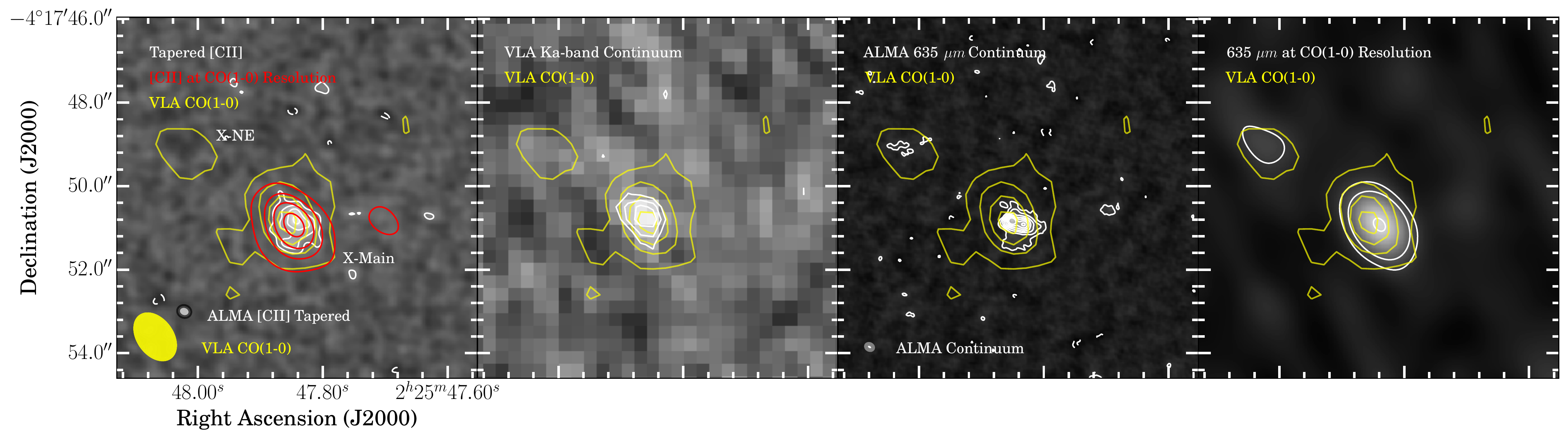}
\caption{Left to right: VLA CO($J$\eq1\rarr0) line emission (yellow contours) overlaid on
ALMA \cii line emission (first panel; integrated over line FWHM of X-Main), and continuum emission at 31.3\,GHz and at 635\,$\micron$ (last two panels).
Contours of CO and \cii line emission are shown
in steps of [$-$3, 3, 6, 9, 12, 14, 15]\,$\times$\sig, where \sig\eq15.3\,mJy\,\kms\,\bmm for the CO line,
0.676\,Jy\,\kms\,\bmm for the tapered \cii (grayscale in the first panel),
and 3.8 Jy\,\kms\,\bmm for \cii convolved to the same resolution as CO (red contours).
Contours for the 635\,$\micron$ continuum emission are shown
in steps of [$-$3, 3, 6, 18, 26, 34, 42, 50, 58, 66, 74, 82]\,$\times$\,$\sigma_{\rm cont}$, where
$\sigma_{\rm cont}$\eq0.22\,mJy\,\bmm in the third panel and
2\,mJy\,\bmm in the last panel (convolved to the same resolution as the CO).
Synthesized beam sizes of the VLA and the ALMA data are shown as gray (ALMA) and yellow (VLA) filled ellipses in the lower left
corners of the first and third panels, and are the same as in Figures~\ref{fig:co10chan} and \ref{fig:mom0}. For the tapered ALMA data, the beam size is 0\farcs31\,$\times$0\farcs26.
Both \cii and dust continuum emission are almost as extended as CO($J$\eq1\rarr0). X-NE is detected in both \aco and 635\,$\micron$
continuum (and \cii emission; see Figures~\ref{fig:co10spec} and \ref{fig:mom0}).
\label{fig:co10cii}}
\end{figure*}

We find comparable deconvolved source sizes for \aco and \cii emission (see \Tab{line}), as
confirmed by the comparable extents found after convolving the high resolution \cii data to the \aco line resolution (\Fig{co10cii}).
At the resolution of the VLA data, the velocity gradient seen in the \aco line emission is
consistent with that of the \cii line, but more sensitive and
higher angular resolution data are required to match the detailed velocity structures of both lines.

\subsection{H$_2$O Line Emission}       \label{sec:h2o}
The H$_2$O\,($1_{11}$\rarr$0_{00}$; 3$_{12}$\rarr$2_{21}$)
lines at redshifted frequencies of 279.383 and 289.367\,GHz are covered by the ALMA \jco line observations.
We do not detect the ground state H$_2$O line in emission or absorption down to
a 3\sig\ limit of $<$\,0.80\,Jy\,\kms\bmm, assuming the same linewidth as the \jco line (760\,\kms).
The H$_2$O(3$_{12}$\rarr$2_{21}$) line is next to the \jco line and is at most weakly detected; we
conservatively report a 3\sig\ upper limit of $<$\,0.87\,Jy\,\kms\bmm,
assuming the same linewidth as for the \jco line.

\subsection{Continuum}    \label{sec:cont}
\begin{figure*}[tbph]
 \centering
\includegraphics[width=0.85\textwidth]{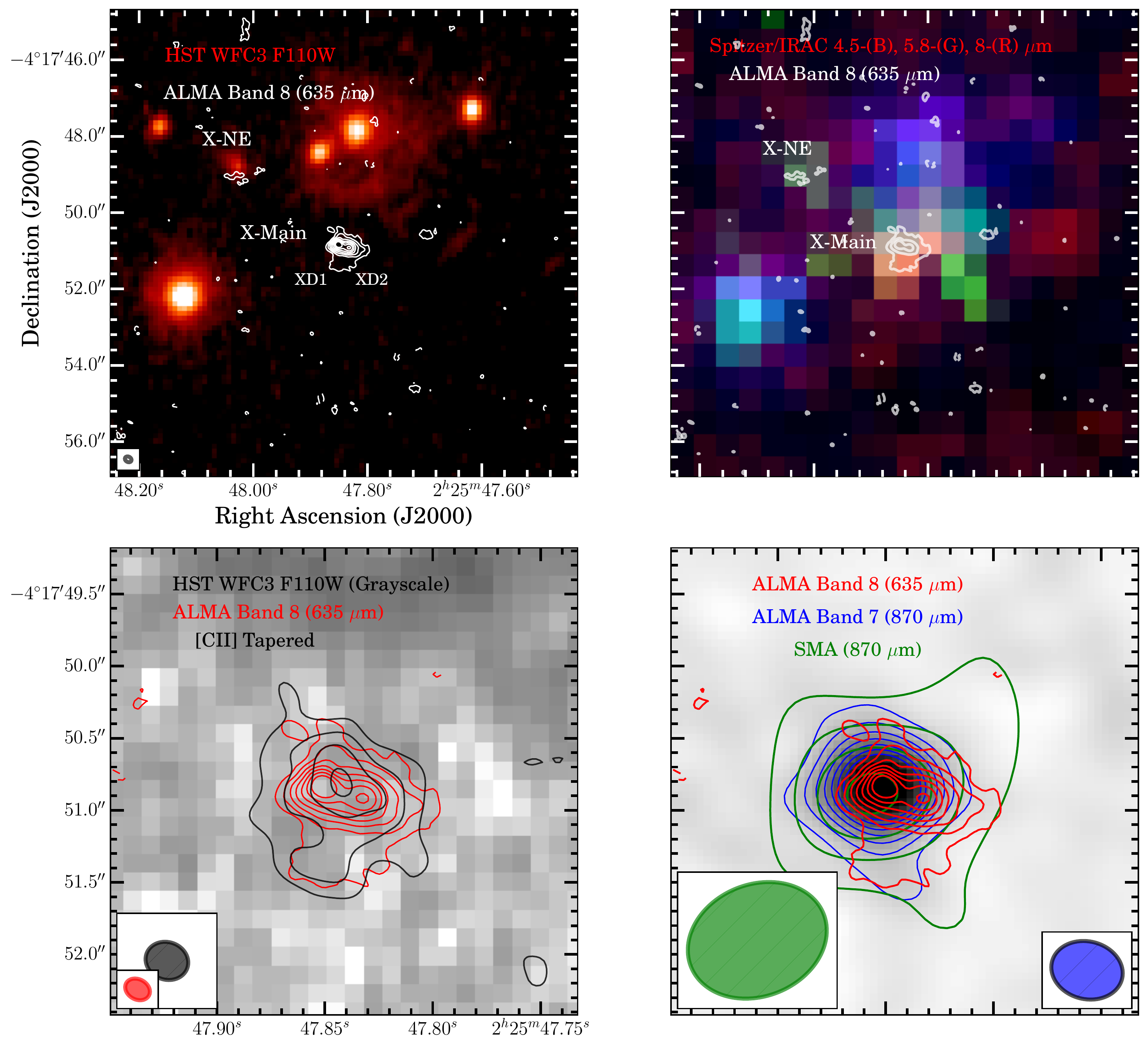}
\caption{
ALMA 635\,$\micron$ continuum emission (white contours) overlaid on an {\it HST} image (color-scale; top left; \citealt{Calanog14a})
and on a composite RGB image created from {\it Spitzer}/IRAC 4.5 (blue), 5.8 (green), and 8\,$\micron$ (red) data (top right).
Emission detected at 4.5\,$\micron$ is dominated by foreground sources (see also \Fig{galfit} in Appendix~\Sec{galfit}), but
emission at 5.8 and 8\,$\micron$ is dominated by \xa.
In addition, X-NE is detected in the UV/optical/NIR wavebands and in the \aco and \cii lines (see Figures~\ref{fig:co10spec} and \ref{fig:mom0}).
Bottom left: Contours of tapered \cii (black) and full resolution continuum (red) emission
overlaid on the same {\it HST} image as the top left panel (grayscale).
Bottom right: ALMA 635\,$\micron$ (red) and SMA 870\,$\micron$ (green) continuum contours overlaid
on the ALMA 870\,$\micron$ image (blue contours and grayscale).
Note the different angular scales shown for the two rows.
Two distinct peaks seen at 635\,$\micron$ are spatially offset from the \cii emission.
One dust peak is detected at 870\,$\micron$ (XD1) and the other is SW of it (which we denote as XD2).
Contours are shown in steps of $\pm$3$n$\,$\times$$\sigma$, where $n$ is an integer,
$\sigma_{\rm 635}$\eq0.22mJy\,\bmm for ALMA 635\,$\micron$,
$\sigma_{\rm 870, ALMA}$\eq0.28mJy\,\bmm for ALMA 870\,$\micron$,
$\sigma_{\rm 870, SMA}$\eq0.92\,mJy\,\bmm for SMA 870\,$\micron$, and
 $\sigma_{\rm [CII]}$\eq0.68\,Jy\,\kms for the tapered \cii emission.
Synthesized beams are color-coded by the contours shown in the lower corner of each panel:
0\farcs18\,$\times$\,0\farcs14 (full resolution \cii),
0\farcs31\,$\times$\,0\farcs26 (tapered \cii),
0\farcs99\,$\times$\,0\farcs78 (SMA 870\,$\micron$),
and 0\farcs5\,$\times$\,0\farcs4 (ALMA 870\,$\micron$).
 \label{fig:newcont}}
\end{figure*}

We show the {\it Spitzer}/IRAC images in \Fig{newcont}.
Sources near \xa are detected in the IRAC IR and CFHT NUV bands (see also Appendix~\Sec{uv}), but \xa remains undetected.

Among the four 3\,mm spectral setups of the CARMA \obs, we do not detect continuum emission in the individual tunings.
A final continuum image created by averaging across all the tunings yields a weak
detection at 4$\sigma$ significance (see \Tab{photometry}).
In the PdBI 2\,mm setups, continuum emission remains undetected.
On the other hand, we detect Ka-band continuum emission underlying the \aco line at 31.3\,GHz
at $\gtrsim$\,5$\sigma$ significance, which remains unresolved at the resolution and sensitivity of the VLA data (\Fig{co10cii}, see \Tab{photometry}).
The centroid of the 31.3\,GHz continuum emission coincides with that of the \aco line emission, and its
flux density is consistent with that obtained from fitting a four-parameter model
(Gaussian plus a first order polynomial) to the \aco line spectrum extracted at the peak pixel.
We also detect unresolved continuum emission at observed-frame $\sim$1\,mm (rest-frame 260\,$\micron$) underlying the \jco line
at $\sim$15\,\sig\ significance (\Tab{photometry}).

Continuum emission underlying the \cii line at observed-frame 635\,$\micron$ is detected at a peak significance of $>$31$\sigma$
(see \Tab{photometry}).
Two dust peaks, separated by 2.4\,kpc, are detected at high significance.
One peak coincides with the 870\,$\micron$ emission centroid (\Fig{newcont})
and with the \aco emission centroid of X-Main (see \Fig{co10cii}), whereas the other dust peak
is offset to the SW (we denote these as XD1 and XD2, respectively, hereafter).
We measure the total continuum flux density using an aperture defined by the 1$\sigma$ contours.
We fit a two-component 2D Gaussian to the continuum image and find deconvolved source sizes of
(0\farcs39\pmm0\farcs05)\,$\times$\,(0\farcs36\pmm0\farcs05) for
XD1 and (0\farcs39\pmm0\farcs06)\,$\times$\,(0\farcs35\pmm0\farcs06) for XD2,
corresponding to physical sizes of about 3\,kpc for both components.
Since the deconvolved source sizes are larger than the beam size, the size measurements
are not limited by the resolution of the \obs.
The peak flux densities are 4.95\pmm0.38\,mJy\,\bmm for XD1
and 2.75\pmm0.28\,mJy\,\bmm for XD2 (\Tab{cont}).
Based on their total flux densities and sizes, their brightness temperatures
in the Rayleigh-Jeans limit are 1.12 and 0.63\,K, respectively,    
corresponding to $T_{\rm B, RJ}$\eq4.5 and 2.5\,K in the rest-frame.

We overplot the 635\,$\micron$ continuum emission with
the SMA and ALMA data observed at 870\,$\micron$ in \Fig{newcont}.
We fit a single-component elliptical Gaussian model to each of the 870\,$\micron$ images.
Only XD1 is detected at 870\,$\micron$.
We also convolve the 635\,$\micron$ data to the native resolution of the ALMA 870\,$\micron$ data, and
find a spatial offset between two peaks emission centroids.
The emission centroids are determined by fitting a two-component Gaussian model to the 635\,$\micron$ data
and a single component Gaussian model to the 870\,$\micron$ data. 
We thus conclude that XD2 is likely to be much fainter than XD1 at 870\,$\micron$, in order for it to remain undetected
down to a 3\sig\ limit of 0.84\,mJy\,\bmm.

While the \cii emission shows a monotonic velocity gradient (\Fig{mom1}), which suggests that \xa is a
rotating disk with ordered motions, the dust continuum is almost
exclusively produced at the two peaks embedded within the kpc-scale \cii disk (\Fig{newcont}).
Likely due to the limited surface brightness sensitivity of our \obs, the \cii line emission appears more irregular compared to the continuum.

We detect low surface brightness emission in the outer region of the 635\,$\micron$ dust continuum map,
which is consistent with the overall extent of the \cii and \aco emission (Figures~\ref{fig:co10cii} and \ref{fig:newcont}).
This diffuse component is likely to be more optically thin compared to XD1 and XD2,
which likely dominate the dust optical depth estimated at 635\,$\micron$ based on the integrated SED model (see \Sec{sed}), given that
they contribute $>$80\% to the total continuum flux at this wavelength.
X-NE (which is detected in CO and \cii line emission) is also weakly detected in the continuum at 635\,$\micron$
at $>$3\sig\ significance, and in the UV, optical, and NIR wavebands
(see the last two panels of \Fig{co10cii}, \Fig{newcont}, and \Fig{cont} in Appendix~\Sec{uv}).

\section{Analysis}  \label{sec:anal}

\subsection{Spectral Energy Distribution Modeling} \label{sec:sed}
We use the extensive multi-wavelength photometric data
available in the XMM-LSS field to determine the IR, dust, and stellar properties of \xa via SED modeling.
Previously, \citet{Wardlow13a} modeled the dust SED of \xa
by fitting a simple modified blackbody
to the photometry measured at (sub-)mm wavebands ({\it Herschel}-SMA-MAMBO),
assuming a dust emissivity index of $\beta$\eq1.5.
This model suggests an IR luminosity (rest-frame
$\lambda_{\rm rest}$\eq8$-$1000\,$\micron$) of \LIR$=$\,(3.2\pmm0.4)\,$\times$\,10$^{13}$\,\Lsun and a dust
temperature of $T_d$\eq(45\pmm1)\,K.
Here, we update the SED with more photometric data obtained since,
covering UV through radio wavelengths (see \Tab{photometry}).
We model the observed dust SED using a modified blackbody (MBB)
and the full SED using the \ncode{magphys} code \citep{Magphys15a}
to derive a stellar mass in a self-consistent way from the dust and stellar emission.

\subsubsection{Modified Blackbody Model}   \label{sec:mbb}
We model the dust SED of \xa by assuming a single-temperature modified blackbody,
which is parameterized by the characteristic dust temperature $T_d$.
We fit MBB-based SED models to 16 photometric points covering rest-frame IR-to-mm wavelengths (observed-frame
24\,$\micron$$-$3\,mm; see \Tab{photometry})
using the code \ncode{mbb\_emcee} \citep[e.g.,][]{Riechers13a, Dowell14a}.
To account for the absolute flux-scale uncertainties associated with the photometry
obtained with ALMA, SMA, PdBI, and CARMA,
we add in quadrature an additional 15\% uncertainty.
The model consists of a MBB component that accounts for the FIR emission 
and a power-law component blue-ward
thereof to describe the warmer dust emission at mid-IR wavelengths. 
The dust optical depth (as a function of wavelength) is
taken into account via the parameter $\lambda_0$, where dust emission at $\lambda<\lambda_0$ (rest-frame) is
optically thick ($\tau_\nu>1$). 
The dust mass is calculated using
\begin{equation}
M_d = S_\nu~D_L^2~[(1+z)~\kappa~B_\nu(T)]^{-1}~\tau_\nu~[1~-~\exp{(-\tau_\nu)}]^{-1},
\end{equation}
where $D_L$ is the luminosity distance and $B_\nu$ is the Planck function.
In estimating the dust mass, we assume an absorption mass coefficient of
$\kappa$\eq2.64\,m$^2$\,kg$^{-1}$ at $\lambda$\,=\,125.0\,$\micron$ \citep{Dunne03a}.
This (general) model is therefore parameterized by five free parameters:
a characteristic dust temperature ($T_d$); emissivity index ($\beta$); power-law index ($\alpha$); normalization
factor ($f_{\rm norm}$); and $\lambda_0$.
We impose uniform priors such that $T_d$\,$>$\,1\,K,
$\beta$\,$\in$\,[0.1, 20.0],
$\lambda_0$\,$\in$\,[1.0, 400.0]\,$\micron$, and
$\alpha$\,$\in$\,[0.1, 20.0].
We adopt the statistical means and 68$^{\rm th}$ percentiles of the
resulting posterior probability distributions as the ``best-fit'' parameters.
For comparison with literature values, we also fit MBB+power-law models
without the wavelength-dependent optical depth parameter (i.e., assuming optically thin dust emission).
All the best-fit parameters are listed in \Tab{mbb}.
We note that the 160\,$\micron$ photometry data is poorly fitted, which may suggest the
presence of a warmer dust component in \xa.
However, with the data at hand, this dust component cannot be constrained.
Fitting models to photometry excluding the 160\,$\micron$ data
yields physical parameters that are consistent with those listed in \Tab{mbb} within the
uncertainties.

\begin{deluxetable}{lccc}[!htbp]
\tabletypesize{\scriptsize}
\tablewidth{0.5\textwidth}
\tablecolumns{4}
\tablecaption{Dust properties of \xa obtained from fitting single-temperature MBB models to its dust SED.}
\tablehead{
\multicolumn{2}{c}{Parameter}      &
\colhead{General} &
\colhead{Optically thin}
}
\startdata
$T_d$                           & (K)                & 64\petm{5}{1}   &
49\petm{20}{14}  \\ [1.05ex]
$\beta$                         &                    & 2.2\petm{0.3}{0.3}   & 1.8\petm{0.5}{0.5} \\ [1.05ex]
$\alpha$                        &                    & 2.2\petm{0.1}{0.2}    & 2.1\petm{0.1}{0.2}  \\ [1.05ex]
$\lambda_0$\tna                 & ($\micron$)        & 170\petm{20}{20}    & ---  \\ [1.05ex]
$f_{\rm norm}$\tnb & (mJy)              &  81\petm{6}{6} &
67\petm{16}{17} \\ [1.05ex]
$L_{\rm FIR}$\tnc                       & (10$^{13}$\,\Lsun) & 2.4\petm{0.2}{0.1} & 2.2\petm{3.1}{0.8}     \\ [1.05ex]
$L_{\rm IR}$\tnd                           & (10$^{13}$\,\Lsun) & 4.6\petm{0.2}{0.3} & 4.5\petm{3.0}{5.0}   \\ [1.05ex]
$M_{\rm d}$\tne                 & (10$^9$\,\Msun)    & 1.4\petm{0.3}{0.3}        & 4.3\petm{0.8}{4.0} \\ [-0.75ex]
\enddata
\label{tab:mbb}
\tablenotetext{a}{Rest-frame wavelength where $\tau_\nu$\,=\,1.}
\tablenotetext{b}{Normalization factor/flux density at observed-frame 500\,$\micron$.}
\tablenotetext{c}{Rest-frame 42.5$-$122.5\,$\micron$ luminosity.}
\tablenotetext{d}{Rest-frame 8$-$1000\,$\micron$ luminosity.}
\tablenotetext{e}{Derived assuming an absorption mass coefficient of $\kappa$\eq2.64\,m$^2$\,kg$^{-1}$ at $\lambda$\,=\,125.0\,$\micron$ \citep{Dunne03a}.}
\end{deluxetable}

\begin{deluxetable}{lccc}[!htbp]
\tabletypesize{\scriptsize}
\tablewidth{0.5\textwidth}
\tablecolumns{4}
\tablecaption{Properties of \xa from modeling its full SED with \ncode{magphys}.}
\tablehead{
\multicolumn{2}{c}{Parameter}      &
\colhead{All Photometry} &
\colhead{Excluding 160\,$\micron$\tna}
}
\startdata
$T_d$                   & (K)                         & 48\petm{9}{1}      &    44\petm{6}{5}      \\ [1.05ex]     
$L_{\rm IR}$\tnb                      & (10$^{13}$\,\Lsun) & 4.1\petm{1.4}{0.4}  & 3.9\petm{0.7}{0.4}   \\ [1.05ex]      
SFR\tnc                     & (\Msun\,yr\pmOne) & 3250\petm{890}{420}  &  2900\petm{750}{595}	 \\ [1.05ex]	
\mstar                   & (10$^{11}$\,\Msun)    & 7.2\petm{9.0}{3.8}       &     12\petm{13}{7} \\ [1.05ex]
sSFR                           & (Gyr\pmOne) & 4.7\petm{4.7}{2.6} & 2.4\petm{4.3}{1.4}   \\ [1.05ex]
$M_{\rm d}$\tnd   & (10$^9$\,\Msun)    & 3.0\petm{0.7}{0.7}        & 3.4\petm{0.4}{0.3} \\ [-0.75ex]
\enddata
\label{tab:magphys}
\tablenotetext{a}{The 160\,$\micron$ photometry forces the dust peak to shorter wavelengths, such that the photometry data long-ward thereof are poorly fitted --- motivating the choice of reporting both fits (see \Sec{magphys}).}
\tablenotetext{b}{Rest-frame 8$-$1000\,$\micron$ luminosity.}
\tablenotetext{c}{Assuming a \citet{Chabrier03a} IMF.}
\tablenotetext{d}{Derived by assuming the same absorption mass coefficient of $\kappa$\eq2.64\,m$^2$\,kg$^{-1}$ at $\lambda$\,=\,125.0\,$\micron$ as in the
MBB models.}
\end{deluxetable}

\begin{figure}[tbph]
\centering
\includegraphics[trim=0 0 0 0, clip, width=0.5\textwidth]{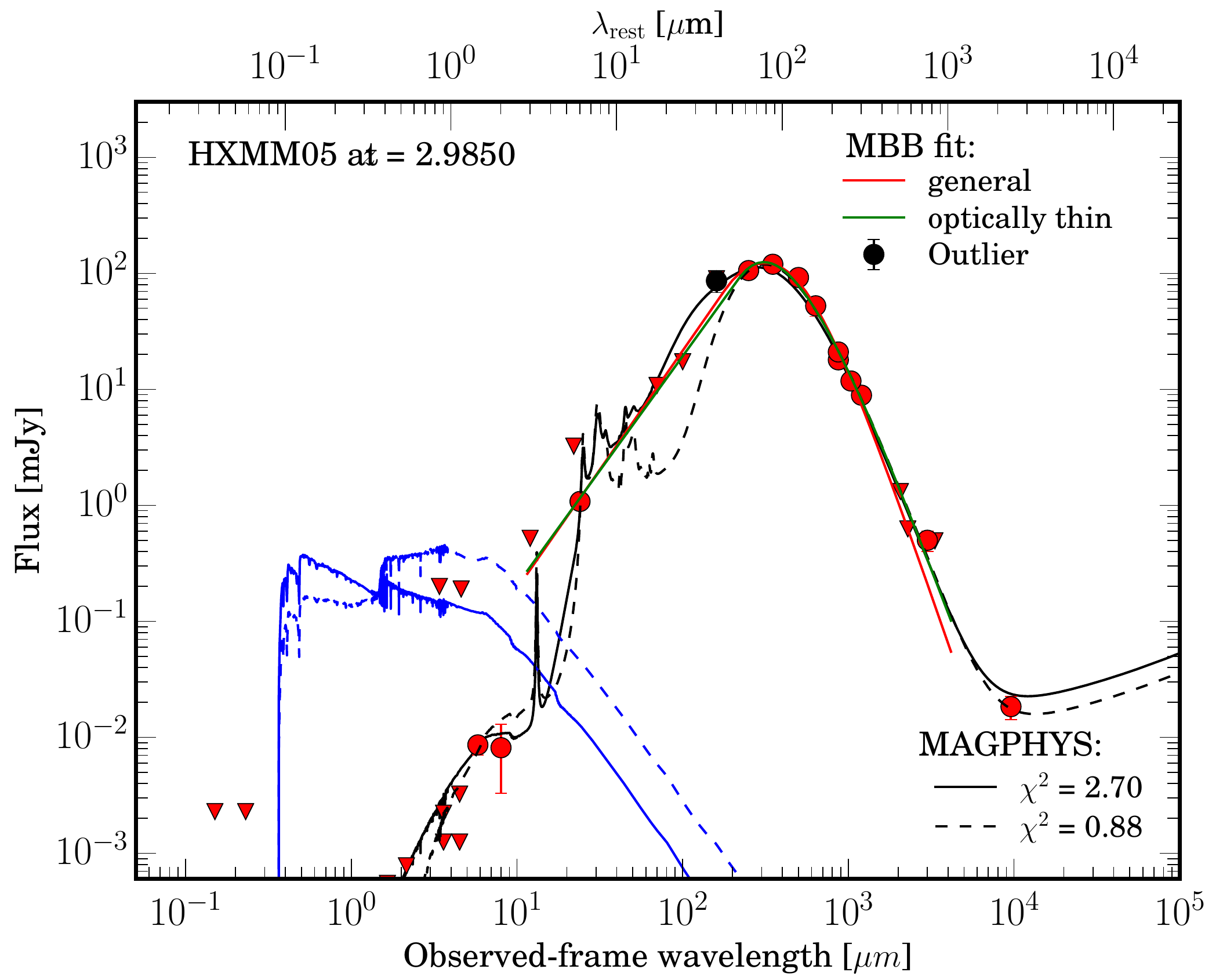}
\caption{
Best-fit MBB models (red and green solid lines) fitted to photometry covering
24\,$\micron$ through 3\,mm (error bars are also plotted, but since flux densities are shown on log-scale, they are not clearly visible).
The red line shows the best-fit general MBB model and the green line shows the best-fit optically thin model.
Solid black line shows the (attenuated) full SED obtained with \ncode{magphys} using photometry from FUV through 1\,cm.
Dashed black line shows the full SED fit excluding the
160\,$\micron$ photometry (see text). Blue lines show the unattenuated stellar spectra.
\label{fig:sed}}
\end{figure}

\subsubsection{MAGPHYS model} \label{sec:magphys}
To determine the stellar mass of \xa, we fit models to its full SED, sampled by the
FUV-to-radio wavelength photometry using
the \highz extension of \ncode{magphys} \citep{Magphys08a,Magphys15a}.
This code exploits a large library of optical and IR templates that are linked together physically through
energy balance, such that the UV-to-optical starlight is absorbed by dust and re-radiated in the FIR.
A detailed explanation of the \ncode{magphys} code and the model priors are given by \citet{Magphys15a}.

Following \citet{Magphys15a}, upper limits are taken into
account by setting the input flux densities to zero and uncertainties to upper limits.
The best-fit SED is shown in \Fig{sed} and
the resulting best-fit parameters
are listed in \Tab{magphys}.

Since in the best-fit model, the {\it Herschel}/PACS 160\,$\micron$ measurement forces the
dust peak to shorter wavelengths and worsens the fit at long wavelengths
(similar to the MBB fit), we re-model the SED excluding this outlier.
The resulting best-fit parameters are listed in \Tab{magphys}.
The dust peak in this fit is in good agreement with the (sub-)mm and radio photometry.
The best-fit parameters determined with and without the
PACS 160\,$\micron$ photometry are consistent within the uncertainties.
We thus adopt the parameters from the latter fit (i.e., excluding the 160\,$\micron$ outlier)
in the following sections.

\subsection{Dynamical Modeling}     \label{sec:dyn}
We fit dynamical models to the 1-kpc resolution \cii data obtained with ALMA to study the gas dynamics of \xa (more specifically, X-Main).
The monotonic velocity gradient observed in \cii suggests that \xa is a rotating disk galaxy,
an interpretation further supported by the analysis of \Sec{rotcur} below.

Assuming that the disk is circular and infinitesimally thin,
we use the kinematic major and minor axes to estimate the inclination angle, which yields $i$\eq46\pmm8$\degr$.
This is slightly different from the value estimated using the morphological axes, which yields 
$i$\eq35\pmm5$\degr$, but the two are
consistent within the error bars. We initialize the inclination angle in the following analyses based on these estimates.

\subsubsection{Harmonic Decomposition and Tilted-ring Model}  \label{sec:rotcur}

\begin{figure}[tbph]
\centering
\includegraphics[trim=0 0 0 0, clip, width=0.5\textwidth]{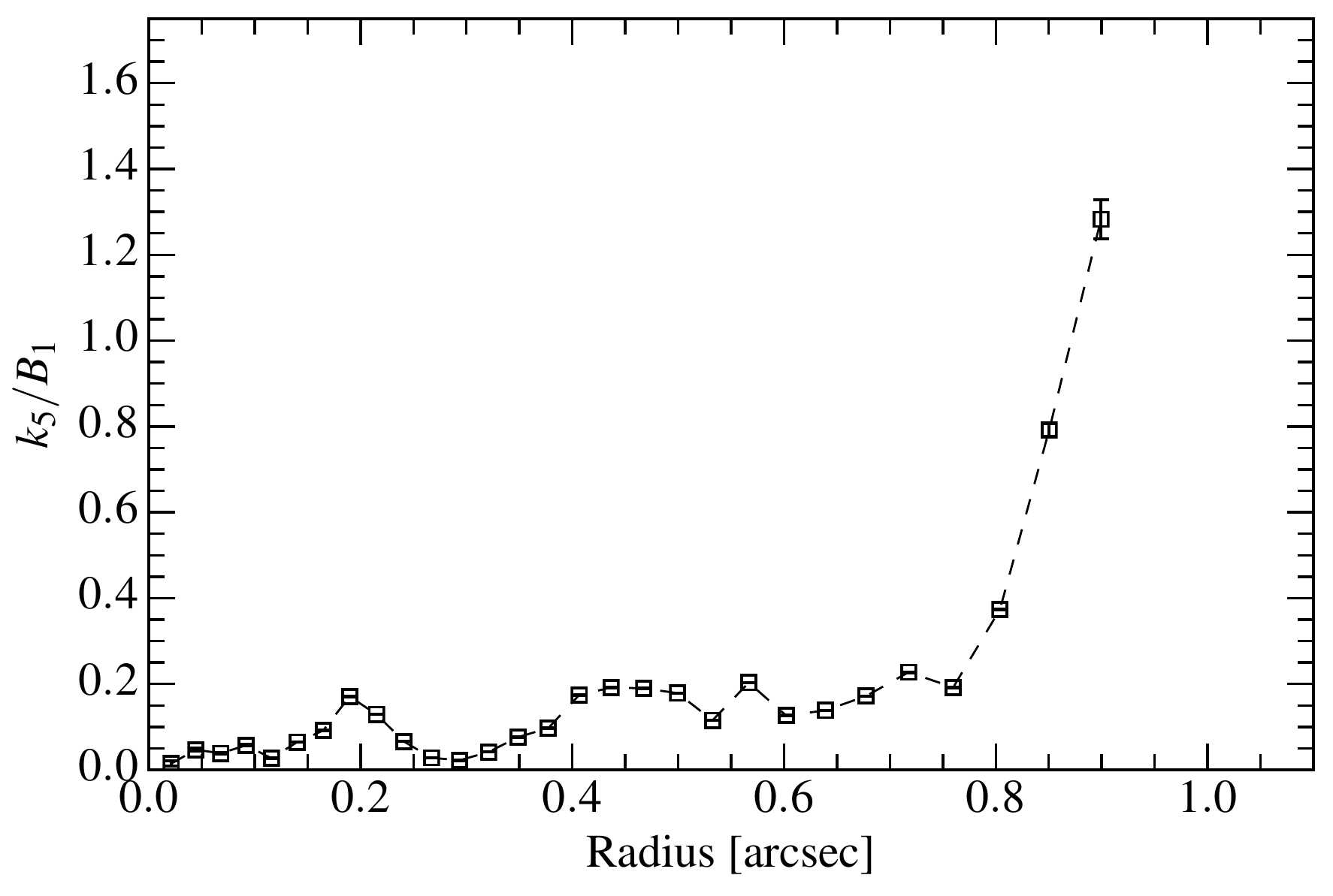}
\caption{
Ratio between the fifth- and first-order terms of the harmonic expansion of the \cii velocity field ($K_5/B_1$) as a function of radius.
The higher order term is insignificant compared to $B_1$ across nearly the entire disk, especially towards the
center ($R$\eq0), where the data have higher S/N.
\label{fig:k5}}
\end{figure}

To assess whether the velocity field observed towards HXMM05 is consistent with its gas being distributed in a disk rather than
effects caused by e.g., merging clumps, tidal debris, or inflows, we apply
harmonic decomposition analysis \citep{Schoenmakers97a}.
Briefly,
this method describes higher order moments, $K$ (e.g., LOS velocity) as a Fourier series:
\begin{align*}
K\left(\psi\right) & = A_0 + A_1 \sin(\psi) + B_1 \cos{(\psi)}~+ \\
& A_2 \sin(2\psi) + B_2 \cos(2\psi) + \dotsb,
\end{align*}
where $\psi$ is the azimuthal angle measured from the major axis.
The above can be recast into the following form:
\begin{equation}
K\left(r, \psi\right) = A_0(r) + \sum_{m} K_m(r)~\cos{\{m[\psi - \psi_m(r)] \} },
\end{equation}
where the amplitude and phase of the $m$-th order term are defined as
\begin{equation}
K_m\equiv\sqrt{A_m^2 + B_m^2}
\quad\mathrm{and}\quad
 \psi_m \equiv\arctan{\frac{A_m}{B_m}}.
\end{equation}
Since the velocity field is expected to be dominated by the cosine term
in the case of an ideal rotating disk; 
in this scenario, $B_1$ should dominate the harmonic terms,
with higher order terms $K_m$ measuring deviations from the ideal case.
Following \citet{Krajnovic06a}, we compare the fifth-order amplitude term to the first-order cosine term ($K_5/B_1$) to
quantify deviations in the \cii velocity map of \xa from a rotating disk,
and thus, differentiate between a rotation-dominated disk and a dispersion-dominated merger.
As shown in \Fig{k5}, the higher order term is insignificant compared to $B_1$ across the majority of the disk, especially towards the
center, where the data have higher S/N.
We thus interpret \xa to be a rotating disk for the remainder of this paper.

Given the modest inclination of \xa, we fit tilted-ring models \citep{Begeman89a}
to the observed velocity field using the task \ncode{rotcur} provided in the \ncode{gipsy} software package to
analyze the gas dynamics of \xa due to bulk motions (i.e., driven by the gravitational potential).
The tilted-ring model
assumes that the gas is in a circular, rotating thin disk, and describes the disk using a series of
concentric rings, where each ring can have an independent inclination angle ($i$),
major axis PA, rotation velocity ($v_{\rm rot}$), and
expansion velocity ($v_{\rm exp}$).
The rotation velocity is related to the projected LOS velocity via
\begin{equation}
v_{\rm LOS} = v_{\rm sys} + v_{\rm rot}\,\cos{(\psi)}\sin{(i)} + v_{\rm exp}\,\sin{(\psi)}\sin{(i)}.
\end{equation}

\begin{figure*}[tbph]
\centering
\includegraphics[width=0.36\textwidth]{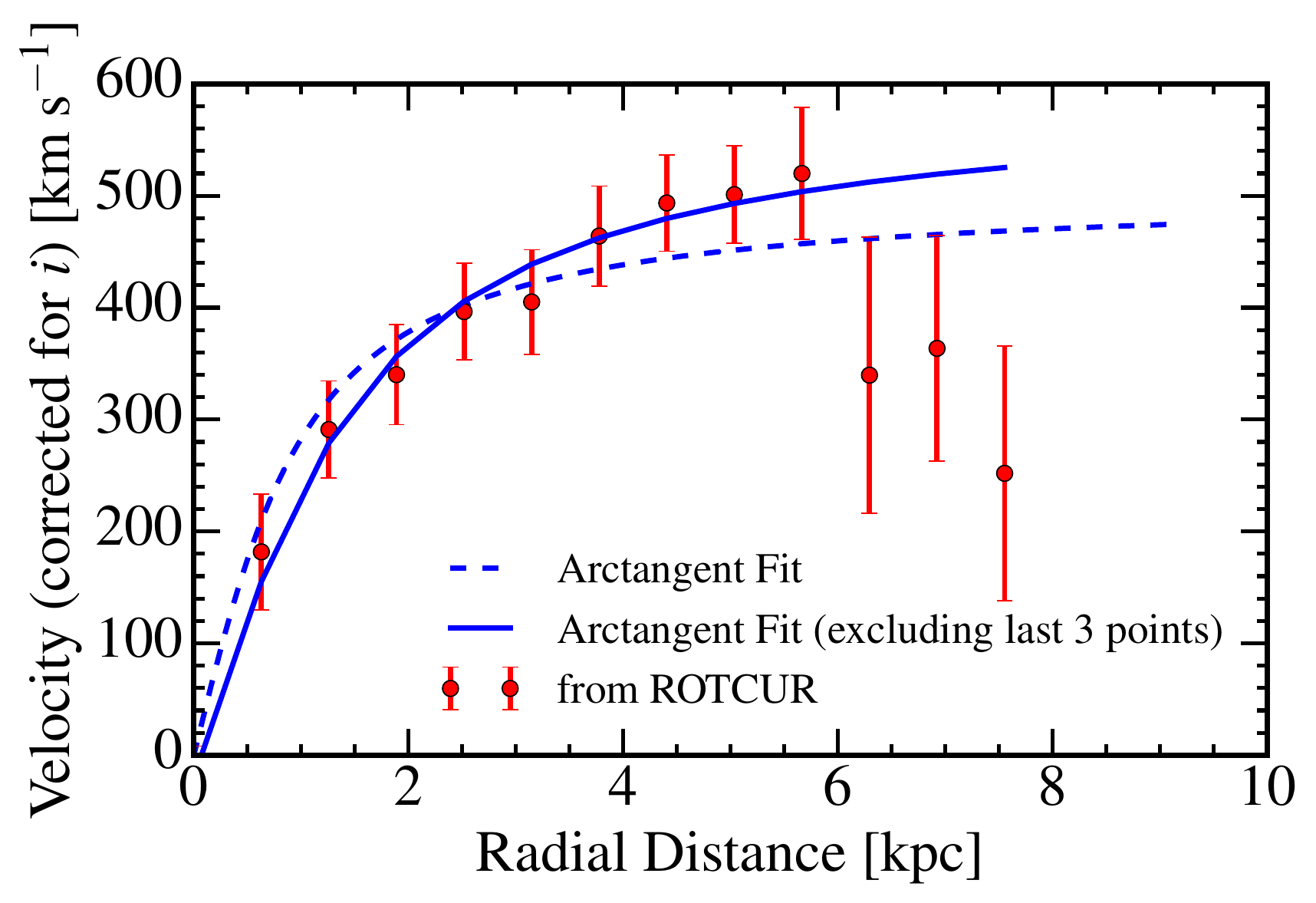}
\hspace{-1.25em}
\includegraphics[width=0.65\textwidth, height=0.18\textheight]{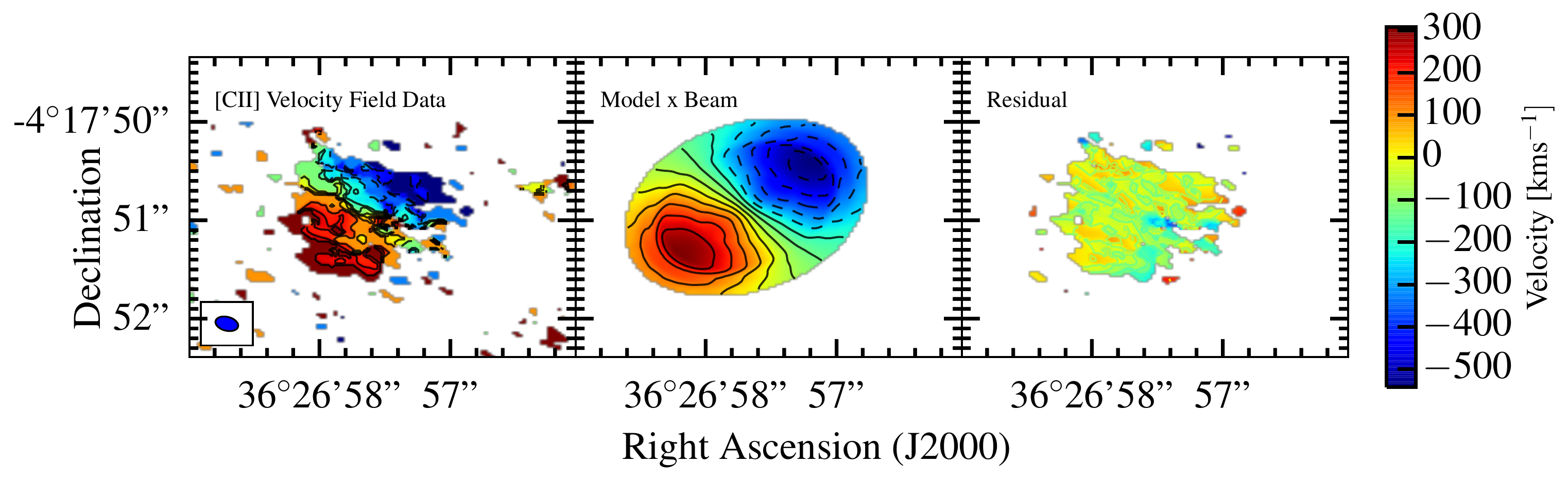}
\caption{
Left panel: Rotation curve of \xa based on a tilted-ring model and an extrapolation using an arctangent function.
The $y$-axis shows the rotation velocity after correcting for an inclination angle of $i$\eq41.3$\degr$.
Beyond a radius of $R$\eq6\,kpc, the rotation velocity appears to drop off, but this decrease is most likely
related to the limited S/N in the reddest velocity channels (see \Fig{chan} in Appendix~\Sec{channel}).
Right panel from left to right: Velocity fields seen in the data, best-fit model, and residual.
Uniform velocities varying between $v\in[0,100]$\,\kms are seen across the residual map.
\label{fig:gipsy}}
\end{figure*}

Here, we assume that
the observed LOS velocity is due entirely to
disk rotation and ignore any radial motions (e.g., due to inflow/outflow) by
setting the expansion velocity to 0\,\kms (i.e., the higher order $K_m$ terms).
We fit the model iteratively with different sets of parameters held fixed, while varying others freely.
We adopted this approach because each ring 
    would have six free parameters otherwise ($x_{\rm cen}$, $y_{\rm cen}$, $v_{\rm sys}$, 
    $i$, PA, $v_{\rm rot}$), which our data do not allow us to fix simultaneously, especially 
    because $v_{\rm rot}$ and $i$ are highly degenerate. Without doing so, models 
    struggle to converge to a solution\footnote{We have tested this by allowing $i$ and PA 
    also to vary across rings.}. The fact that this approach is also adopted 
    in modeling the kinematics of local galaxies, where the data obtained 
    have much higher S/N and spatial resolution, shows that our data do not 
    offer such constraining power \citep[e.g.,][]{Swaters09a, 
    vanEymeren09a, Elson14a, 
    Hallenbeck14a, DiTeodoro15a, Jovanovic17a}. 
    This approach is also adopted in fitting 
    low-S/N and coarser spatial resolution data obtained at high redshift
    (see e.g., \citealt{Shapiro08a}).
Here, we minimize the set of freely varying parameters via least-squares fitting.
Except in the last iteration, the width of each ring is set to the beam size.
In the first iteration, the dynamical center ($x_{\rm cen}, y_{\rm cen}$) and
systemic velocity ($v_{\rm sys}$) vary freely, whereas the inclination angle is fixed to
the average value found from the kinematic and morphological axes,
and the PA is fixed to the photometric/morphological PA.
We then constrain $i$, PA, and $v_{\rm rot}$ while fixing
$x_{\rm cen}, y_{\rm cen}$, and $v_{\rm sys}$ to their weighted-average values
found in the previous iteration.
To better determine the inclination angle, we further fix the PA and fit for $i$ and $v_{\rm rot}$ only.
In the final run, we fix all parameters to the weighted averages found in the previous iterations
and only fit for $v_{\rm rot}$, and the width of each ring is set to half the beam size to sample the rotation curve.
From the model, we find a best-fit PA of 133.6$\degr$\pmm0.6$\degr$
(east of north).
and an inclination of $i$\eq41.3$\degr$\pmm3.9$\degr$.

After this determination, the best-fit parameters are used to form the model velocity field using the \ncode{velfi} task.
A residual image (\Fig{gipsy}) is obtained by subtracting the model (after convolving with the beam) from the data.
The residual is largely uniform across the entire disk, with velocities varying by less than 100\,\kms,
consistent with the velocity dispersion map observed in \Fig{mom1}.
The relatively low residuals indicate
that the best-fit model is a reasonable description of the observed velocity field, and
that non-circular motions (e.g., streaming motions along unseen
spiral arms or bars, or large-scale tidal torquing from galaxy interactions)
are unlikely to be detected in the kpc-scale resolution data.
We note that beam smearing means that velocity information within the inner kpc region will be
largely lost in the data.

We fit an arctangent model \citep[e.g.,][]{Courteau97a} to the \ncode{rotcur} rotation curve (RC).
The model is parameterized as:
\begin{equation}
V_{\rm rot} = V_0 + \frac{2}{\pi} V_{a} \arctan\left(\frac{R}{R_{t}}\right),
\end{equation}
where $V_{\rm rot}$ is the rotation velocity found with \ncode{rotcur},
$V_0$ is the systemic velocity,  
$V_{a}$ is the asymptotic velocity, and $R_{t}$ is the ``turnover''
radius at which the rising part of the rotation curve begins to flatten.
We perform non-linear least-squares fitting using the Levenberg-Marquardt
algorithm to find the best-fit parameters.
We limit the turnover radius to 0$<$\,$R_{t}$\,$<$25~kpc
in order to keep this parameter within a physically meaningful range.
Using this model, we find $V_{a}$\,=\,503\,$\pm$\,83\,\kms,
$R_{t}$\,=\,0.8\,$\pm$\,0.3\,kpc,
and $V_0$\,=\,0\,$\pm$\,28\,\kms (relative to the systemic redshift).
We thus find an inclination-corrected rotation velocity of
$v_{\rm rot}$\eq474\pmm78\,\kms at a spatial offset of 8.8\,kpc   
(the extent of the ground state CO line emission; \Tab{line}).
We note that the model underestimates the velocities at $R\gtrsim4$\,kpc because of the
outermost three data points at $>$6\,kpc, which deviate from the trend of increasing velocity with radius.
Such a trend --- a declining rotation curve with increasing galactocentric radius --- has been reported in some
studies of \highz galaxies (e.g., \citealt{Genzel17a, Lang17a}, cf. e.g., \citealt{Tiley18a}).
In our data, this trend is likely an artifact due to the limited S/N at those PV-positions
(i.e., low number of pixels fitted; see Figures~\ref{fig:pvfit} and \ref{fig:chan} in Appendix~\Sec{channel}).
In other words, the decreasing velocities seen at increasing radius in our target could easily be mimicked by
a lack of sensitivity to low surface brightness emission in the outer regions.
If we instead fit the arctangent model excluding these three data points,
we find an inclination-corrected rotation velocity of $v_{\rm rot}$\eq537\pmm83\,\kms at 8.8\,kpc
and an asymptotic velocity of $V_a$\eq617\pmm97\,\kms. Both models are consistent within the uncertainties.

Rotation curves from both arctangent models do not reach the
terminal velocity\footnote{Terminal velocity is not the same as asymptotic velocity, which
the arctangent model {\it does} constrain.} (i.e., the flat part of the rotation curve).
Therefore, the rotation velocities inferred here may be lower limits only.
On the other hand, part of the rotation curve that is flattening is clearly detected in the PV-diagram (\Fig{pvfit}).
This discrepancy is related to the fact that fitting models to velocity fields can underestimate
true rotation velocities,\footnote{This underestimation occurs because velocity fields are intensity-weighted and the tilted-ring model assumes that all the
gas in a ring is at a unique position along the LOS; however,
gas emission from other velocities along the LOS is blended within the beam.
Thus, the lower the resolution, the more likely the true velocities are underestimated by fitting models to the velocity fields.
}
and that the decreasing velocities seen in the outermost three data points of the rotation curve are of limited S/N.
This flattening part of the rotation curve detected in \xa is likely to be mainly driven by
the dynamics of the parent dark matter halo, as in nearby galaxies;
we see no evidence indicating that \xa is dominated by baryons from the data at hand.
Adopting the inclination-corrected $V_a$ as the maximum rotation velocity, we find that \xa is consistent with
the gas Tully-Fisher relation found for nearby galaxies,
given its gas mass (see \Tab{phy}; \citealt{McGaugh15a}).

\subsubsection{Envelope-Tracing Method} \label{sec:ET}

As an alternative approach to estimate the rotation velocity of \xa, we also use the
envelope-tracing (ET) method, where we fit models to
the PV diagram extracted along the kinematic major axis
(\Fig{pvfit}; see review by \citealt{Sofue01a}).
The ET method attempts to trace out the material
that has the maximum tangential motion along each LOS
(see Figure 5 of \citealt{Chemin09a} for a schematic depiction of this geometric effect).

We fit a third order ($h_3$) Gauss-Hermite
polynomial to a (Hanning-smoothed) spectrum extracted at each position along the PV cut (\Fig{pvfit})
to account for any asymmetries in the spectra.
The rotation curve (traced by the ``envelope'') is derived from the terminal velocity ($v_t^{\rm obs}$) at which
8\% of the total flux under the fitted curve is outside $v_t^{\rm obs}$.
In essence, this approach traces the
isophotes at each position along the kinematic major axis.

The innermost 1.5\,kpc region of the PV diagram is steeply rising (\Fig{pvfit}), which is due in part to
the facts that the velocity gradient in this region is changing rapidly
from positive to negative, and that contributions from multiple radii overlap in the inner roughly 1 kpc
(which remains unresolved at the $\sim$1.2\,kpc resolution of the data).  
Structures within the ``envelope'' modulo inclination and beam smearing effects may result from
the presence of spiral- or ring-like structures, or a clumpy gas distribution in \xa. 

Based on the terminal velocity, we derive the
rotation velocity of \xa using the following equation:
\begin{equation}
v_{\rm rot} = (v_t^{\rm obs} - v_{\rm sys})/\sin{(i)} - \sqrt{(\sigma_{\rm PSF}^2 + \sigma_{\rm ISM}^2)},
\end{equation}
where $v_{\rm sys}$ is the systemic velocity determined from fitting a double-Gaussian to the \cii spectrum (\Fig{spectra}),
$i$ is the inclination angle from \ncode{rotcur},
$\sigma_{\rm PSF}$ is the spectral resolution, and $\sigma_{\rm ISM}$ is the velocity dispersion of the
gas (see e.g., \citealt{Vollmer16a} and \citealt{Sofue17a}).
Here, we adopt
the observed velocity dispersion of $\sigma_v$\eq75\,\kms as the
subtracted term.
We then re-sample the rotation profile every half beam (instead of every pixel) and show the output
rotation curve in \Fig{pvfit}.
We find an inclination-corrected rotation velocity of $v_{\rm rot}$\eq616\pmm100\,\kms at
the last measured radius of the rotation curve ($R$\eq8.0\,kpc), which is consistent with the rotation
velocity of $v_{\rm rot}$\eq537\pmm83\,\kms derived from the arctangent model within the uncertainties.

\subsection{PDR modeling}    \label{sec:pdr}
Photo-dissociation regions (PDRs) are the warm and dense surfaces of molecular clouds exposed to
FUV photons with energies 6\,$<$\,$h\nu$\,$<$\,13.6\,eV escaping from H{\scriptsize II} regions.
In PDRs, gas temperatures and densities are typically
$T$\eq100\,$-$\,500\,K and $n$\eq$10^{2-5}$\,cm$^{-3}$.   
Since the \cii 158\,$\micron$ line is the primary coolant in PDR conditions satisfying
$n$\,$\lesssim$\,10$^3$\,cm$^{-3}$ and $T\lesssim100$\,K,  
\cii and other ISM lines near or in PDRs
are sensitive probes of the physical conditions of the PDR gas and
the intensity of the ambient interstellar radiation field
\citep[$G_0$; conventionally expressed in units of 1.6\E{-3}\,erg\,cm$^{-2}$\,s\pmOne, the Habing flux; e.g.,][]{Hollenbach99a}.
Using the \cii and CO line luminosities
and the PDR model grids from \citet{Tielens85a} and 
\citet{Kaufman99a, Kaufman06a}\footnote{Available through the PDR Toolbox\footnote{http://dustem.astro.umd.edu} described by \citet{Pound08a} and \citet{Wolfire10a}
}
, we constrain the globally-averaged $G_0$, hydrogen density ($n$), and surface temperature
for the PDRs in \xa\footnote{While it is physically unrealistic to model an entire galaxy as a single PDR,
we infer the 
$G_0$ and $n$ values of \xa in a globally-averaged sense to facilitate
comparison with other studies in the literature.}.
We adopt CO grids from an updated version of the code
(\citealt{Hollenbach12a}; M. Wolfire 2017, private communication) that is
merged with the \ncode{Meudon} code \citep{LePetit06a}
for a more detailed treatment of H$_2$ chemistry and thermal balance.

\begin{figure}[tbph]
\centering
\includegraphics[width=0.47\textwidth]{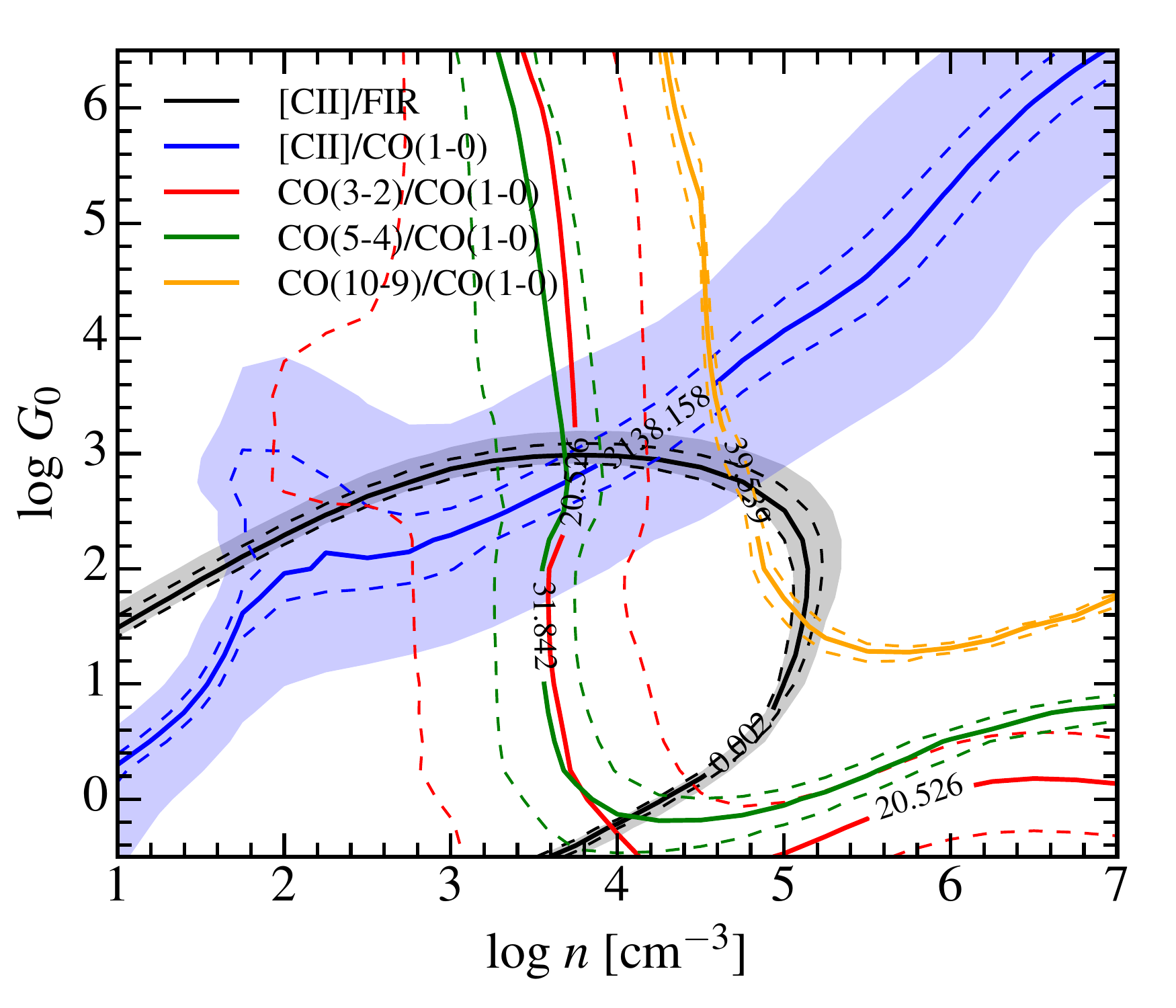}
\caption{Ratios of CO, \cii, and far-IR luminosities (solid) as functions of $G_0$ and $n$.
Model grids are adopted from PDR Toolkit, except for the CO grids (see text).
Dashed lines show the uncertainties associated with these ratios.
Filled regions show larger uncertainties, after including the corrections for
the non-PDR fraction of \cii emission (30\%) and the factor of $2$ in CO due to optically thick CO emission (see text).
\label{fig:pdr}}
\end{figure}

The observed line ratios are shown in \Fig{pdr} as functions of $G_0$ and $n$.
Since a fraction of the \cii emission in the ISM arises outside of PDRs, and we lack other diagnostic lines to
determine this fraction in \xa, we adopt a canonical value of 30\% to account for this non-PDR contribution
(\citealt{Carral94a, Colbert99a, Malhotra01a, Oberst06a}; see also \citealt{Pavesi16a, Pavesi18a, Zhang18b}).
CO line emission is typically optically thick (especially the low-$J$ lines), and so we
multiply their line intensities by a factor of $2$ to account for the emission on the other side of the surface.
Corrections are incorporated into the line ratios as uncertainties (filled regions in \Fig{pdr}).
The best-fit model is determined based on the global minimum $\chi^2$,
corresponding to $\log\,n$\eq4.5\,cm$^{-3}$ and $\log\,G_0$\eq2.25.  
Based on the $\chi^2$ surface, the uncertainties in both $n$ and $G_0$
are approximately an order of magnitude.
As discussed by \citet{Rollig07a}, physical parameters inferred from any PDR models should not be taken too literally,
since they are subject to differences depending on the assumptions adopted and the implementation of microphysics in the code.
Nevertheless, we use the best-fit parameters as simple approximations to compare \xa with other galaxies.

The lower $G_0$ solution ($G_0 < $1) implied by the mid-$J$ CO and
\cii-to-FIR (\Lcii/\LFIR) luminosity ratios disagrees with that implied by the CO (10\rarr9)-to-(1\rarr0)
and \cii-to-\aco 
luminosity ratios.
We reject this low $G_0$ solution since it would require a physically enormous emitting
region to account for the observed high \LFIR in \xa ($G_0\propto\LFIR/D^2$; \citealt{Wolfire90a}).
Assuming the values for M82 ($D\simeq$\,300\,pc, $G_0\simeq$\,1000,~\LFIR$\simeq$\,2.8\E{10}\,\Lsun),
the solution with $G_0\simeq$\,0.2 would require an emitting region 
$D$\eq600\,kpc in size, contrary to what is observed. On the other hand, the best-fit $G_0$\ssim200
corresponds to an emitting region of $D\simeq$\,20\,kpc,
which is more consistent with the sizes observed in \cii and \aco line emission (\Tab{line}).
The FUV radiation field intensity of \xa is thus stronger than the local Galactic interstellar
radiation field intensity by a factor of around 200, comparable to the values found in
nearby normal star-forming galaxies and those found in some other DSFGs \citep[e.g.,][]{Malhotra01a, Wardlow17a}.
The best-fit $G_0$ and $n$ together suggest a surface temperature of $T_{\rm surf}$\eq290\,K for the PDR.
We approximate the PDR pressure using $P\propto nT$, yielding $P/k_B$\eq9.0\E{6}\,cm$^{-3}$\,K.
We note that an offset is found between the CO\,($J$\eq10\rarr9)/CO($J$\eq1\rarr0) line and the other luminosity ratios
in the $\log n-\log G_0$ plane. This
offset likely results from the fact that \jco emission preferentially traces a more highly-excited phase of the ISM
than the other lines (e.g.,
due to mechanical heating or X-ray heating; see also \Sec{sled}).
However, with the data at hand, the presence and properties of an X-ray dominated region (and/or a second PDR component,
and/or shock excitation) are unconstrained and indistinguishable from a simple single PDR.

The PDR properties thus suggest that the
high far-IR luminosity of \xa ($>$10$^{13}$\,\Lsun) may result from extended \SF, with only a modest FUV radiation field intensity.
This is in stark contrast with the compact starbursts seen in the cores of many nearby ULIRGs (less than 
a few hundred parsecs),
which are found to have stronger FUV radiation fields compared to \xa \citep[e.g.,][]{Stacey10a}.
The inferred PDR conditions also suggest that \xa is unlikely to host an AGN or a powerful quasar,
consistent with \Sec{agn}.

\section{Discussion}  \label{sec:diss}
\def\arraystretch{1.2}
\begin{deluxetable}{lcc}
\tabletypesize{\scriptsize}
\tablewidth{0.5\textwidth}
\tablecaption{Physical properties of \xa obtained from dynamical, SED, and PDR modeling.
\label{tab:phy}}
\tablehead{
Parameter  &
Unit            &
Value
}
\startdata
$r_{\rm 3,1}$\tna   &   & 0.76\pmm0.28 \\ [.3em]
$r_{\rm 5,1}$\tna   &   & 0.26\pmm0.10 \\ [.3em]
$r_{\rm 10,3}$\tna  &   & 0.04\pmm0.01 \\ [.3em]
$r_{\rm 10,1}$\tna  &   & 0.03\pmm0.01 \\ [.3em]
$i$                         & deg                         & 41.3\pmm3.9 \\ [.3em]
\aco extent          & kpc                            & 8.8\pmm2.6 \\[.3em]
Rotation curve radius                      & kpc                         & 8.0 \\ [.3em]  
Maximum $v_{\rm rot}$\tnb               & \kms                        & 616\pmm100 \\ [.3em]    
$M_{\rm dyn}$($R<$8.0\,kpc)\tnb               & 10$^{11}$\,\Msun               & 7.7\pmm3.1 \\ [0.3em]      
Observed $\sigma_{\rm \cii}$         & \kms                        & 100\pmm25 \\ [.3em]
\Lcii                       & 10$^{10}$\,\Lsun              & 4.3\pmm0.5  \\ [.3em]
$L_{\rm CO}$                & 10$^{7}$\,\Lsun              & 1.3\pmm0.4  \\ [.3em]
$M_{\rm gas}^{\rm total}$ & 10$^{11}$\,\Msun            & 2.1\pmm0.7  \\ [0.3em]
$M_{\rm gas}^{\rm X-Main}$ & 10$^{11}$\,\Msun           & 1.7\pmm0.4 \\ [0.3em]
$M_{\rm gas}^{\rm X-NE}$  & 10$^{10}$\,\Msun            & 6.5\pmm1.6 \\ [0.3em]
\LIR\tnc                    & 10$^{13}$\Lsun              & 3.9\petm{0.7}{0.4}  \\ [0.3em]
SFR\tnd                     & \Msun\,yr\pmOne             & 2900\petm{750}{595}    \\ [0.3em]
$T_{d}$              & K                           & 44\petm{6}{5} \\ [0.3em] 
$M_d$\tne          & 10$^9$\Msun                 & 2.9\pmm1.5 \\ [0.3em]
$M_*$                       & 10$^{12}$\Msun              & 1.2\petm{1.3}{0.7} \\ [0.3em]
GDR\tne       &                               &  50\,$-$\,145 \\ [0.3em]
$f_{\rm gas}^{\rm dyn}$\tnf        &                 \%               &    33\pmm15   \\ [.3em]
$f_{\rm gas}^{\rm dyn, iso}$\tng &     \%                     & 5.3\pmm2.4 \\ [0.3em]
\mgas/\mstar            &                               & 0.2\petm{0.2}{0.1} \\ [.3em]
\mgas/(\mgas$+$\mstar)            &                               & 0.2\pmm0.2 \\ [.3em]
$\tau_{\rm depl}$              & Myr                         &     72\pmm27  \\ [.3em]
SFE\tnh                         & Gyr\pmOne                   & 13\pmm4 \\ [.3em]
$\Lcii/\LFIR$\tni             &       \%                            &  0.20\pmm0.03    \\ [.3em]
$\Lcii/L_{\rm CO}$\tni             &                         &  3300\pmm1000    \\ [.3em]
$L_{\rm CO}/\LFIR$\tni             & 10$^{-7}$                     &  5.5\pmm1.9    \\ [.3em]
12$+$log(O/H)\tnj               &                               & 9.1 \\ [.3em]
$\alpha_{\rm CO, Z}$\tnk      &    \alphaU          & 1.4$-$1.9 \\ [.3em]
$\Sigma_{\rm SFR}^{\rm XD1}$          & \Msun\,kpc$^{-2}$\,yr\pmOne & 210\\ [0.4em]
$\Sigma_{\rm SFR}^{\rm XD2}$          & \Msun\,kpc$^{-2}$\,yr\pmOne & 120\\ [0.4em]
$\Sigma_{\rm SFR}$          & \Msun\,kpc$^{-2}$\,yr\pmOne & 10 \\ [0.4em]
$\Sigma_{\rm mol}$          & \Msun\,pc$^{-2}$            & 590\pmm410 \\ [-0.5em]
\enddata
\tablenotetext{a}{Brightness temperature ratio between two CO rotational transitions: $r$.}
\tablenotetext{b}{From PV fitting, see \Sec{ET}.}
\tablenotetext{c}{From SED modeling excluding the photometry obtained with {\it Herschel}/PACS at 160\,$\micron$ and
{\it Spitzer}/MIPS 24\,$\micron$ (see \Sec{magphys}). Integrated over rest-frame 8$-$1000\,$\micron$ luminosity.}
\tablenotetext{d}{Averaged over the last 100\,Myr and assuming a \citet{Chabrier03a} IMF.}
\tablenotetext{e}{Dust masses derived from MBB and full SED modeling lie in the range of $M_{d}$\eq(1.44\,$-$\,4.27)\E{9}\,\Msun, which is within
the uncertainties due to intrinsic uncertainties in $\kappa_\nu$ (see \Sec{sed}). Here we quote the average dust mass as the centroid value
and use the extreme values as the uncertainty.}
\tablenotetext{f}{\mdyn determined from dynamical modeling (\Sec{dynmass}).
}
\tablenotetext{g}{Determined using $M_{\rm dyn}$\eq2.8\E{5}~$\Delta v_{\rm FWHM}^2$~$R_{\rm FWHM}$\eq(3.89\pmm1.09)\E{12}\,\Msun to compare
\xa with other \highz galaxies reported in the literature.}
\tablenotetext{h}{SFE\,$\equiv$\,\LFIR/\mgas, where \LFIR\eq2.39\petm{0.16}{0.14}\E{13}\,\Lsun.}
\tablenotetext{i}{Observed quantities. No corrections applied
(see \Sec{pdr}).
}
\tablenotetext{j}{Based on the FMR derived by \citet{Mannucci10a}. On the PP04 scale, the metallicity of \xa is 12$+$log(O/H)$_{\rm PP04}$\eq8.74.}
\tablenotetext{k}{Based on the \alphaco-metallicity relation derived by \citet{Leroy11a} and \citet{Genzel12a} (see \Sec{alpha}).}
\tablecomments{Any quantities in this table relating to the gas mass (and throughout this paper) are derived from \aco emission,
assuming \alphaco\eq0.8\,\alphaU and using
the total molecular gas mass (i.e., both X-Main and the NE component combined) unless otherwise specified.}
\end{deluxetable}

Since X-Main and X-NE remain spatially unresolved from each other in the IR photometry and most of the
spectral line data (except in CO\,$J$\eq1\rarr0 and \cii emission), we discuss the
properties of \xa as a combined system in the following sections.

\subsection{No Evidence of an AGN in \xa} \label{sec:agn}

Given the upper limits imposed on the X-ray luminosity of \xa, we
find no evidence for the presence of a powerful AGN,
but we {\em cannot} rule out the possibility of a heavily dust-obscured AGN in \xa or
a Seyfert galaxy nucleus with modest X-ray emission.
To assess the reliability of the stellar mass derived from SED fitting,
we examine if the mid-IR spectral slope of \xa ($S_\nu$$\propto \nu^{\alpha}$) may be consistent with a low X-ray luminosity AGN
\citep[see e.g.,][]{Stern05a, Donley07a}. 
We fit a power-law to the IRAC 5.8- and MIPS 24-$\micron$ photometry, which correspond to
rest-frame 1.5$-$6.0\,$\micron$.
We find a spectral index $\alpha_{\rm 1.5-6, rest}$\eq1.46\pmm0.58,
which is much flatter than those observed in AGN host galaxies\footnote{
Spectral indices reported in the literature are based on photometry taken at 3.6$-$8.0\,$\micron$, which
correspond to the closest wavelength range used here for \xa in the rest-frame.}
\citep{Stern05a, Donley07a, Donley08a}, suggesting that the NIR emission in \xa
may be dominated by stellar emission.
Thus, we assume in the following that all the NIR emission
detected in the IRAC channels 3 and 4 bands arises solely from the starlight in \xa.
That is, the accuracy of the stellar mass estimated is dominated by the uncertainty on the IMF adopted \citep[see e.g.,][]{Zhang18a}.
If \xa were to host an AGN, however, its stellar mass and SFR would be overestimated.

\subsection{ISM Properties} \label{sec:ISM}

\subsubsection{Stellar Mass and Specific Star Formation Rate} \label{sec:mstar}
We find an unusually high stellar mass of 10$^{12}$\,\Msun for \xa from SED modeling.
The stellar mass estimate relies heavily on IRAC channels 3 and 4 (i.e., rest-frame 1.4 and 2.0\,$\micron$)
photometry.
Previous studies have shown that rest-frame $K$-band (2.2\,$\micron$) photometry appears to be a
reliable proxy\footnote{Since the dust optical depth of \xa
at rest-frame 158\,$\micron$ is $\tau_\nu$\ssim1, its $K$-band emission
could be highly attenuated, unless most of the
starlight is less attenuated than the dust (e.g., if the latter is dominated by compact star-forming knots
and the former is much more extended), which
remains possible given its dust morphology, gas excitation, and $G_0$.} for the
stellar mass of galaxies, since photometry in this band is relatively insensitive to the past star-formation histories of galaxies
(e.g., \citealt{Kauffmann98a, Lacey08a}; cf. \citealt{Kannappan07a}),
and because NIR emission is less affected by dust extinction compared to optical light.
In particular, the difference in the $K$-band luminosity between initial burst and constant
star formation models is less than a factor of 3 (e.g., \citealt{PerezGonzalez08a}).
The main systematic uncertainties associated with \mstar are therefore
the star-formation histories assumed, the IMF and stellar population synthesis model adopted,
and the fact that differential dust extinction is not captured in simple energy balance models (e.g., \ncode{magphys})\footnote{Alternatively, hot dust emission due to a deeply buried AGN could contribute to the observed IR luminosity, and thus
lead to an overestimate of \mstar (but see \citealt{Michalowski14a}, who find insignificant effects of AGN on the SED-derived
\mstar).}.
Nevertheless, the stellar masses inferred from \ncode{magphys} are found to match the true masses of mock galaxies in
simulations fairly well
\citep[e.g.,][]{Michalowski14a, Hayward15a, Smith15a}, unless the dust attenuation in \xa is underestimated by \ncode{magphys}.
Taken at face value\footnote{Note, however, that even assuming no AGN is present in \xa,
the \mstar estimate is accurate to only $\lesssim$0.5\,dex (see also \citealt{Michalowski14a})
on top of the large statistical error bars reported in \Tab{magphys}.}, the high stellar mass suggests that a substantial fraction of
stars may have already formed in some massive galaxies by $z$\eq3 (approximately 2 Gyr after the Big Bang).

The relatively tight ``correlation'' found between SFR and \mstar for
star-forming galaxies at low- and \highz  
suggests that the majority of galaxies are
forming stars over a long duty cycle in a secular mode \citep[e.g.,][]{Rodighiero11a, Lehnert15a}.
The specific SFR of sSFR\eq2.37\petm{4.31}{1.43}\,Gyr\pmOne of \xa
is consistent with the star-forming ``main-sequence'' (SFMS) within the
scatter of the MS relations derived by
\citet{Tacconi13a}, \citet{Lilly13a}, \citet{Speagle14a}, and \citet{Schreiber15a}, if we extrapolate them to higher masses
and include the uncertainties associated with the SFR and stellar mass inferred for \xa.
One possible caveat is the applicability of the SFMS relation, and whether
our current knowledge of
the MS is meaningful at high stellar mass (10$^{12}$\,\Msun).
In this paper, we only consider \xa as a MS galaxy for the sake of comparing its ISM properties
with other \highz MS and starburst systems.

\subsubsection{Gas Mass, Gas-to-Dust Ratio, and Metallicity}    \label{sec:alpha}
Using the \aco line intensities (\Tab{line}) and
assuming a CO luminosity-to-H$_2$ mass conversion factor of \alphaco\eq0.8\,\alphaU \citep[e.g.,][]{Downes98a},
we derive molecular gas masses of
$M_{\rm gas}^{\rm X-Main}$\eq(1.68\pmm0.43)\E{11}\,\Msun for X-Main,  
$M_{\rm gas}^{\rm NE}$\eq(6.52\pmm1.63)\E{10}\,\Msun for X-NE,
and $M_{\rm gas}^{\rm total}$\eq(2.48\pmm0.65)\E{11}\,\Msun for the entire system (\Fig{comom0}).
Using the molecular gas mass of the system, we find a gas-to-dust mass ratio of
GDR~(\alphaco\,$/$\,0.8)$^{-1}$\eq50$-$145, which is
consistent with those measured in the Milky Way, local spiral galaxies,
ULIRGs, and DSFGs \citep[]{Draine07a, Wilson08a, Combes13a, Bothwell13a}.

Based on the fundamental metallicity relation (FMR)
determined by \citet{Mannucci10a}, we infer a gas-phase metallicity of $Z$\eq12$+$log(O/H)\eq9.07 for \xa\footnote{This assumes that the FMR relation holds up to $z$\eq3 and a
stellar mass of 10$^{12}$\,\Msun (see e.g., \citealt{Steidel14a}).},
which is comparable to the that of the $z$\eq4 SMG GN20 \citep{Magdis11a}.
We express the metallicity on the Pettini \& Pagel (\citeyear{PP04}; PP04) scale using the calibration proposed by
\citet{Kewley02a} and \citet{Kewley08a}, yielding $Z_{\rm PP04}$\eq8.74.
The range of GDR derived for \xa
is consistent with the best-fit GDR\,$-$\,$Z_{\rm PP04}$ relation presented by \citet{Magdis11a}, which was determined for
a sample of local galaxies studied by \citet{Leroy11a}. If the CO-to-H$_2$ conversion
factor were \alphaco$>$0.8\,\alphaU, then \xa would lie above this relation.
By applying the \alphaco\,$-$\,$Z$ relations found by \citet{Leroy11a} and \citet{Genzel12a},
we find a range of \alphaco~of 1.4$-$1.9\,\alphaU, which would increase
the molecular gas mass by a factor of 1.7$-$2.4 compared to the value assumed here.

\subsubsection{Dust, Gas, and Stellar Mass Ratios}    \label{sec:gaspoor}
The dust-to-stellar mass ratio (DSR) measures the amount of dust
per unit stellar mass that survives all dust destruction processes in a galaxy (e.g., type II SN explosions).
The DSR of \xa is 2.3\petm{2.7}{1.8}\E{-3}, which is within the range measured in local
star-forming galaxies and ULIRGs, but is among the lowest measured in
intermediate-\z ULIRGs and quasars \citep[e.g.,][]{Dunne11a, Combes13a, Leung17a}.
This ratio is also lower than those measured in
DSFGs at similar redshifts \citep[e.g.,][]{Magdis11a, Calura17a}.

The molecular gas-to-stellar mass ratio of \xa is $M_{\rm gas}$/\mstar\eq0.2\petm{0.2}{0.1}, which is
higher than those observed in local SFGs and early-type galaxies \citep[e.g.,][]{Leroy08a, Saintonge11a, Young14a}.
Previous studies
report a positive evolution in this ratio with redshift \citep[e.g.,][]{Tacconi10a, Dave12a}.
The $M_{\rm gas}$/\mstar ratio of \xa is lower than those typically measured in
other \highz SFGs and DSFGs at \z$>$\,1.2, and is
the lowest\footnote{The $M_{\rm gas}$/\mstar ratio is susceptible to
uncertainties in the \alphaco~conversion factor and in stellar mass.
If we were to adopt a conversion factor of 4.6\,\alphaU, the gas-to-stellar mass ratio of \xa would be consistent with the
expected redshift evolution of the molecular gas mass content in galaxies \citep{Geach11a}.}
found among massive galaxies at
\z$\simeq$\,3 to date \citep{Leroy08a, Tacconi10a, Daddi10a, Geach11a, Decarli16b}.

The low DSR and gas-to-stellar ratio of \xa may indicate that it is a relatively evolved system, in which
a large fraction of its gas has been converted into
stars and a large fraction of dust has been locked up in stars.
That said, as discussed in \Sec{agn} and \Sec{mstar}, it is possible that the stellar mass maybe overestimated.

\subsection{Star Formation Efficiency and Gas Depletion Timescale}       \label{sec:sfe}
To first order, the star formation efficiency (SFE) measures the \SF rate per unit mass of molecular gas available in a galaxy.
The SFE of \xa is \LFIR/\Lp\eq91\pmm25\,\Lsun\,(\LpU)\pmOne (or 13\pmm4\,Gyr\pmOne), which is  
slightly higher than but consistent with the range found in
nearby active star-forming spiral galaxies
\citep[$z$\,$<$\,0.1;][]{Gao04a, SV05a, Stevens05a, Leroy08a, Wilson09a, Leroy13a} and
high-$z$ massive disk-like galaxies \citep{Daddi08a, Daddi10a, Aravena14a}.
Assuming that the \SF in \xa continues at its current rate without gas replenishment,
the gas will be depleted in $\tau_{\rm depl}$\,=\,72\pmm27\,Myr\footnote{However, the gas reservoir would last 6 times longer if we
had instead adopted \alphaco\eq4.6\,\alphaU in deriving \mgas.},
comparable to the depletion times in starburst systems.
\xa thus lies between SFMS and starburst galaxies in the so-called ``transition region''
on the integrated version of the ``star-formation law'' (i.e., \LFIR$-$\,\Lp relation; \citealt{Daddi10a, Magdis12a, Sargent14a}).
We conclude that the gas depletion timescale in \xa is short compared to those of SFMS galaxies at high redshift.

\subsection{Dynamical Mass}     \label{sec:dynmass}

The rotation curve of a galaxy reflects
its dynamics due to the total (i.e., baryonic and dark matter) enclosed mass.
We estimate the total dynamical mass enclosed within 8\,kpc
using \mdyn$= v_{\rm rot}^2~R/G$.    
We find an inclination-corrected dynamical mass of \mdyn\eq(7.7\pmm3.1)\E{11}\,\Msun.
Taken at face value\footnote{The dominant systematic uncertainties in \mdyn are the
uncertainties in the rotation velocity due to
the potential presence of inflows or outflows, in the
velocity dispersion, and in our assumption that \xa is a thin disk with negligible scale height.}, we find a molecular gas mass fraction of
$f_{\rm gas}^{\rm dyn}$\eq$M_{\rm mol}$/$M_{\rm dyn}$\eq18\pmm8\%
using the gas mass of
the main component of \xa only ($M_{\rm gas}^{\rm X-Main}$) and
33\pmm15\% using the total molecular gas mass of the system ($M_{\rm gas}^{\rm total}$).
The dynamical mass is consistent with the stellar mass within the considerable uncertainties.

Since the dynamical masses derived for most other \highz galaxies in the literature are based on marginally
resolved or unresolved \obs,
we also estimate the dynamical mass of \xa using the isotropic estimator
$M_{\rm dyn}^{\rm iso}$\eq2.8\E{5}~$\Delta v_{\rm FWHM}^2$~$R_{\rm FWHM}$ \citep[e.g.,][]{Engel10a},
where $\Delta v_{\rm FWHM}$ is the \aco line FWHM measured by fitting a single-Gaussian to the line profile
in units of \kms, and $R_{\rm FWHM}$ is the FWHM extent of the galaxy measured from \aco line emission in units of kpc.
Here, we adopt the linewidth of \xa excluding X-NE as $\Delta v_{\rm FWHM}$
and the average between the major and minor axes
as the extent ($R_{\rm FWHM}$\eq7.6\,kpc).
We thus find an inclination-corrected dynamical mass of $M_{\rm dyn}^{\rm iso}$\eq(3.89\pmm1.09)\E{12}\,\Msun,
yielding a molecular gas mass fraction of
$f_{\rm gas}^{\rm dyn, iso}$\eq4.3\pmm2.9\% using $M_{\rm gas}^{\rm \xa}$
for X-Main only and 5.3\pmm2.4\% for the \xa system.
However, given the evidence of disk-like rotation for \xa, we consider
the first dynamical mass estimate (i.e., \mdyn) to be more reliable.

\subsection{Star Formation Rate and Gas Surface Densities and the Spatially-resolved Star-formation Law}    \label{sec:sd}

The Schmidt-Kennicutt relation (i.e., the star-formation law) is an empirical
relation relating SFR and gas surface densities as
$\Sigma_{\rm SFR}\propto\Sigma_{\rm gas}^N$, where $N\simeq$1.4 is established
from measurements of different nearby galaxy populations
\citep[e.g.,][]{Schmidt59a, Kennicutt98a, Kennicutt08a}.
Based on the SFR of 2900\petm{750}{595}\,\Msun yr\pmOne and the sizes and flux ratio of XD1 and XD2 at 635\,$\micron$,
we find star formation rates of SFR\eq1500 and 860\,\Msun\,yr\pmOne and
SFR surface densities of $\Sigma_{\rm SFR}^{\rm XD1}$\eq210\,\Msun yr\pmOne kpc$^{-2}$ and
$\Sigma_{\rm SFR}^{\rm XD2}$\eq120\,\Msun\,yr\pmOne\,kpc$^{-2}$
for XD1 and XD2, respectively.
These SFR surface densities are elevated compared to those measured in the circumnuclear \SB regions
of nearby galaxies \citep{Kennicutt98b}, consistent with the overall somewhat shorter gas depletion timescale
but are much lower than those observed in \highz
``maximal starburst''-like galaxies, such as 
the $z$\eq5.3 SMG AzTEC-3, the $z$\eq5.7 HyLIRG ADFS 27, and the $z$\eq6.3
HFLS3 \citep[][]{Riechers13a, Riechers14a, Riechers17a, Oteo17b}.

For the low surface brightness diffuse dust component, which is
almost as extended as the \cii line emission (\Fig{newcont}), we find
a source-averaged SFR surface density of
$\Sigma_{\rm SFR}\eq$10\,\Msun\,yr\pmOne\,kpc$^{-2}$
(or 60\,\Msun\,yr\,\pmOne\,kpc$^{-2}$ including the pair of nuclei).
Based on the CO($J$\eq1\rarr0) line-emitting source size of (8.8\pmm2.9)\,$\times$\,(6.4\pmm3.5)\,kpc and the
total molecular gas mass measured in the \xa system,
the molecular gas surface density
is $\Sigma_{\rm gas}$\,$=$\,(590\pmm410)\,$\times$\,(\alphaco/0.8)\,\Msun\,pc$^{-2}$.
We thus find that \xa lies along the ``starburst sequence'' of the Schmidt-Kennicutt relation reported
by \citet{Bouche07a}. 
Accounting for uncertainties in SFR and gas surface densities,
we find that \xa lies in the same region of parameter space as the sub-regions of GN20 and the
$z$\eq2.6 SMG SMM\,J14011$+$0252 \citep{Sharon13a, Hodge15a}.

\subsection{CO Gas Excitation}       \label{sec:sled}
Due to the different physical conditions
required to excite the various rotational transitions of CO, flux ratios between the low- and high-$J$ CO
lines are sensitive to the molecular gas volume densities and kinetic temperatures.
With the data at hand (i.e., only four CO lines spanning the CO ``ladder'' up to $J$\eq10\rarr9),
we do not attempt to fit radiative transfer models to the observed line fluxes. Instead,
we compare the line ratios measured in \xa (\Tab{phy}) with those of other galaxy populations to
study the connection between the SFR surface density and the gas excitation in \xa
(since SFR surface density is tightly linked to gas density, temperature, and line optical depths).

\begin{figure}[tbph]
\centering
\includegraphics[width=0.47\textwidth]{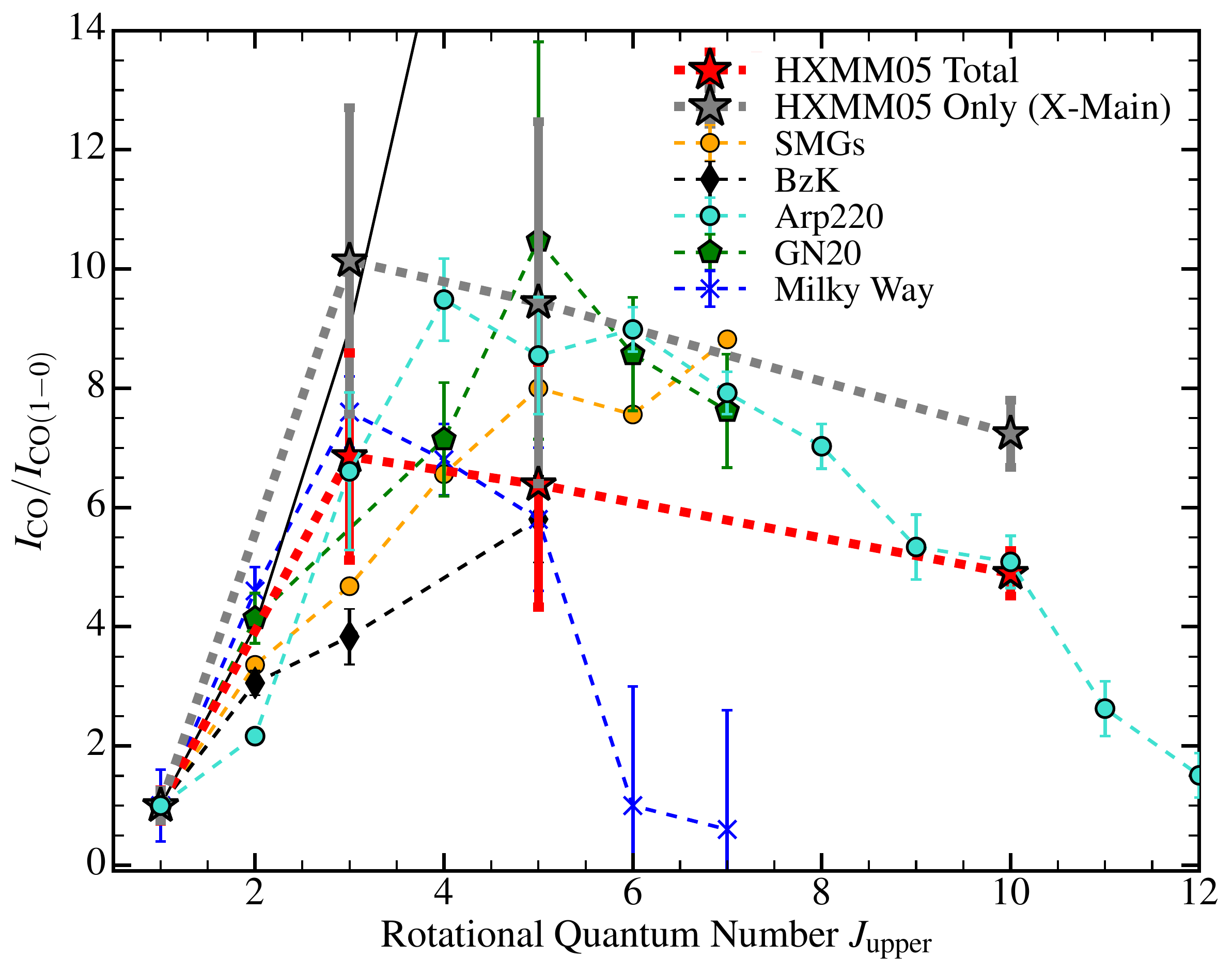}
\caption{
CO SLEDs of \xa and other low- and \highz galaxies reported in the literature.
Line fluxes are normalized to the \aco line.
Red stars show the SLED of \xa based on the total \aco flux (i.e, X-Main and X-NE; see \Fig{comom0}).
Grey markers show the SLED of X-Main only, relative to its \aco flux.
Solid black line shows the SLED expected for thermalized excitation and optically thick emission.
Literature data are compiled from:
\citet{Fixsen99a, Greve09a, Carilli10a, Rangwala11a, Danielson11a, Riechers13a, CW13, Bothwell13a, Kamenetzky14a}; and \citet{Daddi15a}.
\label{fig:sled}}
\end{figure}

The global SFR of the \xa system is comparable to those of the most luminous DSFGs known, but
their different CO spectral line energy distribution (SLED)
shapes indicate that the underlying physical conditions in their ISM may be different.
As shown in \Fig{sled}, the gas excitation of the \xa system probed by transitions up to $J_{\rm upper}$\eq5
is lower than those typically observed in nuclear starbursts, SMGs, and quasars,
but is comparable to those observed in the outer disk of the Milky Way (despite \xa's much higher SFR), and
those observed in \highz BzK disks. Such a relatively modest gas excitation
is in accord with the modest source-averaged SFR surface density of \xa and
its PDR gas conditions (\Sec{pdr}) --- i.e.,
its total SFR of 2900\,\Msun\,yr\pmOne is spread across the entire disk (as seen in the co-spatial gas and FIR dust distribution),
and its extended \SF is embedded within a medium with only moderate radiation flux and pressure.

Including the highest-$J$ line probed with the data at hand ($J$\eq10\rarr9), we find that the overall
SLED shape (and thus gas excitation) of \xa resembles that of the local merger-driven ULIRG Arp\,220.
This may suggest that the molecular ISM of the \xa system is composed of (at least) two gas-phase components ---
a diffuse extended cold component and a dense warm component.

If we exclude X-NE, we find that the molecular gas in X-Main (i.e., grey symbols in \Fig{sled}) is more highly excited
than the system overall, which is comparable to other \highz DSFGs \citep[e.g.,][]{Riechers11e, Sharon16a}.
As noted in \Sec{co}, the true \eco flux may be a factor of two higher.
In this case, the excitation conditions of \xa and X-Main would be consistent with
(and possibly more excited than) those of other DSFGs.
However, higher fidelity data are needed to confirm this scenario.

\subsection{Morphology and Kinematics of the \cii and CO Emission}
\label{Sec:kin}

The cold molecular gas reservoir of \xa is approximately 9\,kpc\,$\times$6\,kpc in diameter, which is comparable
to the mean size of nearby disk-like U/LIRGs
\citep[]{Ueda14b}, 
and of ULIRGs in general (\citealt{Gao99a}).
Similarly extended gas reservoirs have also been observed
in some other high-$z$ galaxies \citep[][]{Daddi10a, Ivison10d, Riechers11a, Ivison11a, Hodge12a}.

The \aco FWHM linewidth of \xa is much broader than those typically observed in
``normal'' star-forming galaxies at low- and high-redshifts,
ULIRGs, 
and \highz gas-rich galaxies \citep{Solomon97a, SV05a,
Daddi10a, Danielson11a, Ivison11a, Riechers11a, Riechers11d, CW13, Combes13a, Sharon16a},    
although galaxies with similarly broad line do exist \citep[e.g., J13120$+$4242;][]{Hainline06a, Riechers11a}.
Similarly, the \cii line of \xa ($\Delta v$\eq667\pmm46\,\kms)
is also broader than those seen in many other \highz galaxies apart from
major mergers \citep[e.g.,][]{Ivison10c, Walter12a, Ivison13a, Riechers13a,
Riechers14b, Neri14a, Rhoads14a}.

The velocity dispersion traced by the \cii line emission in \xa is the highest in the central 0\farcs2 region,
as seen in \Fig{mom1}.
The higher dispersion at the center may indicate gas dynamics
affected by late-stage merger activity, intense cold gas accretion/inflows, or
enhanced turbulence caused by an undetected AGN,
or the fact that systemic motions/radial velocities change abruptly in this region,
where the velocity curve is also the steepest. 
We therefore estimate the
gas dispersion in the extended part of \xa based on the velocity dispersion observed in its outskirts,
yielding $\sigma_v\simeq$75\,\kms.

We estimate the $v/\sigma$ stability parameter for \xa using the maximum rotation velocity and
the observed velocity dispersion of $\sigma$\eq75\,\kms (see \Sec{cii}), yielding $v/\sigma$\eq7\pmm3.
This ratio is closer to those measured in nearby disk galaxies ($\simeq$10) than in
other \highz galaxies (e.g., \citealt{Genzel06a, ForsterSchreiber06a, Cresci09a, Gnerucci11a, Schreiber17a}).
This distinction may suggest that the ISM of \xa is not as turbulent as other \highz galaxies studied to date \citep[e.g.,][]{Law09a, Jones10a, Swinbank11a} ---
perhaps a result of its lower gas mass fraction\footnote{Compared based on $f_{\rm gas}^{\rm dyn, iso}$.} compared to other \highz galaxies.
However, in most \highz studies with reliable $v/\sigma$ estimates (requiring spatially resolved information),
the ratio is typically derived from {\em stellar} kinematics
(examples based on CO line emission are still limited in number; e.g., \citealt{Swinbank11a}).
Determining the gas stability of galaxies by imaging their molecular gas reservoirs
is more meaningful for characterizing their prospects for \SF, since molecular gas is the raw fuel for \SF.
In other words, the stability of gas against gravitational collapse is more closely linked to \SF
than the velocity structure of the existing stellar component, which may (re-)settle on a different timescale
from the gas after perturbations.

\subsection{Dust}

\subsubsection{Morphology and Optical Depth}
\xa remains undetected in deep UV and optical images,
indicating that it is highly dust obscured, consistent with its rest-frame 158$\micron$ optical depth of $\tau_\nu\simeq$1
(determined from SED modeling; \Sec{sed}).
This optical depth exceeds those of most ``normal'' star-forming galaxies and nearby
disk galaxies, but is similar to that seen in
Arp\,220 and \highz starburst galaxies --- e.g., HFLS3,  AzTEC-3, and ADFS 27 \citep{Riechers13a, Riechers14a, Riechers17a}.

The dust emission morphologies at 635 and 870\,$\micron$ appear different (\Fig{newcont}).
While two compact dust components are found to be embedded within
an extended component at 635\,$\micron$, only one compact component
coincides with XD1 at 870\,$\micron$.
The second 635\,$\micron$ dust peak, XD2, is 1.8 times fainter than XD1 at its peak flux (see \Tab{cont}).
If XD1 and XD2 were to have the same peak flux ratio at 635 and 870\,$\micron$, we
would expect
a peak flux density of 6.2\,mJy\,\bmm for XD2 at 870\,$\micron$, which we would have detected at $>$20\,$\sigma$
significance.
Hence, the non-detection of XD2 at 870\,$\micron$ may be a result of
the lower dust column density at 870\,$\micron$, where the emission is optically thin on average based on\footnote{The
optical depth derived from
a galaxy-averaged SED model is luminosity-weighted, and thus, biased toward compact dust components.
The true optical depth is likely to be even lower in the outskirts of a galaxy. }
the best-fit dust SED model ($\tau_\nu$\eq0.54).

\subsubsection{Interpretation of the compact dust components} \label{sec:clump}
 The compact dust components, XD1 and XD2, detected at 635\,$\micron$
 could be two regions of intense \SF, or the remnant cores from
a previous merger \citep[e.g.,][]{Johansson09a}.
At the positions of the double nuclei,
the velocity field of \xa is the steepest (see markers in \Fig{mom1}), 
but we find no obvious signs of a misaligned velocity gradient at their positions,
which would be expected for the latter scenario.
However, the velocity field is only a first-order representation of the kinematics of a galaxy,
since it is calculated based on intensity-weighted LOS velocities and is affected by the limited
spatial resolution of the data (similarly for the velocity dispersion map).
Thus, it will not capture the full kinematics in the system.
As such, we cannot rule out the possibility that the
double nuclei may be the cores of a pair of progenitor galaxies,
where the gas disk may have reconfigured itself into rotation already \citep[e.g.,][]{Springel05a, Robertson06a,
Narayanan08b, Robertson08a, Hopkins09a}.
Such a scenario would be reminiscent of the nearby ULIRG Arp\,220 \citep[e.g.,][]{Sakamoto08a, Scoville17a},
but with a greater separation between the pair in \xa.

Alternatively, if the dust peaks were truly giant star-forming ``clumps'' that are virialized, we would expect their velocity
dispersions to be $\sigma_v\simeq$\,40\,\kms or $\simeq$\,400\,\kms
based on the size-linewidth relations found for
local GMCs in a quiescent environment or the Galactic center, respectively \citep[][]{Larson81a}.
As shown in \Fig{mom1}, the observed velocity dispersion in the nuclei of \xa is 160$-$200\,\kms. Thus, the
dust peaks are unlikely to be virialized clumps.

Similarly, a scenario in which XD1 and XD2
correspond to the ``{\it twin} peaks'' produced in response to an
$m = 2$ (i.e., bar or oval)
perturbation \citep[see e.g.,][]{Kenney92a}
is disfavored for two main reasons:
the lack of obvious non-circular motions (\Sec{rotcur}), and
the pronounced asymmetry in the 635/870\,$\micron$ flux ratio of XD1 and XD2 (i.e., implying differences in
their optical depths and dust column densities).

\subsection{\cii and FIR Luminosity Ratio}   \label{sec:ratio}

The \Lcii/\LFIR ratio measures the fraction of far-UV photons
that is heating up the gas versus that deposited onto dust grains.
We find a \Lcii/\LFIR ratio of 0.20\pmm0.03\% for \xa.
Thus, \xa lies in the same region of parameter space as nearby star-forming galaxies and LIRGs,
despite its two orders of magnitude higher \LFIR \citep[e.g.,][]{Stacey10a}.
This ratio is consistent with those measured in other \highz \SF-dominated
galaxies with similar far-IR luminosities in the \Lcii/\LFIR$-$\,\LFIR plane
\citep[cf.~nearby ULIRGs and \highz quasars; ][]{Malhotra01a, Hailey-Dunsheath10a, Stacey10a, Wang13a, Diaz-Santos13a,
Zhang18b},
suggesting that \xa is dominated by extended \SF rather than
a compact starbursts or AGN (see also \Sec{xmm}).
This evidence is consistent with the extent observed in its gas and dust emission.

\subsection{Spatially Resolved \Lcii/\LFIR Map and Star Formation} \label{sec:sfrtracer}

\begin{figure}[tbph]
\centering
\includegraphics[width=0.47\textwidth]{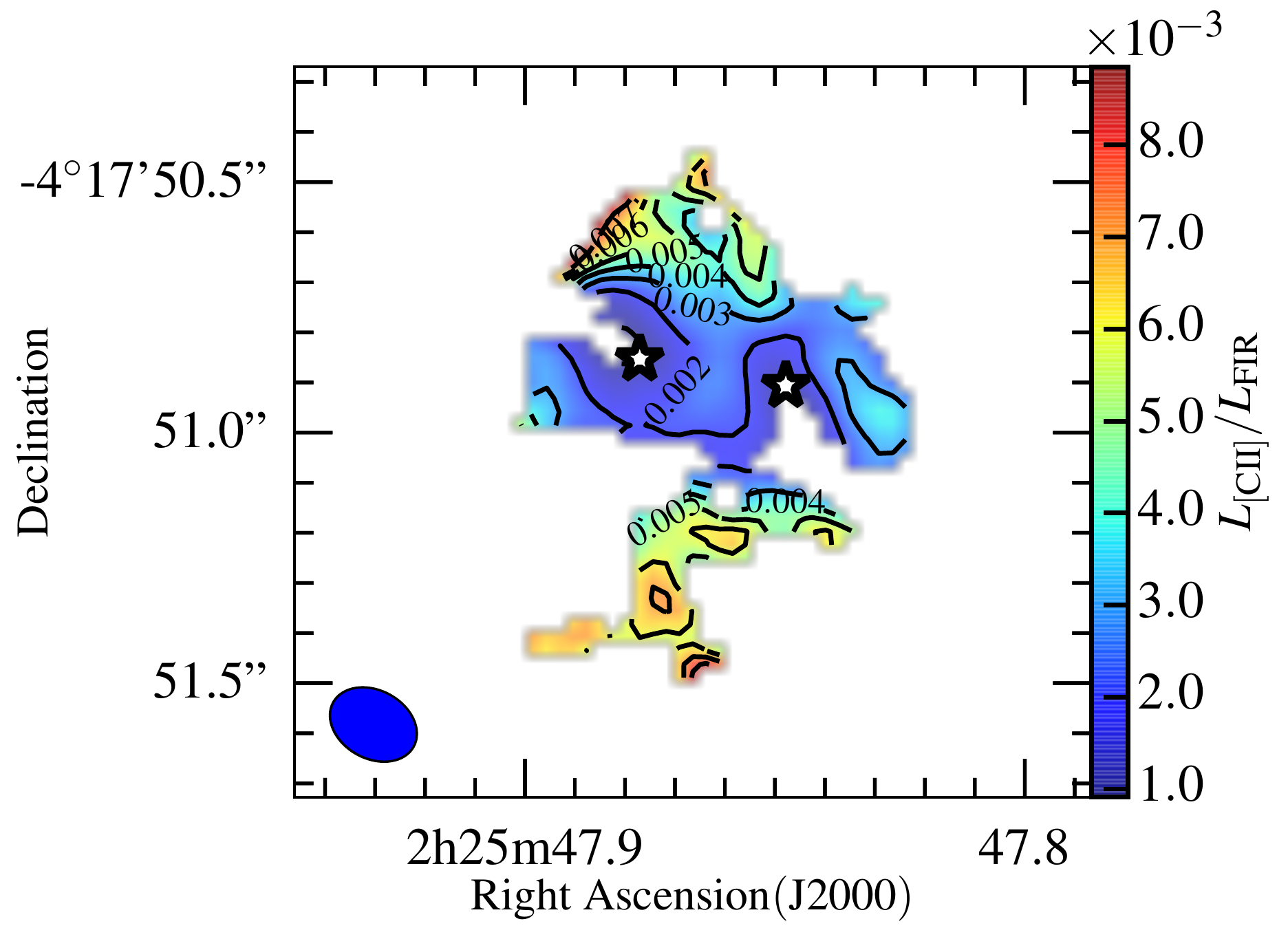}
\caption{\cii-to-FIR luminosity ratio map of X-Main (X-NE is outside the field of view shown here).
We clipped the \cii and continuum maps at a 3\sig\ level here.
The spatially resolved \LFIR is derived using the 635\,$\micron$
continuum image.  A negative gradient is observed toward the center of \xa
and is most likely caused by an increase in the radiation field intensity (see \Fig{ratio} and \Sec{sfrtracer}).
Star symbols mark the positions of XD1 and XD2.
\label{fig:ratiomap}}
\end{figure}

\begin{figure*}[tbph]
\centering
\includegraphics[width=0.47\textwidth]{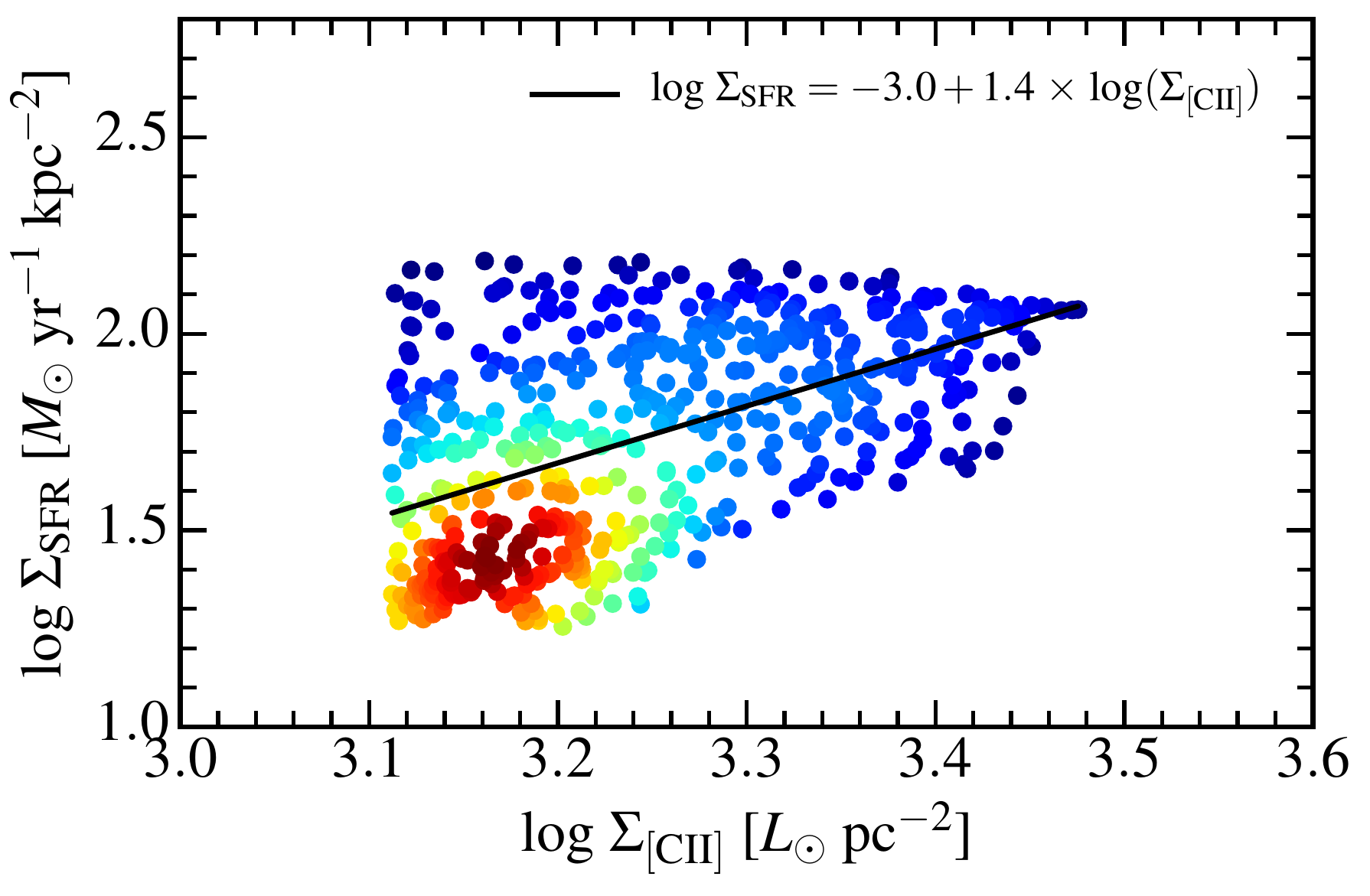}
\includegraphics[width=0.47\textwidth]{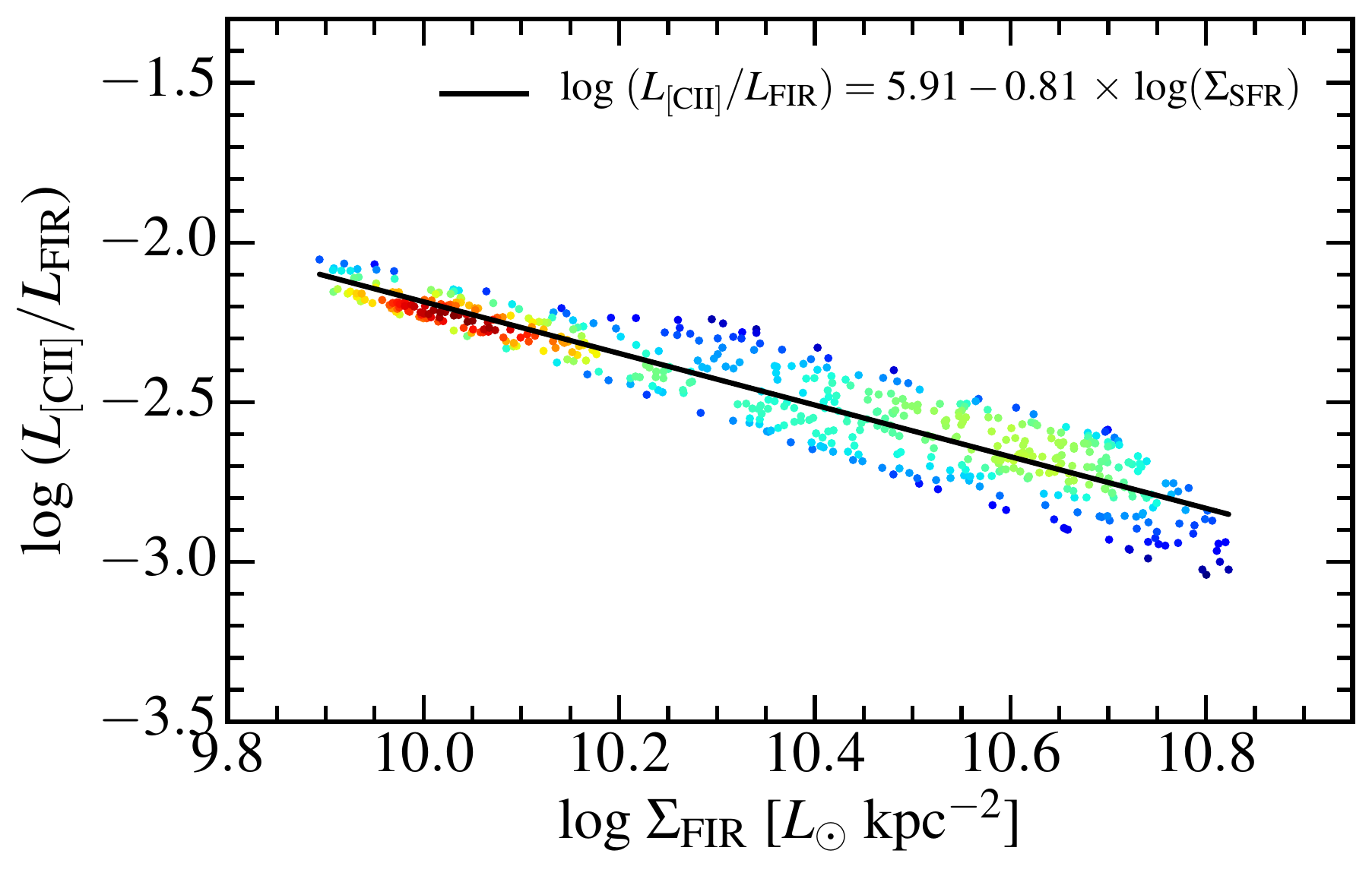}
\caption{
Relations based on a pixel-by-pixel analysis, scaling the quantities to their respective units shown.
Left: $\Sigma_{\rm SFR}$ and $\Sigma_{\rm [CII]}$ relation from pixel measurements from the
[C{\scriptsize II}]$/$635\,$\micron$ continuum ratio map (\Sec{sfrtracer});
a large scatter is seen between the two quantities.
Right: \Lcii/\LFIR as a function of \LFIR surface density ($\Sigma_{\rm FIR}$).
To derive this surface density map, we clipped both \cii and continuum maps at a 3\sig\ level.
Colors in both panels represent the density of points;
approximately bins of 50 points are correlated, based on the beam size of the data.
\label{fig:ratio}}
\end{figure*}

We investigate the spatially resolved \cii-to-FIR luminosity ratio in \xa on 1-kpc scale to examine the connection
between SFR and \cii surface densities.
To create the surface density plots in Figures~\ref{fig:ratiomap} and \ref{fig:ratio}, we clipped
both the \cii and the 635\,$\micron$ continuum maps at 3\sig\footnote{
We also test our results with less clipping (at 1\sig) to
confirm that the trends and relationships found are not artificial or biased because of ``excessive'' clipping.}.
Essentially, the notion of using the \cii luminosity as a proxy for SFR relies on the assumption
that \cii dominates the cooling budget of the neutral ISM,
in which heating is dominated by the photoelectric effect of UV photons from young and massive stars.
We show the $\Sigma_{\rm SFR}-\Sigma_{\rm \cii}$ relation for \xa in the left panel of \Fig{ratio}.
The large scatter suggests that \cii emission traces both
star-forming regions and ``diffuse'' gas reservoirs.
The trend of decreasing \Lcii/\LFIR at high $\Sigma_{\rm SFR}$ suggests that
the former is suppressed in compact, high-SFR surface density regions (\Fig{ratiomap}).
On the other hand, we find a tighter relation between
$\Sigma_{\rm \cii}/\Sigma_{\rm FIR}$ and $\Sigma_{\rm FIR}$, as shown in the right panel of \Fig{ratio}.

We fit power-laws of the forms $\Sigma_{\rm SFR}$\eq A$\Sigma_{\rm \cii}^N$ and
$L_{\rm \cii}/L_{\rm FIR}$\eq A$\Sigma_{\rm FIR}^N$ to our data for \xa, and
find the following best-fit relations:
\begin{equation}
\begin{split}
\log{\left(\frac{\Sigma_{\rm SFR}}{M_\odot~\textrm{yr}^{-1}~\textrm{pc}^{-2}}\right)} & = -3.0~(\pm 0.3)~+~\\
&  1.4~(\pm 0.1) \times \log{\left(\frac{\Sigma_{\rm \cii}}{L_\odot~\textrm{pc}^{-2}}\right)};
\end{split}
\label{eqn:e1}
\end{equation}
and
\begin{equation}
\log\left({\frac{L_{\rm \cii}}{L_{\rm FIR}}}\right) = 5.91~(\pm 0.14) - 0.81~(\pm 0.01)\times\log{\left(\frac{\Sigma_{\rm FIR}}{L_\odot~\textrm{kpc}^{-2}}\right)}.
\label{eqn:e2}
\end{equation}
The slope of the former relation coincidentally resembles the slope of the
Schmidt-Kennicutt relation for the \aco line on kpc scales, albeit with a large scatter \citep{Kennicutt98b}.
The slope of the latter relation
is steeper than those reported by \citet{Diaz-Santos13a, Diaz-Santos17a} for nearby ULIRGs, perhaps
due to the different tracers used.
The latter authors use \LIR and source size measured at 24\,$\micron$ to derive the FIR surface density,
whereas we use \LFIR and FIR size (or pixels) measured at rest-frame 158\,$\micron$ for \xa.
The steeper relation found in \xa can also be understood if we were to assume that its outer region,
where the \Lcii/\LFIR ratio is the highest, has a
lower metallicity (and thus, a higher photoelectric heating efficiency).
In any case, the tight $L_{\rm \cii}/L_{\rm FIR}-\Sigma_{\rm FIR}$ relationship
is consistent with the notion that the two quantities are connected through the the local FUV radiation field intensity.

We find that the \Lcii/\LFIR ratio of \xa decreases toward the center, as shown in the
spatially resolved map in
\Fig{ratiomap}. Such a negative gradient has been observed in nearby star-forming
galaxies \citep[e.g.,][]{Kramer13a, Smith17a}
and U/LIRGs (\citealt{Diaz-Santos14a}).
The deficit at the center may be explained by a higher dust temperature (see \Sec{cont}) and a
more intense radiation field at the center (given that $G_0\propto R^{-2}$).
In addition, the gas density at the center is likely to exceed the critical density of \cii,
where collisional de-excitation dominates and saturates the
\cii emission 
(\citealt{Luhman98a, Malhotra97a, Goldsmith12a}).
This effect also explains the decreasing $L_{\rm \cii}/L_{\rm FIR}$ ratio found with increasing FIR surface densities
in the $L_{\rm \cii}/L_{\rm FIR} - \Sigma_{\rm FIR}$ relation (\Fig{ratio}).

\subsection{\xa in the Context of High-$z$ Galaxy Populations}   \label{sec:others}
\xa is one of the most IR-luminous galaxies known at high redshift.
Given its IR luminosity of \LIR\,\eq4\E{13}\,\Lsun, it can be classified as a HyLIRG.
However, we find that its ISM properties differ from those observed in some other unlensed HyLIRGs studied to date.
For instance, both the gas and SFR surface densities of \xa are much lower than those observed in
GN20 and the $z$\eq5.7 binary HyLIRG ADFS 27 \citep{Hodge12a, Riechers17a},
but are comparable to those of the \z\eq2.4 HyLIRG merger HATLAS\,J084933
and the sub-regions of GN20 \citep{Ivison13a, Hodge15a}, suggesting that the \SF in \xa
is relatively modest compared to ``maximum''-starburst-like HyLIRGs.

Given the dynamical mass, stellar mass, and sSFR of \xa, it is among the most massive
galaxies known at $z$\eq3.
However, as discussed in \Sec{cont},
\xa was discovered with {\it Herschel}/SPIRE \obs at submillimeter wavelengths
and remains undetected in deep UV and optical observations.
It therefore differs from other \highz massive disk galaxy populations, such as those typically selected in the UV, optical, and NIR wavebands
by applying the $U$-, $B$-, $G$-, $R$-, $z$-, $K$-band color-selection and the Lyman Break ``dropout'' technique
(i.e., BzK, BM/BX, and LBG; \citealt{Steidel96a, Adelberger04a, Steidel04a, Daddi04b}), in that it has a larger dust content, which may
suggest different evolutionary histories for these \highz populations.

The molecular gas extent, kinematics, gas excitation, SFR,
dust mass, SFE, SFR surface density, and metallicity of \xa are similar to those of GN20.
This agreement suggests that \xa and GN20 may belong to the same class of DSFG.
The finding of such rare massive disk galaxies at $z$\ssim3 could be consistent with model predictions that disk-wide
\SF plays an important role for some of the
most massive DSFGs at early epochs, whether as a phase
in a merger event or \citep{Hayward13a} independent of
a major merger altogether\footnote{This statement does not explicitly address the relevance of merger activity in
the overall evolution of massive disks.}.

\section{Implications on the Formation Scenarios of \xa\ --- Major Merger and Cold Mode Accretion}     \label{sec:implication}

With a SFR of 2900\,\Msun\,yr\pmOne and a stellar mass of $10^{12}$\,\Msun,
distributed across a rotating disk 9\,kpc in diameter,
\xa is a massive rotation-dominated star-forming galaxy.
One of the main goals in studying \highz star-forming galaxies
is to examine and understand what drives their high SFRs.
A critical question concerns whether an interaction is required
to drive the high SFRs observed in \highz starbursting DSFGs --- which would put them in a transient phase ---
or whether DSFGs are just a massive galaxy population
undergoing ``quiescent'' star formation, but at higher rates due simply to their higher masses and/or gas mass fractions
compared to nearby and low-mass galaxies.
Previous theoretical and observational studies have suggested that \SF in the most massive
starburst-dominated DSFGs is likely triggered by major mergers, whereas less massive systems
could be triggered by
gravitational instability as a result of their high gas mass fractions
\citep[e.g.,][]{Chapman03b, Engel10a, Narayanan10a, Hayward11a, Hayward13a, Riechers17a}.

In cosmological N-body zoom-in and hydrodynamic simulations,
massive galaxies with stellar masses similar to that of \xa, 
albeit rare, can be formed quickly by $z$\eq6 
via multiple gas-rich major mergers
 \citep[][see also \citealt{Ruszkowski09a}]{Li07a, Dave10a}.
From a theoretical point of view, it is thus conceivable that \xa may have recently experienced a
major merger that would explain its broad CO lines, high SFR, large molecular gas mass,
and its 3\,kpc size double nuclei observed at 635\,$\micron$.
In this scenario, the double nuclei may correspond to two compact obscured \SB regions
triggered by massive gas inflows, or to the remnant cores of two similar mass progenitor galaxies \citep{Johansson09a}.
On the other hand, the observed spatial extent, velocity gradient, $G_0$, and gas surface density
observed in \xa are more consistent with a rotation-dominated ``normal'' star-forming galaxy, suggesting that additional mechanisms
may be at work to form a system like \xa.

In the standard model of dissipational disk formation,
infant disk galaxies form from the gas that is infalling into hierarchically growing dark matter halos.
However, since a substantial fraction of the angular momentum of gas is lost to the surrounding
halo through dynamical friction (up to 90\%) 
while it configures itself into a rotationally supported disk in the inner
portion of the dark matter halo, disks are an order of magnitude
smaller than those observed \citep[also known as the angular momentum ``catastrophe''; e.g.,][]{Steinmetz99a}.
In this formation paradigm, a massive extended disk like \xa is quite unexpected
at $z$\eq3 (only about 2\,Gyr after the Big Bang).
While feedback and the continuation of tidal torquing and
accretion of satellite galaxies/minor mergers
have been proposed to resolve the disagreement between models and observations, as they
can prevent the gas from over-cooling and losing its angular momentum
\citep[e.g.,][]{Sommer-Larsen01a, Robertson04a, Scannapieco08a, Zavala08a}, it remains
unclear whether minor mergers\footnote{Major mergers would take a few Gyr to form
an extended disk again from two progenitor disks, if ever \citep[e.g.,][]{Governato09a}.}
alone could increase the angular momentum sufficiently to explain the properties and number density
of disk galaxies observed
(e.g., \citealt{Vitvitska02a}).

In recent years, the cold mode accretion formation model has been put forward as an alternative mechanism capable of driving the
high SFRs seen in \highz gas-rich star-forming galaxies,
which may also explain the discrepancy with major mergers
(i.e., there are not enough major mergers in models to explain all DSFGs as merger-driven starbursts; \citealt{Keres05a, Dekel09a, Dekel09b, Dave10a}; see also
\citealt{Narayanan15a} and \citealt{Lacey16a}).
Since cold streams 
can provide additional angular momentum, extended gas-rich disk galaxies with kpc-scale \SF can be explained naturally
\citep[][]{Keres05a, Dekel09a}. 

Given that some properties of \xa are consistent with the major merger scenario and others are
consistent with
the cold mode accretion scenario, it is conceivable that both
mechanisms together are important to give rise to a galaxy like \xa, which perhaps is similar to the
case of GN20 (see \citealt{Carilli10a}).

\section{Summary and Conclusions} \label{sec:sum}
We determine the redshift and gas excitation of the
{\it Herschel}-selected DSFG \xa at \z\eq2.9850\pmm0.0009
by observing its CO($J$\eq1\rarr0; 3\rarr2; 5\rarr4; 10\rarr9) line emission.
We image its gas reservoir and dust-obscured \SF on 1.2 kpc scales using
\cii line and dust continuum emission.

We detect a companion galaxy (hereafter X-NE) about 20\,kpc NE of the main component of \xa
(hereafter X-Main) in \aco and \cii line emission at a redshift close to X-Main ($\Delta v$\eq$-$535\pmm55\,\kms).
X-NE is also detected in the UV, optical, and NIR continuum emission.
Based on the \aco line flux, we infer a
total molecular gas mass of $M_{\rm gas}^{\rm total}$\eq(2.12\pmm0.71)\,$\times$\,(\alphaco/0.8)\E{11}\,\Msun residing in the \xa system (composed of X-NE and X-Main),
yielding a gas mass fraction of $f_{\rm gas}^{\rm dyn}$\eq33\pmm15\%.

Based on the \aco and \cii line data, the velocity structure of X-Main is consistent with a rotating disk, with a diameter of $\sim$9\,kpc.
Thus, the gas reservoir of \xa is more extended than those typically observed in \highz
DSFGs and quasars, but comparable to those observed in \highz ``main-sequence'' galaxies and the $z$\eq4 starburst galaxy GN20 \citep{Carilli10a, Hodge15a}.   
We find that the widths of its CO($J$\eq1\rarr0; 10\rarr9) and \cii lines are broader
than those typically observed in ``normal'' star-forming galaxies, ULIRGs, and \highz SMGs, but comparable to
those observed in the more extreme systems (e.g., J13120$+$4242 and G09v124; \citealt{Riechers11a, Ivison13a}).
We find that the overall
gas excitation of \xa resembles that of the nearby galaxy merger Arp\,220.
The shape of the CO excitation ladder (i.e., SLED) suggests that the molecular ISM of \xa may consist of (at least) two
gas phases --- a diffuse extended cold component and a dense compact warm component.

The X-Main component of the \xa system
remains undetected in deep UV and optical \obs, indicating that it is highly dust obscured.
We find a pair of compact dust components (XD1 and XD2) in the dust continuum emission at 635\,$\micron$,
which are about 3\,kpc across each and are separated by 2\,kpc.
The pair is embedded within an extended dust component, which also appears to be as extended
as the \aco and \cii line emission.
The brightness temperatures of the nuclei suggest that they may be
warmer, more optically thick, and/or
with higher beam filling factors than the extended dust component.
We find that the source-averaged FUV radiation field intensity of \xa is around 200 times stronger than
that of the local Galactic interstellar medium,
but is comparable to those observed in nearby star-forming galaxies and other DSFGs.
The PDR properties of \xa together with its gas properties and excitation are indicative of galaxy-wide \SF,
consistent with its extended gas and dust emission observed
(as opposed to those typically observed in compact starburst galaxies).

We find a stellar mass of \mstar$\simeq$\,10$^{12}$\,\Msun and an SFR of $\simeq$\,2900\,\Msun\,yr\pmOne
for \xa from SED modeling, consistent with it being one of the most massive star-forming galaxies at $z$\eq3.
We also find source-averaged SFR and molecular gas surface densities of
$\Sigma_{\rm SFR}$\eq10\,$-$\,60\,\Msun\,yr\pmOne\,kpc$^{-2}$ and
$\Sigma_{\rm gas}$\eq590\,$\times$\,$($\alphaco$/0.8)$\,\Msun\,pc$^{-2}$.
Thus, \xa lies along the ``starburst sequence'' of the Schmidt-Kennicutt relation \citep[e.g.,][]{Bouche07a},
similar to the
sub-regions of GN20 and the $z$\ssim2.6 SMG SMM J14011$+$0252
(\citealt{Sharon13a, Hodge15a}). This locus corresponds to an elevated SFE
compared to
other SFMS galaxies.
The SFR surface densities for the double nuclei are elevated compared to those observed in
the circumnuclear starburst regions of nearby galaxies \citep{Kennicutt98b}, but are much lower than those observed in other
(not strongly lensed) \highz HyLIRGs \citep[``maximum starbursts''; e.g., ][]{Riechers13a, Riechers14a, Riechers17a, Oteo17b}.

A large scatter seen in the $\Sigma_{\rm SFR} - \Sigma_{\rm \cii}$ relation for \xa on 1\,kpc scale
suggests that its \cii emission traces both star-forming regions and ``diffuse'' gas reservoirs.
We find a tighter relation between \Lcii /\LFIR and $\Sigma_{\rm SFR}$ across \xa, which
is consistent with our understanding that the two quantities are connected through the local FUV radiation field intensity \citep[e.g.,][and references therein]{Tielens85a}.
We find that the \Lcii/\LFIR ratio is ``suppressed'' at high SFR surface densities
(e.g., near the center of \xa), which is suggestive of a stronger UV radiation field and warmer dust emission there.
On the other hand, the source-averaged \Lcii/\LFIR ratio of \xa is comparable to those of nearby star-forming galaxies and
LIRGs rather than nearby ULIRGs and quasars, despite its two orders of magnitude higher \LFIR.

The scatter observed in the spatially resolved and galaxy-integrated \cii and FIR luminosity relations for \xa
are consistent with our understanding that \Lcii and SFR are not related linearly.
The spatially resolved data presented in this paper thus confirm the speculation put forward by
\citet{Stacey10a} based on unresolved \obs: that \highz DSFGs are not simple scaled-up ULIRGs and that
\SB-dominated DSFGs can be much more extended than ULIRGs,
which is also consistent with previous findings of spatially extended CO emission \citep[e.g.,][]{Riechers11a,Ivison11a}.

While rotationally-supported clumps may yield velocity 
    gradients (\Sec{clump}), we find no evidence of such with the data at hand, in spite of the 
    pair of dust peaks identified. Even in 
    the merging clump scenario (e.g., in late stage merger), it is unlikely for the clumps 
    to have a huge impact on the global scale across the entire galaxy as to cause a 
    monotonic velocity gradient over $\sim$9\,kpc across, 
    especially given the observed centrally peaked velocity 
    dispersion observed in the \cii data, which is relatively uniform outside the central $\sim$1.2\,kpc\footnote{
    Approximately the beam size.}.
    Another piece of evidence disfavoring \xa from being strictly a dispersion-dominated merger 
    system comes from the fact that the potential merger candidates 
    (the pair of dust peaks) are oriented almost-perpendicular to the 
    velocity gradient. We further quantified the disk-like kinematics of 
    \xa based on the higher order Fourier coefficients of the harmonic 
    decomposition (\Sec{rotcur}), which are found to be insignificant compared 
    to the $m$\eq0 term. We thus interpret \xa to be a rotating disk\footnote{Note that 
    this does not rule out the possibility that the disk is part of a merger.}.
    
The disk-like kinematics, extended \SF, high SFR and \mstar, and
gas and SFR surface densities of \xa
are quite similar to those of GN20 \citep{Hodge12a, Hodge15a}, suggesting that
the two may correspond to  the same class of DSFG --- massive extended rotating disks with highly dust-obscured \SF.

\xa can be classified as a HyLIRG, making it one of the most IR-luminous galaxies known.
In a sample of the brightest \highz DSFGs discovered in the 95\,deg$^2$ surveyed by HerMES,
only around 10\% appear to be intrinsically comparably luminous, corresponding to a surface density of only 0.03\,deg$^{-2}$ \citep{Bussmann15a}.
In fact, the stellar mass function also suggests that massive galaxies like \xa are very rare at $z$\eq3 (\citealt{Dave10a, Schreiber15a}).
In the framework of the hierarchical formation model, one would expect a
massive galaxy like \xa to form via major mergers, given its high SFR and \mstar.
The two compact dust nuclei and enhanced central velocity dispersion as well as the detection of a companion galaxy at only 20\,kpc away may be consistent
with such a scenario.
However, its extended massive gas disk, monotonic velocity gradient, $G_0$, and gas and SFR surface densities at $z$\eq3
suggest additional mechanisms such as proposed in the cold mode accretion model may also play an important role
in shaping the existence and subsequent evolution of massive DSFGs.
\xa could thus be a rare example of such system showing both mechanisms at play.

\acknowledgements

We thank the referee for providing detailed and constructive comments that have significantly improved the clarify of this manuscript.
We thank Mark Lacy and the data analysts at the North American ALMA Science Center (NAASC)
for assistance with the ALMA data reduction.
T.K.D.L. thanks Amit Vishwas and Drew Brisbin for helpful discussions,
Carlos G\'omez Guijarro for assistance with the IRAC flux extraction and setting up the \ncode{galfit} software,
and Gregory Hallenback and Luke Leisman for helpful discussions on dynamical modeling.
We thank Shane Bussmann for leading a proposal to obtain some of the data presented in this paper.
T.K.D.L. acknowledges support from the NSF through award SOSPA4-009
from the NRAO and support from the Simons Foundation.
D.R. acknowledges support from the NSF under grant
number AST-1614213 to Cornell University.
A.J.B acknowledges support from NSF grant AST-0955810, to Rutgers, The State University of New Jersey.
I.P.-F. acknowledges support from the Spanish research grants ESP2015-65597-C4-4-R and ESP2017-86852-C4-2-R 
J.L.W acknowledges support from an STFC Ernest Rutherford Fellowship (ST/P004784/1 and ST/P004784/2), 
and additional support from STFC (ST/P000541/1).
The Flatiron Institute is supported by the Simons Foundation.
This work is based on observations carried out
with the IRAM PdBI Interferometer. IRAM is supported by INSU/CNRS (France), MPG (Germany), and IGN (Spain).
Support for CARMA construction was derived from the Gordon and Betty Moore
Foundation, the Kenneth T. and Eileen L. Norris Foundation, the James S.
McDonnell Foundation, the Associates of the California Institute of
Technology, the University of Chicago, the states of Illinois, California, and
Maryland, and the National Science Foundation.
CARMA development and operations were supported by the National Science Foundation under a
cooperative agreement and by the CARMA consortium universities.
The National Radio Astronomy Observatory is a facility of the National Science
Foundation operated under cooperative agreement by Associated
Universities, Inc.
This publication makes use of data products from the {\it Wide-field Infrared Survey Explorer},
which is a joint project of the University of California, Los Angeles, and the Jet Propulsion Laboratory/California
Institute of Technology, funded by the National Aeronautics and Space Administration.
This paper makes use of the following ALMA data:
ADS/JAO.ALMA\# 2016.2.00105.S;
ADS/JAO.ALMA\# 2013.1.00749.S; and
ADS/JAO.ALMA\# 2011.0.00539.S. ALMA is a partnership of ESO (representing its member states), NSF (USA) and NINS (Japan), together with NRC (Canada) and NSC and ASIAA (Taiwan), in cooperation with the Republic of Chile. The Joint ALMA Observatory is operated by ESO, AUI/NRAO and NAOJ.
This research has made use of NASA's Astrophysics Data System Bibliographic
Services.
The Submillimeter Array is a joint project between the Smithsonian Astrophysical Observatory and the Academia Sinica Institute of Astronomy and Astrophysics and is funded by the Smithsonian Institution and the Academia Sinica.
This work is based in part on observations
made with the NASA/ESA {\it Hubble Space Telescope}, and obtained from the {\it Hubble}
Legacy Archive, which is a collaboration between the Space Telescope Science
Institute (STScI/NASA), the Space Telescope European Coordinating Facility
(ST-ECF/ESA), and the Canadian Astronomy Data Centre (CADC/NRC/CSA).
This research made use of \ncode{Astropy}, a community-developed core Python package for Astronomy \citep{astropy}.
This research made use of \ncode{APLpy}, an open-source plotting package for Python hosted at \url{http://aplpy.github.com}.
This work is based on observations made with the {\it Spitzer Space Telescope}, which is operated by the Jet Propulsion Laboratory, California Institute of Technology under a contract with NASA.
This work is also based on observations obtained with the MegaPrime/MegaCam instrument, a joint project of CFHT and CEA/IRFU, at the Canada-France-Hawaii Telescope (CFHT), which is operated by the National Research Council (NRC) of Canada, the Institut National des Science de l'Univers of the Centre National de la Recherche Scientifique (CNRS) of France, and the University of Hawaii. This study is also based in part on data products produced at Terapix available at the Canadian Astronomy Data Centre as part of the Canada-France-Hawaii Telescope Legacy Survey, a collaborative project of NRC and CNRS.
Based on observations made with the NASA {\it Galaxy Evolution Explorer}.
{\it GALEX} is operated for NASA by the California Institute of Technology under NASA contract NAS5-98034.
This work uses data based on observations obtained with
the {\it XMM-Newton}, an ESA science mission with instruments and contributions directly funded by ESA Member States and NASA.
%
Facilities: IRAM PdBI, CARMA, VLA, {\it HST}(WFC3), SMA, ALMA, {\it Spitzer} (IRAC, MIPS), {\it WISE}, {\it Herschel}(PACS, SPIRE), VISTA, CFHT(MegaCam), {\it XMM-Newton}(EPIC), {\it GALEX}

\bibliographystyle{apj}
\bibliography{ref}

\appendix

\section{CO and \cii Channel Maps} \label{sec:channel}
\begin{figure*}[tbph]
\centering
\includegraphics[trim=0 0 0 0, clip, width=0.75\textwidth]{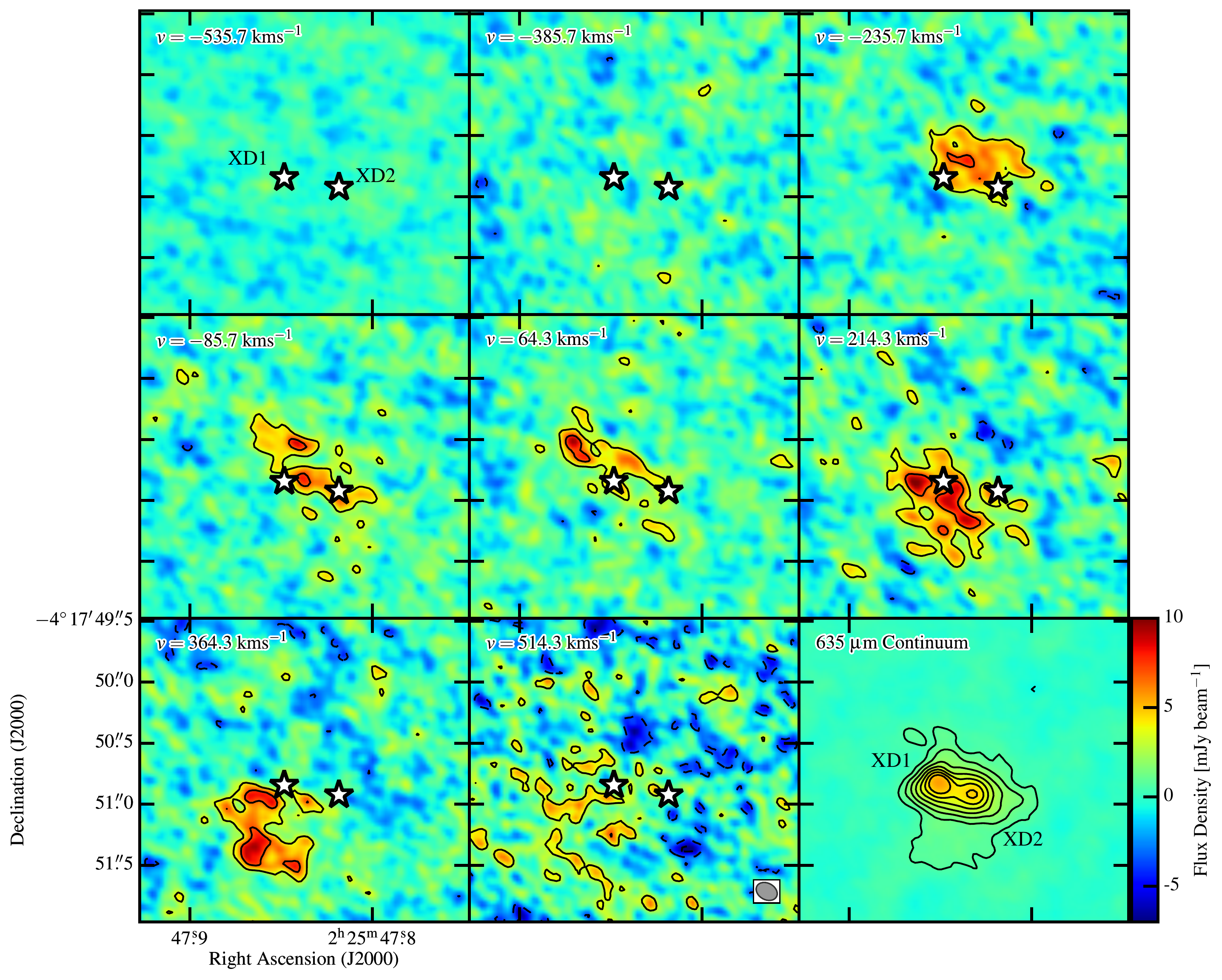}
\caption{Channel maps of the \cii line emission in X-Main shown at
full spatial resolution with velocity bins of $\Delta v$\eq150\,\kms. Last panel shows the 635\,$\micron$ continuum underlying
the \cii line. The central velocity of each panel is indicated at the upper left corner.
Contours are shown in steps of [$-$3, 3, 6, 9]\,$\times$\,$\sigma_{\rm ch}$, where $\sigma_{\rm ch}$\eq1.15\,mJy\,\bmm
(0.22\,mJy\,\bmm for the continuum).
The last channel (i.e., second last panel) is dominated by noise since it is
near the edge of a spectral window, where an atmospheric feature is present.
Black markers indicate the positions of the 635\,$\micron$ dust peaks (XD1 and XD2; see last panel).
Synthesized beam size is shown in the lower right corner of the second last panel (same as the leftmost panel in \Fig{mom0}).
X-NE is outside the field-of-view shown here.
\label{fig:chan}}
\end{figure*}

\begin{figure*}[tbph]
 \centering
\includegraphics[width=0.75\textwidth]{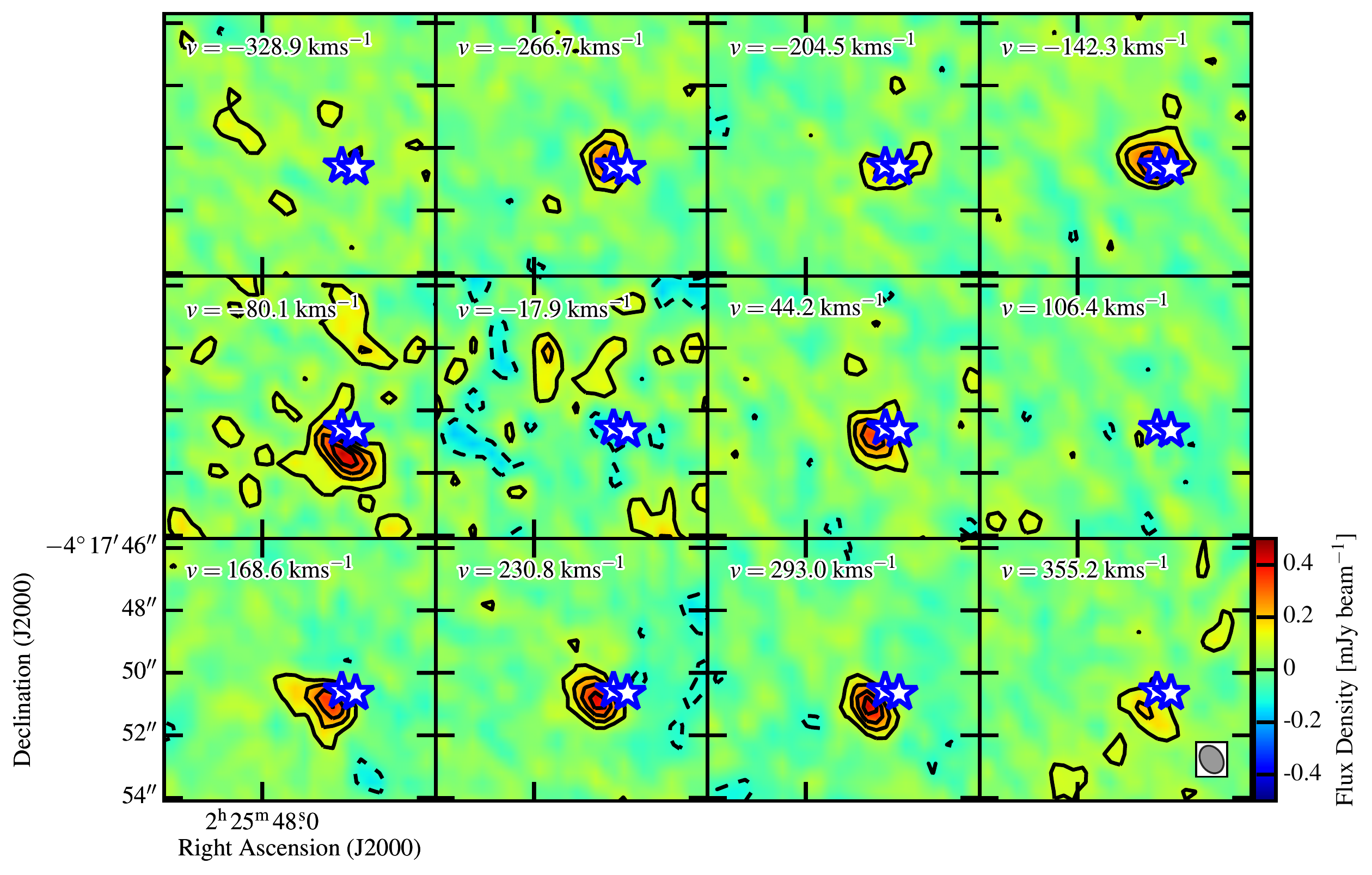}  
\caption{Channel maps of \aco emission imaged with Briggs weighting, covering a
velocity range of $\Delta v$\,$\in$\,[$-$626, 712] \kms. 
Number in the upper left corner of each panel indicates the central velocity $v_{\rm LSR}$ of each map, where
the emission is integrated over $\Delta v$\eq145\,\kms.
The \aco emission is marginally spatially resolved.
The emission centroid shifts from NW to SE with increasing velocity.
Contours are shown in steps of [$-$3, 3, 4, 5, 6, 9]$\times$\,$\sigma_{\rm ch}$,
where $\sigma_{\rm ch}$\eq 0.031\,mJy\,\bmm. The star symbols indicate the positions
of the two compact dust components detected at 635\,$\micron$ (rest-frame 158\,$\micron$; XD1 and XD2; see last panel of \Fig{chan}).
The synthesized beam is shown in the lower right corner of the last panel (0\farcs94\,$\times$0\farcs71 at PA\eq31$\degr$).
\label{fig:co10chan}}
\end{figure*}

Since we are investigating the gas kinematics, it is essential to show and acknowledge
the limited significance of the detected signal {\em per velocity bin}. We show the channel maps for the
\aco and \cii lines in Figures~\ref{fig:chan} and \ref{fig:co10chan}.
In the \cii maps, structures on the scale of the angular resolution ($\lesssim1.2$\,kpc) are seen, but at low S/N significance.
We therefore do not discuss the properties of potential star-forming ``clumps''/structures in this paper.
Exploring such direction with higher resolution and better sensitivity data would be useful to better understand the physics behind
the high SFR of \xa.

As noted in \Sec{rotcur},
a drop off is seen in the rotation velocity beyond a radius of $R$\eq6\,kpc (in \Fig{gipsy}).
This is most likely a result of the limited S/N in the reddest velocity channels, as illustrated in \Fig{chan}.

\section{Non-detection of X-Main at UV/optical Wavebands} \label{sec:uv}
\begin{figure}[htbp]
 \centering
\includegraphics[width=0.4\textwidth]{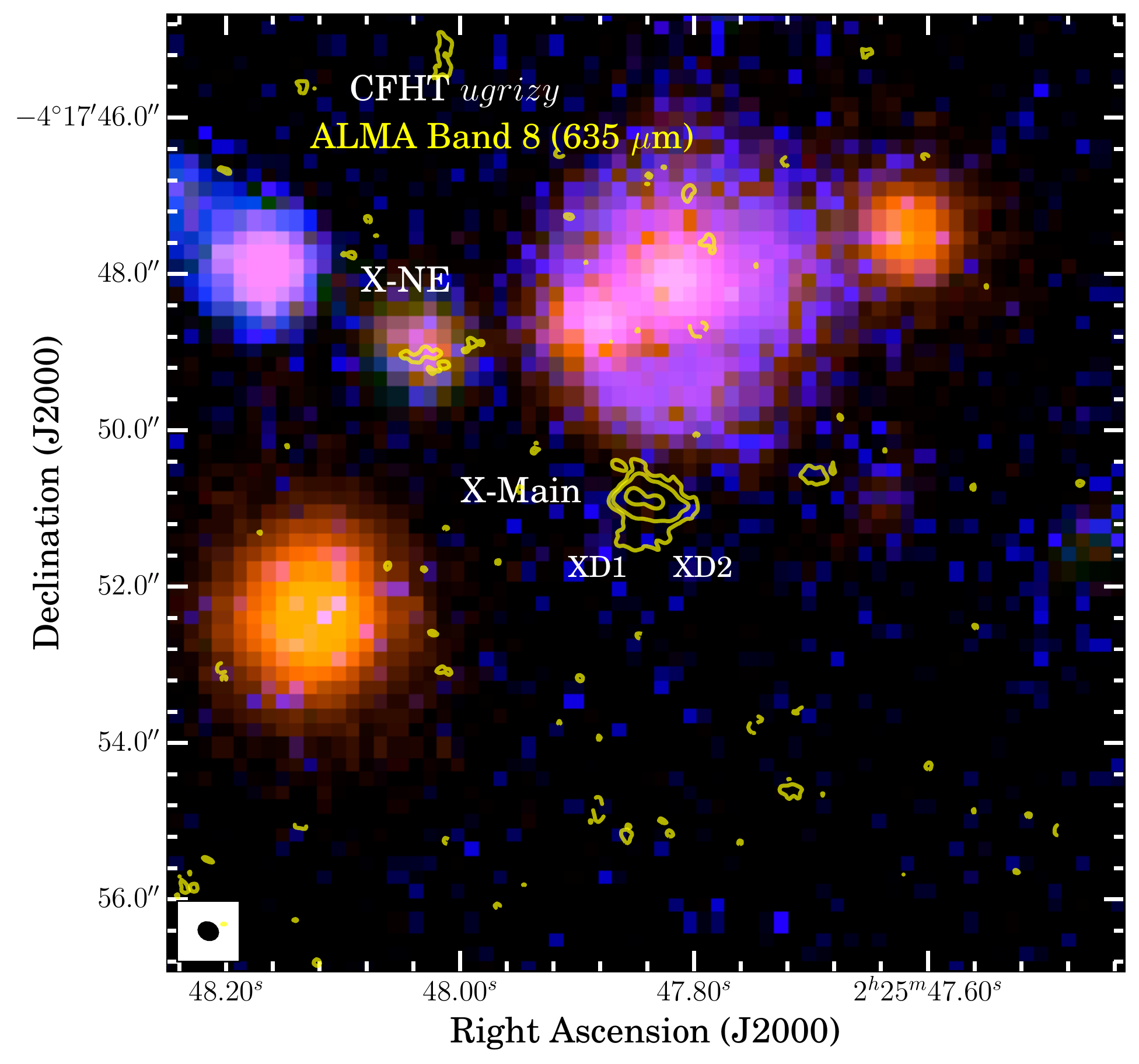}
\caption{
ALMA 635\,$\micron$ continuum emission (yellow contours)
overlaid on UV/optical/IR images: $u^*$- (blue), $g'r'i'$- (green), and $z'$-bands (red) obtained with the CFHT at 0\farcs8 resolution.
Contours are shown in steps of [$-$3, 3, 6, 18]$\times$$\sigma_{\rm 635}$, where $\sigma_{\rm 635}$\eq0.22\,mJy\,\bmm.
The synthesized beam for the ALMA data is shown in the lower left corner.
The main component of \xa (X-Main) remains undetected, whereas X-NE is detected in the UV/optical/NIR wavebands and in the
\aco and \cii lines (see Figures~\ref{fig:co10spec} and \ref{fig:mom0}).
 \label{fig:cont}}
\end{figure}

As shown in the RGB image created from {\it Spitzer}/IRAC 4.5 (blue), 5.8 (green), and 8\,$\micron$ (red) data (\Fig{newcont}),
emission detected at 4.5\,$\micron$ is dominated by foreground sources (see also \Fig{galfit}), but emission at 5.8 and 8\,$\micron$ is dominated by \xa.
We therefore model the surface brightness profiles of the sources near \xa
based on their morphologies seen in the CFHT and VISTA images in order to de-blend the emission observed at
3.6 and 4.5\,$\micron$ (see Appendix~\Sec{galfit}).

On the other hand, X-NE is detected in the UV, optical, and NIR wavebands (as shown in Figures~\ref{fig:newcont} and \ref{fig:cont}).
As discussed in \Sec{results}, this component is also detected in \aco and \cii line emission (see Figures~\ref{fig:co10spec} and \ref{fig:mom0}).
With the available data, we cannot discriminate and obtain reliable constraints on the stellar 
masses and SFRs for X-NE and X-Main separately. We thus infer the properties of the system as 
a whole in \Sec{diss} and subsequent sections.
That said, optically-selected \highz sources (e.g, 
BzKs, LBGs) appear to be different populations from these highly dusty 
starburst galaxies (possibly due to different evolutionary stages), and 
surveys done at only one wavelength are likely to miss other high-z 
candidates in the field. Given that X-NE is optically visible, and  
thus may have less dust than X-Main, it may be a young nearby galaxy 
soon to be engulfed by X-Main. We report the pair's gas mass ratios in 
\Tab{phy}. 
More \obs will be useful to better understand the physical properties of 
X-NE, and thus, its nature in relation to X-Main in the 
HXMM05 system.

\section{De-blending {\it Spitzer}/IRAC Emission} \label{sec:galfit}
\begin{figure}[htbp]
\centering
\includegraphics[width=0.5\textwidth]{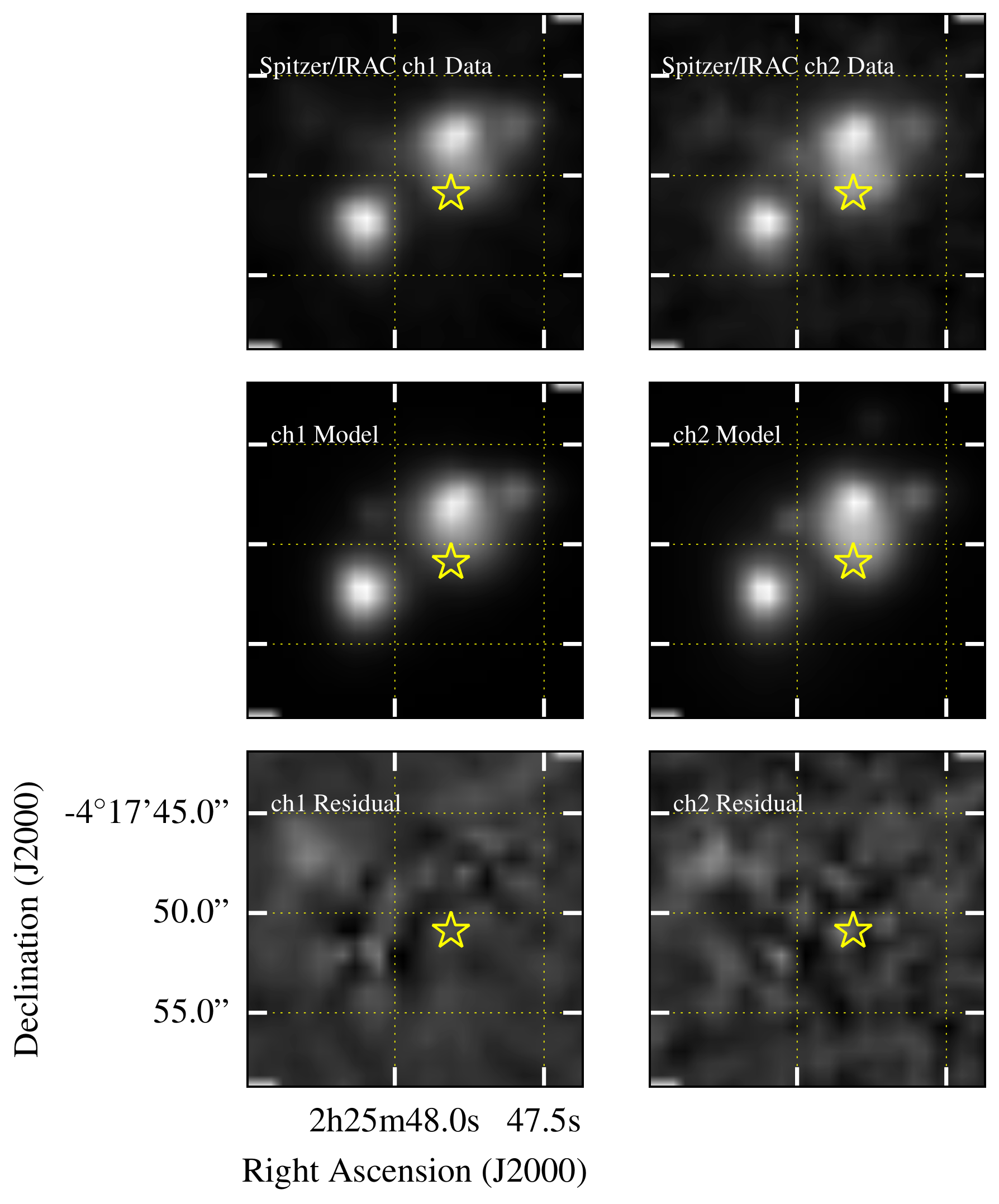}
\caption{Top: {\it Spitzer}/IRAC images at 3.6 (ch1) and 4.5\,$\micron$ (ch2).
Middle: \ncode{galfit} models.
Bottom: Residual maps, showing that \xa remains undetected after de-blending.
Yellow symbols indicate the position of the \xa system.
\label{fig:galfit}}
\end{figure}

Multiple sources are detected near \xa
at 3.6 and 4.5\,$\micron$ (channels 1 and 2; Figures~\ref{fig:cont} and \ref{fig:galfit}).
We examine whether part of the emission detected at 3.6 and 4.5\,$\micron$ may arise from \xa by
using the publicly available software \ncode{galfit} \citep{Peng02a} to de-blend the emission.
We initialize the fitting parameters based on the positions, brightnesses, and morphologies
of the sources near \xa as observed in the higher resolution NIR images
({\it HST}/WFC3 F110W, VISTA, and CFHT; see e.g., \Fig{cont}).
We use a total of six components and a sky background
to account for all the emission detected in the high resolution NIR images.
We model the surface brightness distributions of the two brightest components using Sersic profiles, each with seven free
parameters: $x$, $y$, $I$, $R_e$, $n$, $b/a$, and PA, where $x$ and $y$ describe the position of the component,
$I$ is the integrated flux,
$R_e$ is the effective radius, $n$ is the Sersic index, $b/a$ is the axial ratio, and PA is the position angle.
We model the remaining four components as point sources, for which
we adopt the point response functions (PRF), described by three free parameters $x$, $y$, and $I$ per source.
We allow all parameters to vary without imposing any priors in order to avoid biasing the best-fit parameters.
The PRFs
are adopted from the IRAC calibration routines\footnote{\url{http://irsa.ipac.caltech.edu/data/SPITZER/docs/irac/calibrationfiles/psfprf/}}.
We do not detect any statistically significant emission at the position of \xa in the residual maps (\Fig{galfit}).
We thus adopt the SWIRE survey depths at the two IRAC wavebands
as 3$\sigma$ upper limits (\Tab{photometry}).

\end{document}